# Voynich Manuscript (The "Book of Dunstan") coding and decoding methods.

Alexander G. Ulyanenkov, alex.ulyanenkov@protonmail.ch

Abstract: The Voynich manuscript (VMS) is the book dated as 15th - 16th century, written using specific and smart coding methods. This article describes the methods how it was analyzed and how coding keys were found. It also shows the VMS author's method of coding.

## 1. Introduction

This is an ancient manuscript, familiar to many researchers of medieval riddles under the names "Voynich Manuscript" (also known in abbreviation – **VMS**, which I will use in the article), it was also called "The book, which no one can read," because neither the language used for writing it or its author, nor place, nor even the date of writing was unknown.

Its current name the manuscript received from the name of one of the last owners - Michael Voynich (1865 -1930) - Polish revolutionary (pseudonym "Wilfred"), bibliophile and antiquarian, husband of well known writer Ethel Lilian Voynich.

Voynich acquired the mysterious manuscript for his collection in 1912. And although that did not affected the success in deciphering the manuscript, but finally influenced its future fate - in 1959 Voynich family heirs sold it to bookseller Hans Kraus, who in 1969 gave it to Yale University Benecke rare book library, where it is still being held.

On the official website of Benecke library of rare books of Yale University [1] the manuscript description is the following:

Call Number: Beinecke MS 408 (Request the physical item to view in our reading room)

Alternate Title: Voynich Manuscript
Date: [ca. 1401-1599?]
Genres: Manuscripts
Botanical illustrations
Astronomical charts
Drawings
Hand coloring
Illustrations

Type of Resource: mixed material

Description: Parchment. ff. 102 (contemporary foliation, Arabic numerals; not every leaf foliated) + i (paper), including 5 double-folio, 3 triple-folio, 1 quadruple-folio and 1 sextuple-folio folding leaves. 225 x 160 mm.
Abstract: Scientific or magical text in an unidentified language, in cipher, apparently based on Roman minuscule characters.

Physical Description: 1 vol.
color illustrations
23 x 16 cm. (binding)

The main text and illustrations look like image on Fig.1 (all the images of VMS pages were taken from VMS web-page of the library of rare books Benecke at Yale University [1])

The most of plants on VMS pages are not exists in the forms represented on VMS pages.

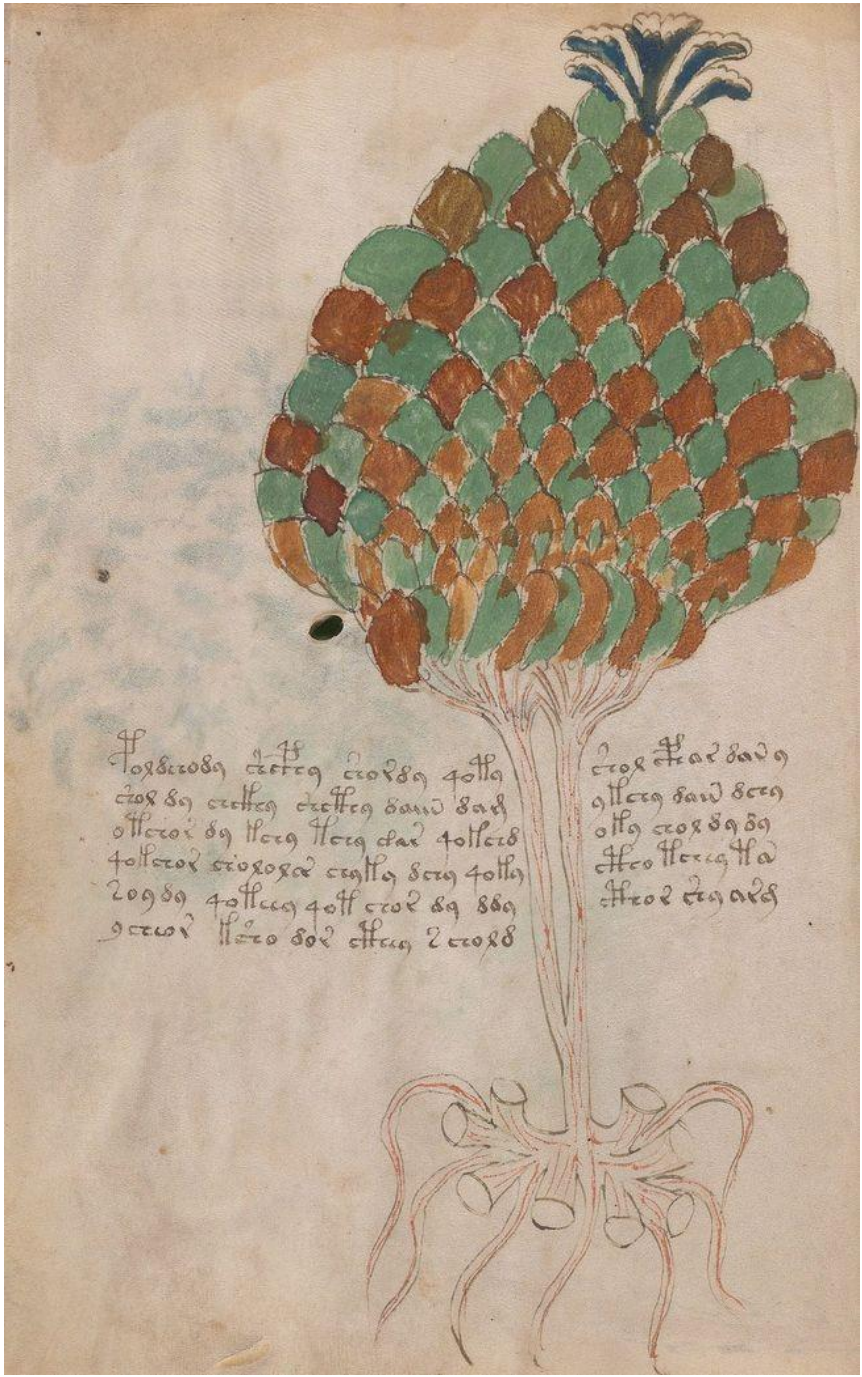

Fig. 1  The typical VMS image of page containing plant (VMS page 11v in accordance to the library page identification).

## 2. Initial approach and assumptions for the VMS analysis.

1) **No stereotypes** - After reading some corresponding to VMS articles the decision was to start from "zero" – to try to avoid any stereotypes which can finally lead to critical errors. Because of that you will not find a lot of citing links in this article.

2) **Trust you own eyes** - what you can see from the initial VMS data:

- VMS is well structured (you can see botanical, astronomical, anatomical and pharmaceutical data)- something similar to Avicenna's Canon
- it consists of encrypted and non-encrypted text
- it contains allegory images
- it contains Arabic numerals
- Written by clear and stable handwriting and using at least 4 different handwriting fonts – 1x stenography like (for main text), 1x Latin gothic (in Astronomical part for zodiac constellations marking), 1x italic (for ciphering), 1x – ancient (probably "black English" like used for hornbooks - also for ciphering) - all are present on the 4r page (in accordance with library page identification).
- It consist of the text in different colors
- VMS is written on most expensive type of parchment

In more details - the manuscript is visually divided onto the following sections:

 - Botanical - this is one of the most voluminous chapters. It contains the description and the drawings of plants, the vast majority of which can't be recognized even approximately.

 - Astronomical - contains charts on which the symbols of the sun, the moon, stars and zodiac constellations are used. Zodiacal part consists of 12 related images, among which there are no signs of Capricorn and Aquarius, but twice repeated signs of Aries and Taurus. In addition, the zodiacal section contains signatures in Gothic letters in the Latin alphabet, which are relatively easy to read and which relates to the corresponding months.

   - Biological - a chart with allegorical images of human organs - in the form of tanks and channels (swimming pools etc), filled with different colored liquids. In the channels and in pools there are strange dames, or, as some call them - nymphs.

- Cosmological - Contains pictures and diagrams which are not fully understandable.

- Pharmaceutical - section containing plants reduced drawings on the background of some hypothetical pharmaceutical vessels, which may indicate the section as a guide for the compilation of certain herbal mixtures.

- Prescription - the last section, consisting of a rich set of short text paragraphs, the beginning of each is indicated by the image of a star. There are no other pictures in the section.

The first my assumption was that VMS is not a forgery.

If so – that means that the VMS author (there were no computers in the time when he lived) had the opportunity to simply read without necessity to use the special complicated cheats.

If so – the VMS author used rather simple coding method.

If so – the VMS author possibly left some hints for himself and for other dedicated readers

### 3. Basic tasks for codes & keys investigation.

1) What ciphering methods were used in the time when VMS was written?

At first let's pay our attention to the VMS dating. The radiocarbon analysis of the VMS parchment was done by University of Arizona and the result shows period 1404 – 1438 with the probability of 95% [2].

**Important note!** We should not mix the dating for parchment and for manuscript, because parchment analysis gives us only the information when the animals, whose skins were used for parchment manufacturing, were killed. That Arizona analysis tells us only that manuscript can not be written earlier than in the beginning of 15th century.

According to my assumption it will be necessary to extend the probable period when VMS was written – at least till ~ 1500.

This assumption re 1500 was based on the information of average period of maximal aging of quality medieval paper before it was used – 10-15 years. That info has been found on the e-forums of experts who investigating the watermarked paper.

As we are talking about parchment – the more stable material – the maximum aging can be much larger. I used 45-50 years, because not found any corresponding info re parchment aging. But if we will compare the preservation of medieval paper and the parchment – my assumption looks more-less accurate.

Second point re parchment in VMS – it contains some footprints of erasing previous text and some pages definitely were used before VMS was written.

In that period (in about the year 1500) were several ciphering methods in use, and, the most modern for that time was - steganographia – you will find the description below.

Where should we look for codes & keys?

Definitely not in the main text because it just coded by unknown alphabet.

That means somewhere in the pictures.

There are only a few possibilities left by VMS author how to find the keys:

- a. In some images which can contain masked letters
- b. In image captions
- c. In Latin letters scattered on the pages (we also need to know the order how to construct the words)
- d. Using identified numerical data

## 4. VMS keys and codes searching and analysis.
### 4.1 Analysis of the images to search for the masked letters.

For the analysis the last handwritten page of the VMS was
used.

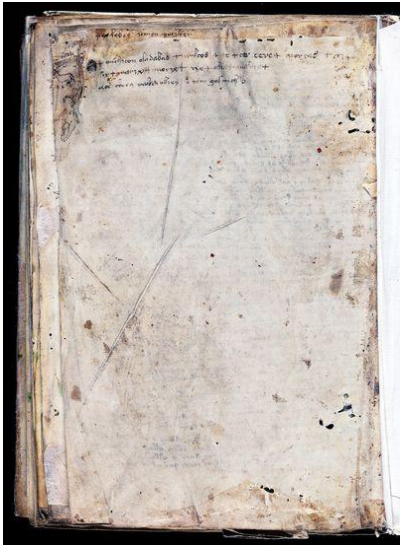

Fig. 2   The image of the last handwritten page of VMS

In the top part of the page you can find the pictures and the
text.

The reason to start the image analysis from the last
handwritten VMS page was the following – the images location
and images compilation are very unusual in comparison with
other parts of VMS.

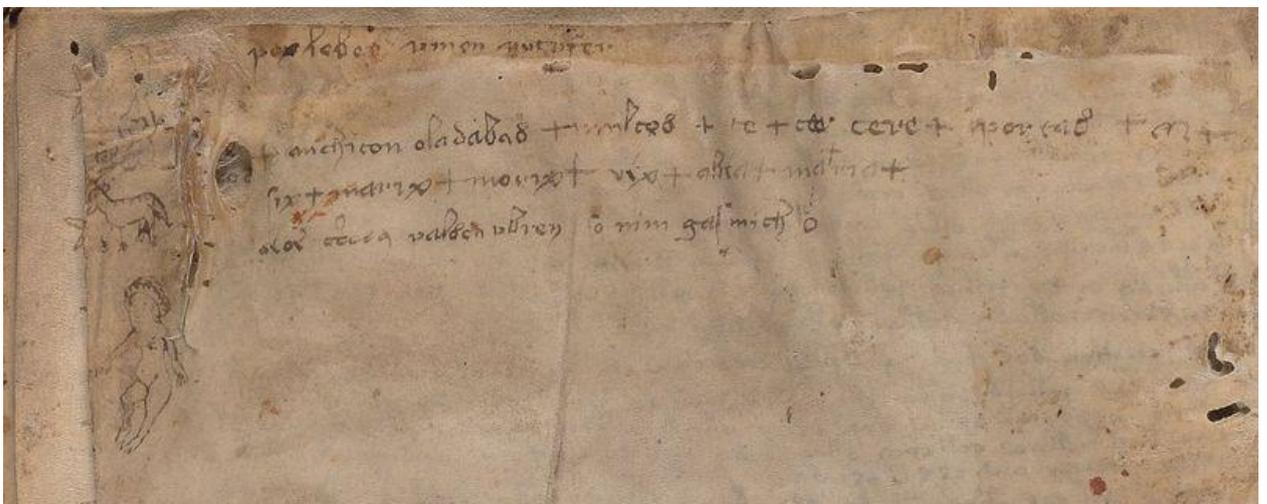

Fig. 3    The images and text on the final page of VMS.

Let's focus on the images first. The contrast was enhanced by image processing software [3].

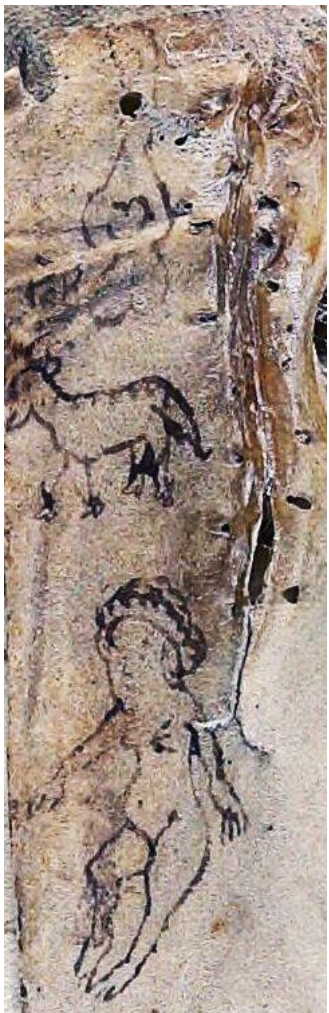

That image looks like an animal such as a Duck or a Dragon (D). The animal body consists of symbols similar to Latin letters like (from left to right) - F, E and L.

That image looks like a string of letters starting from turned and crossed by vertical bar letter V (V)

That image looks like an animal such as a Goat or Capricorn (C)

That image looks like Madame or Maiden (M).

Not Lady or Virgo (explanation is below)

Fig. 4    Images from the last page of VMS.

Why Dragon, Capricorn and Madame (or Maiden) titles were used?

Because all other animal images in VMS corresponds to zodiac constellations. Only concerning "Madame" there was a question. Regarding "Capricorn" some clarification you will find in the chapter dedicated to examples deciphering.

If we follow the logics that all the images corresponds to zodiac constellations the crossed V letter should also denote a constellation.

It was easily identified - the sign of Ophiuchus, which lies on the side. Now crossed "U" letter is used for Ophiuchus, but in medieval period "V" letter was used.

I am especially fixed the Latin letters related to each constellation – the first letter of each name of the constellations :

D, F, V, C and M.

That partly explains the reason why the name "Madame" was used - this sequence can also give us some additional info, like date (for example), because these letters (excluding F) can be also used as roman numerals.

We knew that the date analysis gave us an approximate period of VMS writing as 1404 – 1438. Means 15$^{th}$ century.

"Madame"(not Lady or Virgo) – as "M" comes from the following assumption:

-if VMS author left the date and if we are talking about date corresponding with 15$^{th}$ century - at least st we should see the letter "M" – meaning "Millenium". The only image which can be related to "M" letter is the image of naked girl…

The letter looking like "F" we will investigate below.
Let's start to analyze each image in details.

Let's start from the Dragon. The dragon head is looking to the left with open mouth.

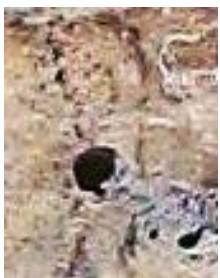
Fig. 5   Head of Dragon (original)

It consist of a couple of visible roman letters

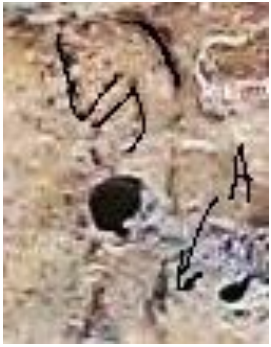

Fig. 6   Head of Dragon with masked letter.

Now a few words about the order how to read the text (same applicable for all other images): you should read it from the left to the right, but  - **the upper letter (even if it is located in right position) has a priority in front to the bottom letter.**

Here we see the most upper letter is **V**.

Next (also on the top) is **I**.

Next is **V**.

Last is **A**

The word hidden here  - **V I V A**

Now let's have a look to strange body of that animal...

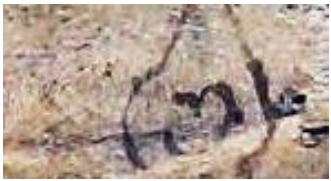

Fig. 7   The part of Dragon body (original image)

It definitely consists of **L E** and most probably **X** (not F).

By the way - in that particular case the sign of letter "E" can be also similar to the letter "u" (look like upside down), but in all other cases for letter "u" author used signs "V" or "U"

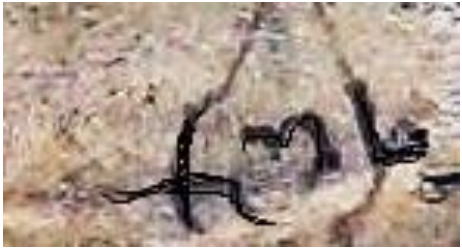
Fig. 8    The part of Dragon body with hidden text.

That means **L E X**

Now we see correct roman numerals. That also gives us the right data for possible date analysis - **D X V C M = 500 10 5 100 1000**. Look like incorrect writing of **1615**...
That strange order of roman numerals may also be a sign that not all of it we have to add in the sum, 2 of it - 10 and 5 can be subtracted... That gives us another date - **1 5 8 5**

Why **X** and **V** were excluded?    **D, C, M** -  have a graphical interpretation from related images of zodiac constellations. **X** and **V** are written as a letters.

First of all it is interesting what the meaning of VIVA LEX is.
Let's try to understand the string of strange letters.

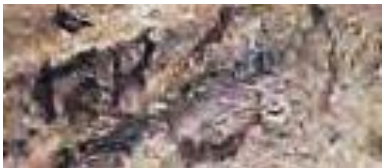
Fig. 9    The string of masked letters.

My  decoding  (the  resolution  of  the  images  from  Yale unfortunately   not   enough   for   more   precise analysis) will be:

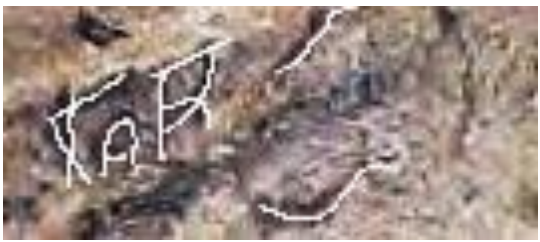

OR

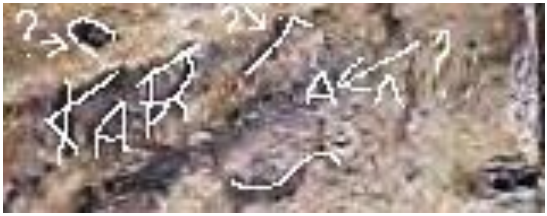

Fig. 10 and Fig. 11    The string of letters with possible identification.

At least in my assumption that look like **V I A R I U S or V I A R I A S**

There are also other candidates - VICARIU(A)S and ESQUIRIU(A)S
Now possible sentence here is "**V I V A L E X V I A R I U(A)S**"

Now one of most interesting animals - Capricorn.

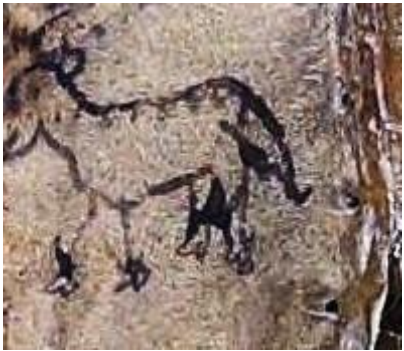

Fig. 12   The original image of Capricorn

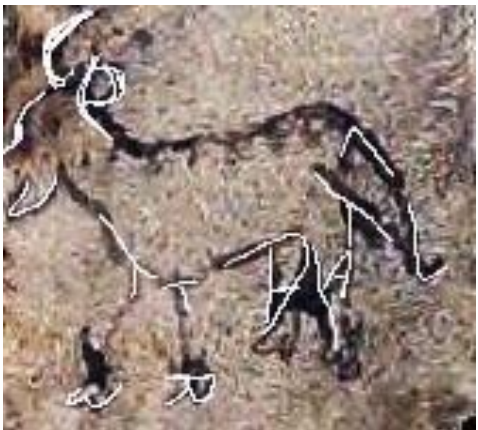

Fig. 13   The image of Capricorn with hidden letters

Look like the name of the author –

**C A L L Y A D V A R T E Y U S** .

Simply – Edward Kelly…

But let's finalize our analysis.

The last figure – is  Maiden.

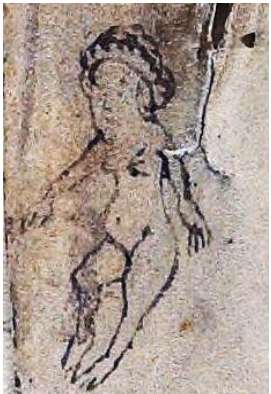

Fig. 14   The original image of Madame or Maiden

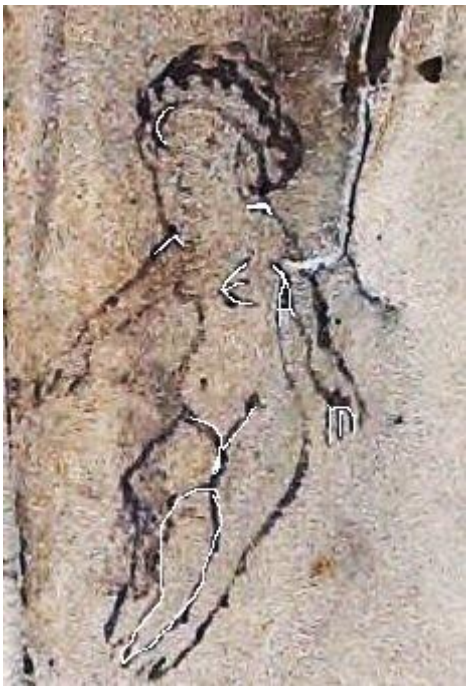

Fig. 15   The image of Madame or Maiden with hidden letters

Let's read...

**<u>C R E A T Y O R</u>**

or

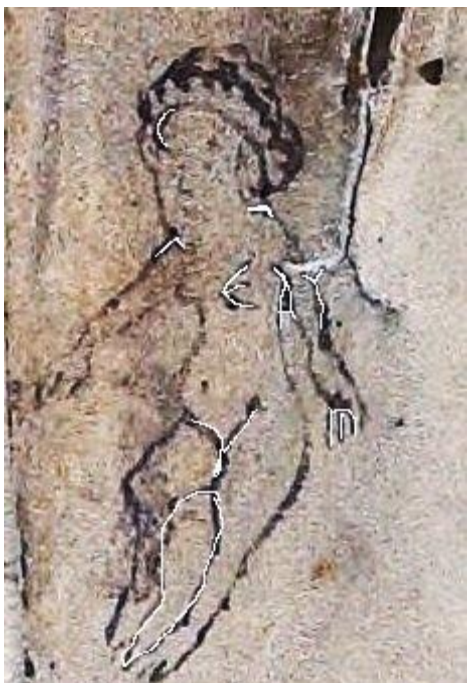

Fig. 16   The image of Madame or Maiden with hidden letters

**C R E A Y T Y O R**

We know that Kelley is died not later 1597.

**VIVA LEX VIARIUS**
**EDWADRD KELLEY**
**CREATOR**
**1585**

Of course this record can slightly vary, but in general it looks that one big VMS riddles now resolved...

If my assumption is right  (let's try to do your own investigation) – we can do the following conclusions (see paragraph 4.2).

The record should be investigated in more details using specific microscopy methods – UV, IR, DF, POL ect.

## 4.2   Initial results.

1) The VMS author probably is Sir Edward Kelly (1555 – 1597?) – well known English alchemist.
2) The manuscript was written in late 16$^{th}$ century.
3) The manuscript was written on the parchment which is more than 100 years older than manuscript was written and for such cases the radiocarbon dating is not good enough.

4) The manuscript was written in England.
5) The languages used for hidden text are Latin and English

So, we possibly found the author, the date, the location where VMS were created and the languages used for it writing.

### 4.3 Analysis of captions to the image on selected pages.

For further analysis several VMS pages [1] from different VMS parts were selected:

- Anatomical
- Botanical
- Astronomical
- Pharmaceutical

For the astronomical part the page used for analysis were selected just because of one significant feature – double star also well known in medieval period.

There were also several additional pages used for verification of the codes and keys were identified.

A) **Selected Page#1** (78r according to library identification)

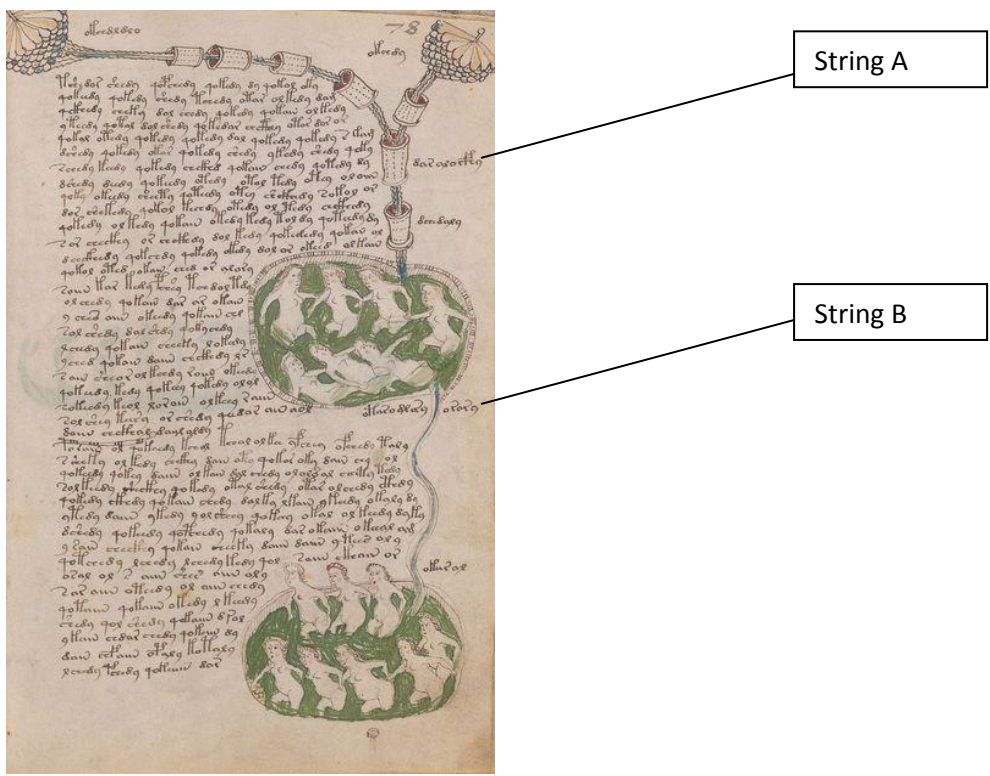

Fig. 17  The selected page from anatomical part of VMS with marked locations of analyzing symbols

B) **Selected Page# 2**(33v according to library identification)

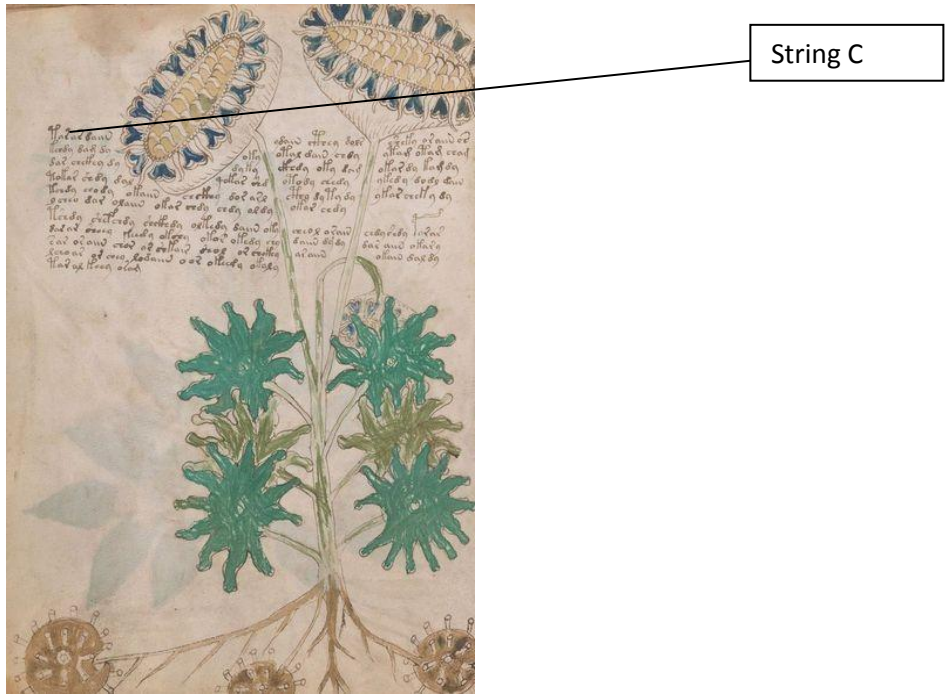

Fig. 18  The selected page from botanical part of VMS with marked location of analyzing symbols

C) **Selected Page# 3**(70v according to library identification)

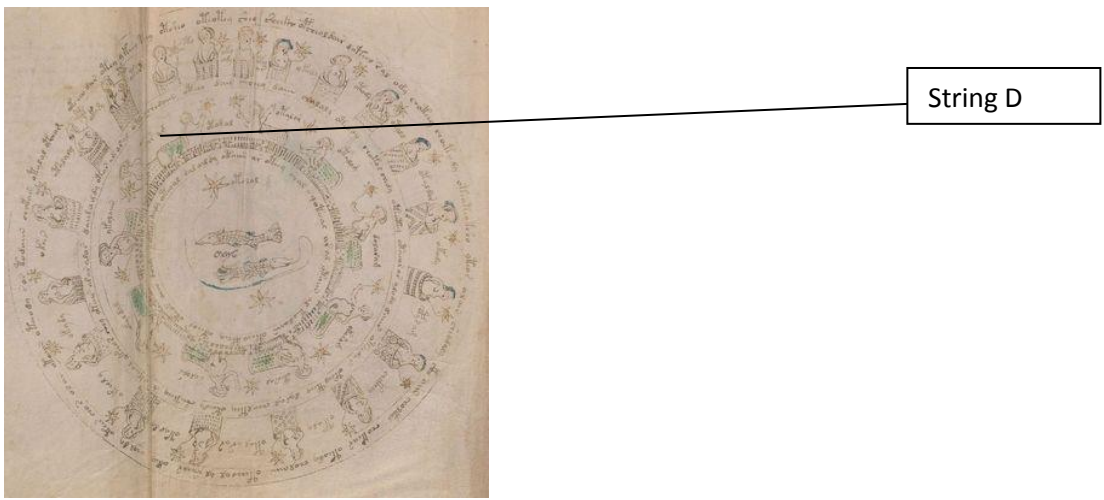

Fig. 19  The selected page from astronomical  part of VMS with marked location of analyzing symbols

D) **Selected Page#4**(100r according to library identification)

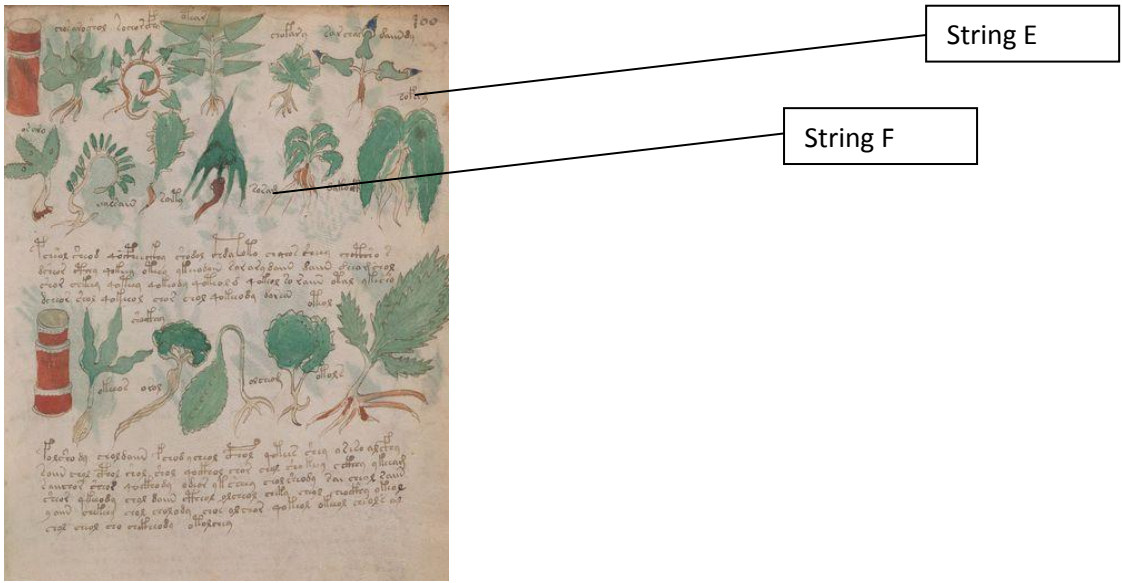

Fig. 20   The selected page from pharmaceutical part of VMS

### 4.3.1 Results

#### 4.3.1.1   Analysis of string A.

On the page# 1 we found the following candidates for encrypted keys:

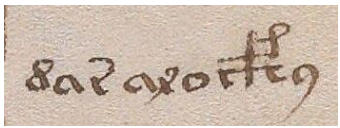

Fig. 21   Image of string A.

That caption to the image visually corresponds with  more-less realistic  image of human **AORTA**.

After final analysis (with use of different Latin alphabet based languages) the following symbols (see below) were identified.

Important remarks:

 - all meanings of identified codes and keys first of all related to modern pronunciation and spelling, because of that you will see various related co-sounds like "B"-"P", "D"-"T", "S"-"SH" etc;
- the symbols meanings are also related of the symbols locations or symbols combination or if symbol of rule - "o"- is used.
In a case of possible several meaning the bold font used to mark the main one.

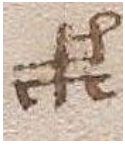
Symbol - **ORT** [ɔːrt]

Fig. 22 Identified symbol #1 (ORT)

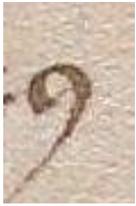
Symbol – **AT** [ət] or [ɑt]   and  it definitely has a numeric value "**15**" on the diagram on VMS page 57v (see chapter 6.1 below).
**May be also  - A** [ɑː] – if it location in the end of the symbolic string
It can be also with meaning "COMMA"

Fig. 23 Identified symbol #2 (AT)

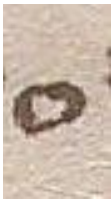
Symbol – **O** [oː], **TO** [toː] and also  in the meaning – to <do>.  It definitely has a numeric value "1" on the diagram on VMS page 57v (see chapter 6.1 below).

It is also sets the rule how to read nearby  symbols.

Fig. 24 Identified symbol #3 (O,TO or "symbol of rule" or spacebar)

## 4.3.1.2   Analysis of string B.

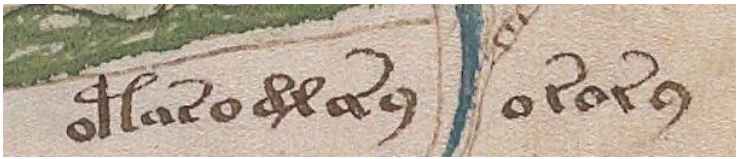

Fig. 25  Image of string B.
That caption to the image visually corresponds with  more-less realistic  image of human gall biliary vesicle or cholecyst.

After final analysis of different Latin alphabet based languages the preferences were done for old English – here it was called as **GALLBLADER** (that was done for longest part of symbols above).

That gave us other keys:

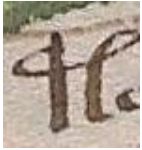
Symbol  – **GEV** [g ə v] in the basic meaning like modern <give> in a case if the following symbol starts from vowel letter,  **GE** [g ə] if the following symbols starts from consonant letter. >. It definitely has a numeric value "11" on the diagram on VMS page 57v (see chapter 6.1 below).

Fig. 26 Identified symbol #4 (GEV)

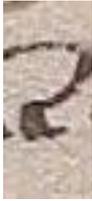

> Symbol – **ALL** [ɑːl],
>
> **Also can be spelled as LI** [lɪ], if symbol location in between 2 symbols similar to "a" letter

Fig. 27 Identified symbol #5 (ALL)

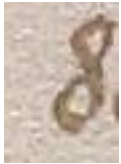

> Symbol – **PLA** [plɑː] , **BLA** [blɑː], it definitely has a numeric value "**15**" on the diagram on VMS page 57v(see chapter 6.1 below).

Fig. 28 Identified symbol #6 (PLA)

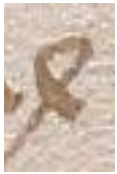

> Symbol – **AND** [ænd] or **END** [ənd] depending of location, it definitely has a numeric value "**2**" on the diagram on VMS page 57v(see chapter 6.1 below).

Fig. 29 Identified symbol #7 (AND)

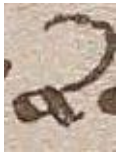

> Symbol – **ER** [ər], **IR** [ir]

Fig. 30 Identified symbol #8 (ER)

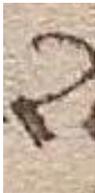

> Symbol – **INN** [inn], it definitely has a numeric value "**4**" on the diagram on VMS page 57v (see chapter 6.1 below).

Fig. 31 Identified symbol #9 (INN)

The rest of the caption containing **GALLBLADER** (in VMS – **GALLBLADERA**) can be read as **IN** or **INN** or **INNE** (in meaning inside) or like **NINA,** because the picture of **GALLBLADER** contains of images of small dames – in Spanish it sounds as NINA (baby-girl). It is also consist of symbol of rule "o" and we will verify it later.

### 4.3.1.3 Analysis of string C.

On the page# 2 were found the following candidates for encrypted keys:

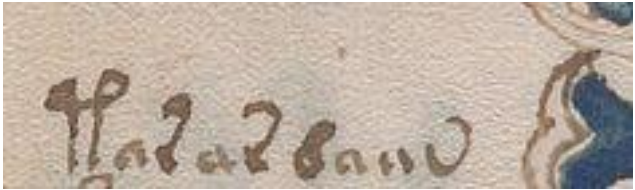
Fig. 32   Image of string C.

This caption was linked to the image of the flower which corresponds to the real flower of Heliantus or Sunflower. If we will divide the name "SUNFLOWER" for 2 simple words we will receive "SUN FLOWER"  wich can be also translated as HELIA PLANT

Some symbols we already Identified  above.

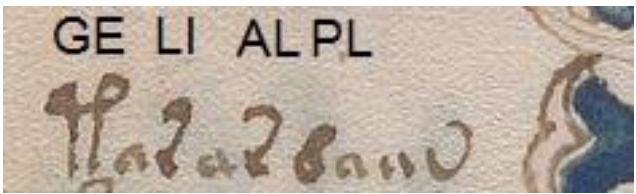
Fig. 33   Image of string C with marked symbols identified above.

Now we found new one:

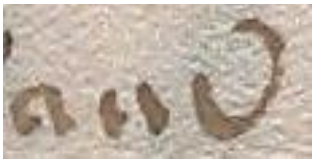

Symbol – **ANT** [ɑːnt]

Fig. 34 Identified symbol #10 (ANT)

### 4.3.1.4   Analysis of string D.

On the page# 3 only one candidate was  used to analyze the encrypted keys:

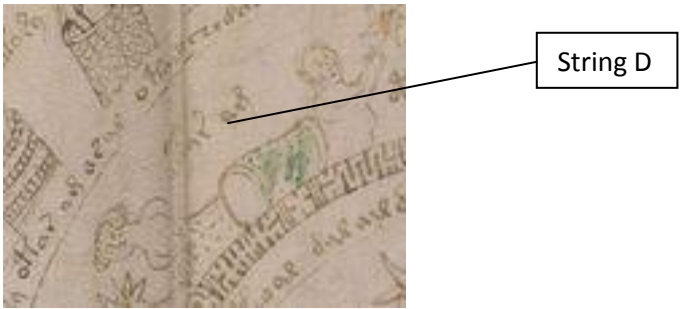
Fig. 35   Image of string D

There is a symbolic image of a double star in the Fish constellation.  It is ID is Alpha Piscium (Alpha Psc,

α Piscium, α Psc) but the traditional name of that star is
Alrescha (Al Rescha, Alrischa, Alrisha)

As a result:

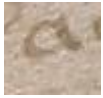
> Symbol – **RI, RY** - [ri] & [rri]. That symbol is another symbol of rule. If there are the symbols string in the following sequence "a" "other symbol" "a" - symbols "a" are not readable, other symbol cutted till last consonant letter, and that last consonant letter should be readable with letter "i" ([i:]) following next. I.e. if we have symbol "INN" in between symbols "RI" the final spelling will be "NI" ([ni:])

Fig. 36 Identified symbol #11 (RI)

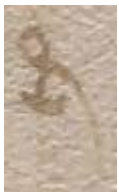
> Symbol – **AS** [æs] or [æz]**, or IS** [is] or [iz],

Fig. 37 Identified symbol #12 (AS)

### 4.3.1.5   Analysis of string E.

On the page# 4 were found the following candidates for
encrypted keys:

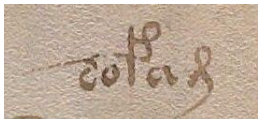

Fig. 38  Image of string E

That symbols string contains symbol of rule "o"(but in the
same time that symbol also can be readable (as T, TO,TA))  and
…RIND. After analyzing the whole modern meaning was found –
TAMARIND. After verification using symbol in the middle  we
can confirm the following:

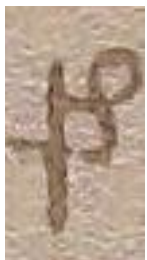
> Symbol – **MA** [m ɑː]**, MAT** [mat] or [mət]**, in  dependence of the next symbol starts from vowel or consonant letter.** it definitely has a numeric value "**9**" on the diagram on VMS page 57v (see chapter 6.1 below).
>
> General meaning is not finally clear for me in some locations it logically means – TO HAVE, TO Obtain.

Fig. 39 Identified symbol #13 (MA)

## 4.3.1.6 Analysis of string F.

Next selected word.

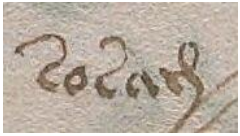

Fig. 40   Image of string F

It contains LA, RI, S. After more detailed verification using other pages the modern meaning of that word is ACICULARIS, the original (above) can be read as CULARIS.

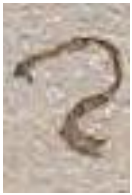

Symbol – **CUT** [kuːt], **C** [k] and [[ts] like tz]

Fig. 41 Identified symbol #14 (CUT)

Finally after use of symbols identified (above), some other symbols were identified and verified.

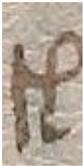

Symbol – **HEAT** [hiːt], it definitely has a numeric value "**7**" on the diagram on VMS page 57v (see chapter 6.1 below).

Fig. 42 Identified symbol #15 (HEAT)

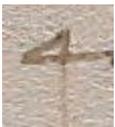

Symbol – **LIF** [liːf] in modern meaning (depends of location):
   1) leaf

Fig. 43 Identified symbol #16 (LIF)

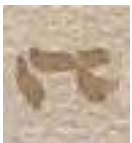

Symbol – **TRE** [træ] or [triː]
In separate always means TREE

Fig. 44 Identified symbol #17 (TRE)

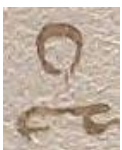

Symbol – **RET** [rət]
In separate location always means RED color

Fig. 45 Identified symbol #18 (RET)

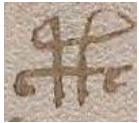
Symbol – **PUT** [puːt]

Fig. 46  Identified symbol #19 (PUT)

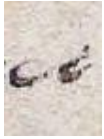
Symbol – **ERN** [ɛə rn] or [ɛərnæ]

Fig. 47  Identified symbol #20 (ERN)

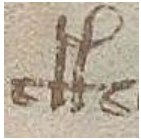
Symbol - **POT** [poːt]

Fig. 48  Identified symbol #21 (POT)

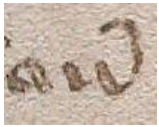
Symbol–**SOM** [saːm] (as modernt "some") or **SUM** - [saːm]

Fig. 49  Identified symbol #22 (SOM)

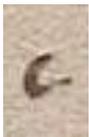
Symbol – **HERE** or **HEA** [həːaː]
or **HER** – in the modern meaning **HERE** [həːaː]

Fig. 50   Identified symbol #23 (HERE)

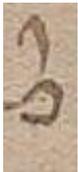
Symbols  – **ICE or ISE** [aɪs ]

Fig. 51   Identified symbol #24 (ICE)

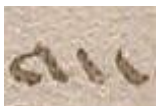
Symbol – **RUN  or RAN** [ran]  - in modern meaning "run"

Fig. 52   Identified symbol #25 (RUN)

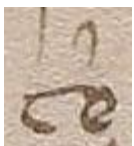
Symbol – **REST** [rəst]

Fig. 53   Identified symbol #26 (rest)

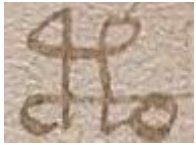
Symbol – **TOP** [to:p] used also as modern "cover"

Fig. 54   Identified symbol #27 (TOP)

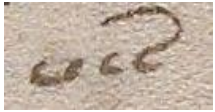
Symbol – **EYE** [aɪ]

Fig. 55   Identified symbol #28 (EYE)

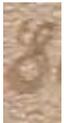
Symbol – **EPL** [əpl]
In modern meaning - apple

Fig. 56   Identified symbol #29 (EPL)

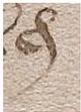
Symbol – **SA** [sa:] or [za:]/ or **SALT** [salt], it definitely has a numeric value "**8**" on the diagram on VMS page 57v (see chapter 6.1 below).

Fig. 57   Identified symbol #30 (SA)

Not all symbols were identified, but now we can read the text more-less easily.

### 4.3.2 Words and spelling.

In that part you will find the small dictionary based on identified codes.

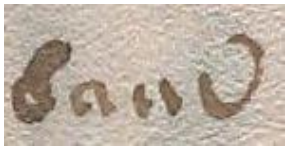
Symbols combination **PLANT** [plɑnt] (in only one modern meaning – plant)

Fig. 58 Identified combination PLANT

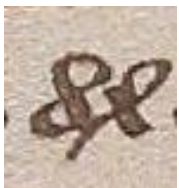
Symbols combination – **BLAAD** [blɑ:ɑ:d] or **PLAAD** [plɑ:ɑ:d] (in only one modern meaning – BLOOD

Fig. 59 Identified combination BLAAD

The investigation of different words which constructions follows the rule: if previous symbol ends on the letter from which next symbol starts – that letter will be not doubled and should be used only once.

So:
PLA + ANT = PLANT
BLA (PLA) + AND = BLAD (may be "N" is not excluded, but it absence  or it presence is not influence the word meaning )
PLA + AT =  PLAT ( modern plate)
 There is an exclusion for that rule – word "riin" (in modern meaning "fresh running water")

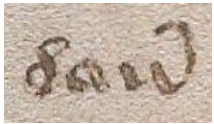
Symbol combination – **PLASOM (as blossom)** –[pla:a:sa:m]

Fig. 60 Identified combination PLASOM

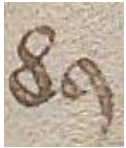
Symbols **PLAT** (plate)

Fig.61 Identified combination PLAT

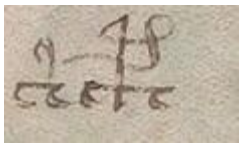
Symbols combination meaning only   – **RETORT** [rətort] (retort)

Fig. 62A Identified combination RETORTS

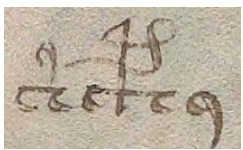
Symbols combination meaning only   – **RETORTAT** [rətortat] (retorts)

Fig. 62B Identified combination RETORTS

Here is an example that symbol "AT" used in the end of word "retort" showing us that author mentioned plural number of that item.

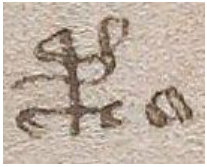
Symbols combination– **ORTRI** [ortri:]
In possible modern meaning – "straightforwardly" or "once"

Fig. 63 Identified combination ORTRI

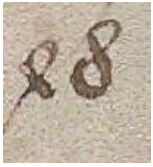
Symbols combination – **DABLA** [da:bla:] **or DAPLA** [da:pla:]
(in modern meaning - double or twice)

Fig. 64 Identified combination DABLA

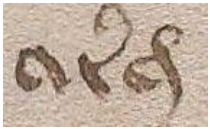
Symbols combination – **RINS** [ri:nz] or **RINSA** [ri:nza:]

Fig. 65 Identified combination RINSA

### 4.3.3 General symbol of rule "o".

One of the most interesting riddles left by VMS author.
It is definitely can be readable as "TO" or "O", but in different combinations and/or locations it can change even meaning of the word.

Let's investigate some examples.

Coming back to symbol#7 – AND

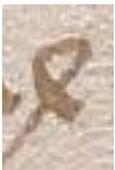
Symbol – **AND** [ænd] or [ənd],

Fig. 66 Symbol #7 (AND)

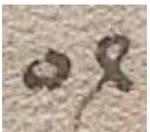
Symbol – **AD** [æd] or [əd], in meaning ADD

Fig. 67   Symbols meaning AD

That example gives us an idea how to find some rules.
Basic "AND" now transformed to "AD" – means symbol of rule probably tells us that if it location in front of some other

symbols – that means you have to exclude the letter/sound in the middle – "A**N**D". So "AND" transforms to "AD".

More complicated case.

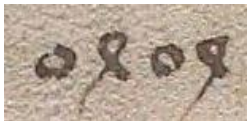

Fig. 68   Symbols combination required interpretation.

Look that it consist of duplet symbols as on fig.67 and finally should sound ADAD. May be that also works somewhere, but according the text meaning – it sounds like "**AD TO END**" (in modern meaning "add to finalize")

Let's now verify the idea above.

Let's use the symbol #9 (below)

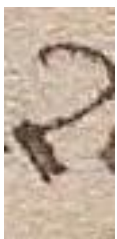

Fig. 69   Symbol #9 (INN)

INN in the manuscript means internal space of something.

In a case if symbol of rule is used

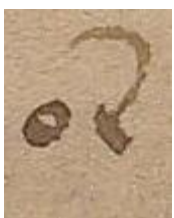

Fig. 70   Symbol INN in combination of symbol of rule.

According our idea that combination (above) sounds "IN".

"AD" and "IN" very common words and that explains why we see these symbolic combination so often in the alchemical manuscript.

Now more complicated case.

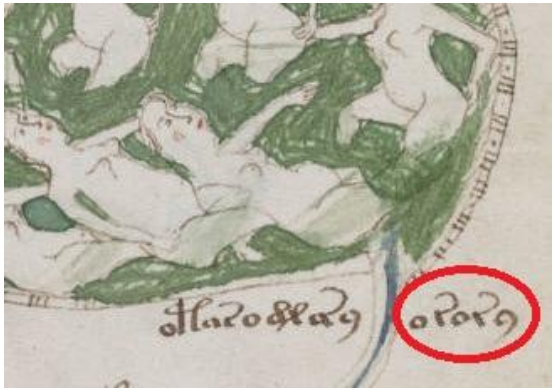
Fig. 71  2x Symbols "INN" in combination of 2x symbols of rule with followed symbol "AT" (marked in red).

That example is known for us from initial codes search – it is a gallbladder from VMS page 78r.

The selected combination  sounds like "IN TO INN A",  the right description will be "INTO INNA"  in meaning "inside". In total meaning "gallbladder inside".

There is a probability that symbol "AT" (at the end of the string) used as a masking symbol to hide the word. But that needs further investigation.
Now I'll give you most interesting example of simple symbols combinations. Let's go to the VMS page 17v…

Here you will find the following
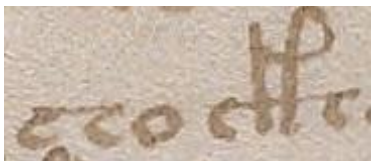
Fig. 72  Symbol TRE in combination with symbol of rule and with symbol POT.

Without usage of symbol of rule – the phrase will be the following – "TRE TO POT" or "TREO POT". First one sound strange, second one  contradicts with usual for VMS author "TREPLA" (three or tree time).

Symbol of rule located here after symbol of TRE, but it function is the same as for A~~N~~D and I~~N~~N.

If that rule still works as mentioned above - the final translation of these combination will be "TE POT". Hope you remember the spelling – [ti:pot]… In meaning – **teapot**!

The location of the symbol of rule was unusual and because of that the example above was verified. Further investigations gave the result that if symbol of rule located after other symbol and both symbols are linked to each other we should read the first symbol in versa direction.

In our case the right result will be **"ERT POT"**, where **"ERT"** is equal to the modern **"EARTH"**, i.e. all together equal to «POT WITH EARTH».

And finally one more example:

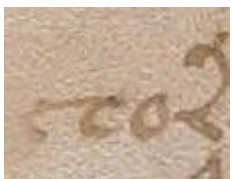

Fig. 73  Symbol TRE in combination with symbol of rule and with symbol INN.

That is difficult example which requires more investigation because both TRE and INN can be influenced by symbol of rule. It can be 3 possible translations:
- TRE TO INN
- TREIN (I am not sure was that word known in $16^{th}$ century or not)
- ERTINN or ERT IN(may be in meaning "on or in the earth", or it is also possible that meaning is numeral = 18)

There is also probability that the symbol of rule located after any other symbol and linked by handwritten lines to it gives the instruction to read the previous symbol backwards: i.e. symbol "RET" with following symbol of rule "o" means that you have to read symbol "RET" as "TER".

In addition it is possible to say that the symbol of rule "o" in location before the symbol on which it sets role are never used for the words consisting more than 3 letters.

## 5. General conclusions.

After the analysis above we can do the following conclusions:

- We can confirm the general coding method – STEGANOGRAPHIA. I am talking about STEGANOGRAFIA as about method described by Johannes Trithemius (1462 - 1516) in his work "Steganographia" (1500).

- It is the modified STEGANOGRAPHIA using much more powerful encrypting mechanism

- We can confirm the main VMS language - is **English** with inclusions of some specific terminology from Latin (like plants names, zodiac constellations, anatomy terminology etc.) and Arabic (especially in Astronomical part – like "Al-Risha"), probably also Greek common terminology.

- The language of the main text is look like especially modified or even simplified English or like one of English dialects.

- For coding by modified STEGANOGRAPHIA VMS author used symbols - each one corresponding with common simple and often singleroot words

- The VMS author have used special symbol – symbol of rule "o" with gives more opportunities for strong ciphering. The symbol of rule "o" were used generally for main text coding.

- The VMS author is also used additional symbol of rule "a". The symbol of rule "a" were used to hide the hints in pictures captions.
- According some (not fully identified) order the symbols can be used as a letters in some single words (like symbol "AT" in the end of symbols strings mostly means letter "A"[a:]).

- The combination of symbols in the main text can represents different structures – phrases and words.

- The combinations of symbols were created sometimes randomly and it look like single words, but in reality – there are a short phrases

- The VMS author used the also used the absence of punctuation as an element of coding.

After analysis of the spelling of identified symbols we will immediately find the next level of subcoding of the main VMS text - it is the semantic one, using 3 main principles:

- The symbols spelling is mostly equal to Latin spelling (not similar to modern English spelling of the same letters and words)

- The main role of the singleroot words (symbols) is to convey is sounds (not grammar!)and it is not directly linked to correct grammar.

- In all cases it is enough to know the word or phrase spelling to understood it significant meaning, even grammar is looks incorrect. Same principle is used for symbols spelling in dependence of its locations.

- The pronunciation related to Latin language can vary - related to co-sounds like "B"-"P", "D"-"T", "S"-"SH" etc.

One of the reason why nobody of VMS researchers not came fto any success is simply the modified steganoraphy, but not classical as was described by Trithemij – the VMS author used not letters but words hidden by corresponding symbols. As well the groups of symbols were created somehow randomly and created erroneous opinion that you are looking to the encrypted words (but in principle that are phrases). Because of that for any statistical methods used for VMS text analysis researchers are used incorrect input data. If somebody wants to use the more correct method – he should at least count each symbol as a word consisting (in average) 2 consonant and 1 or 2 vowel letters.

## 6. Use of results and conclusions

Of course anyone can start now to read the text of VMS. But it is still requires to understood the complete principles of coding.

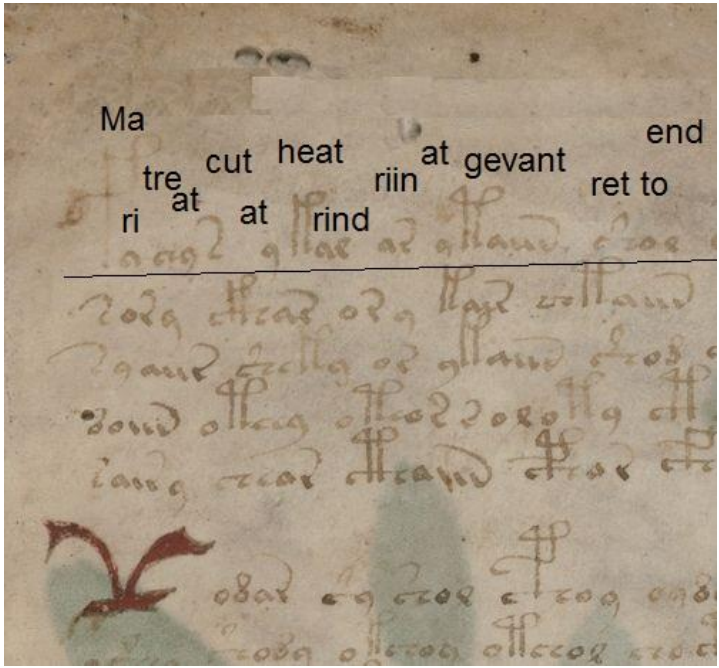

Fig. 74 The example how to read the VMS (used page 1r of VMS)

The book is starting from the phrase "Meri treat cut at hrid riin at gevant ret ad".
That means something like "Combine the treated acuted pieces on the grid, add red…."

Here you can also see how nice the idea was to use simple words for coding.
"TRE" – separately will be "TREE"
"AT" – separately like modern "AT"
But all together is writing is "TREAT"…

## 6.1 General coding method

Few words about general coding – about STEGANOGRAPHIA.

At that time it was the most modern and sophisticated ciphering methods. The name "Steganographia" was for the first time described by Johannes Trithemius (1462 - 1516) in his work: "Steganographia" (written in 1500, but officially published only in 1608, in 1609 included in the list of prohibited books (Index Librorum Prohibitorum))

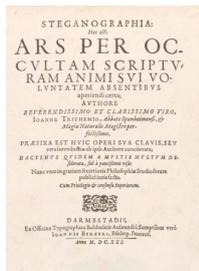

Fig. 75 The title page of book "Steganographia" of Joh. Tritemius [4].

There is an interesting picture in the Trithemius book ([4], p.55)

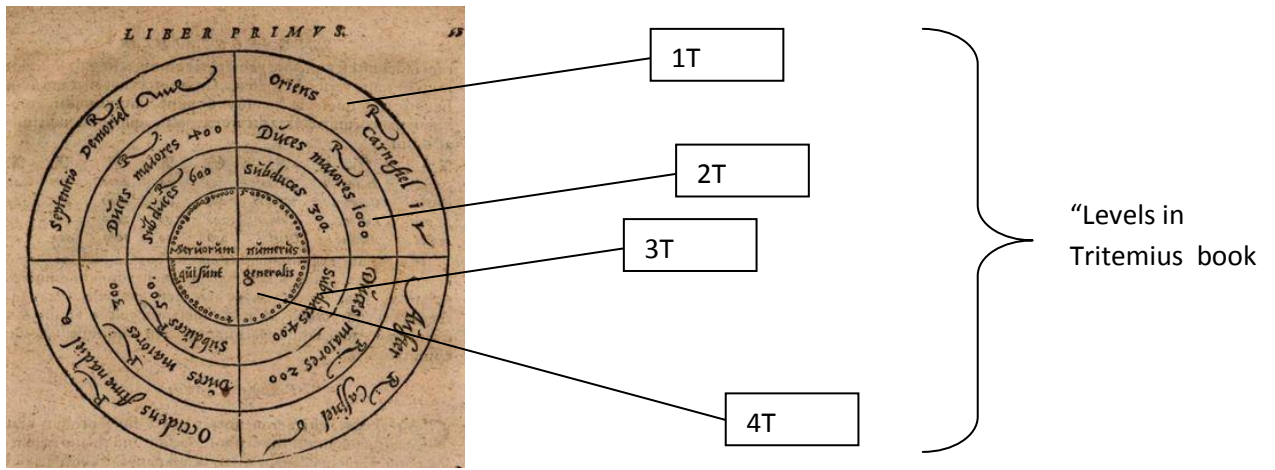

Fig. 76 The part of coding principles from "Steganoraphia"

If we will pay our attention to the VMS page # 57v (according to library identification)

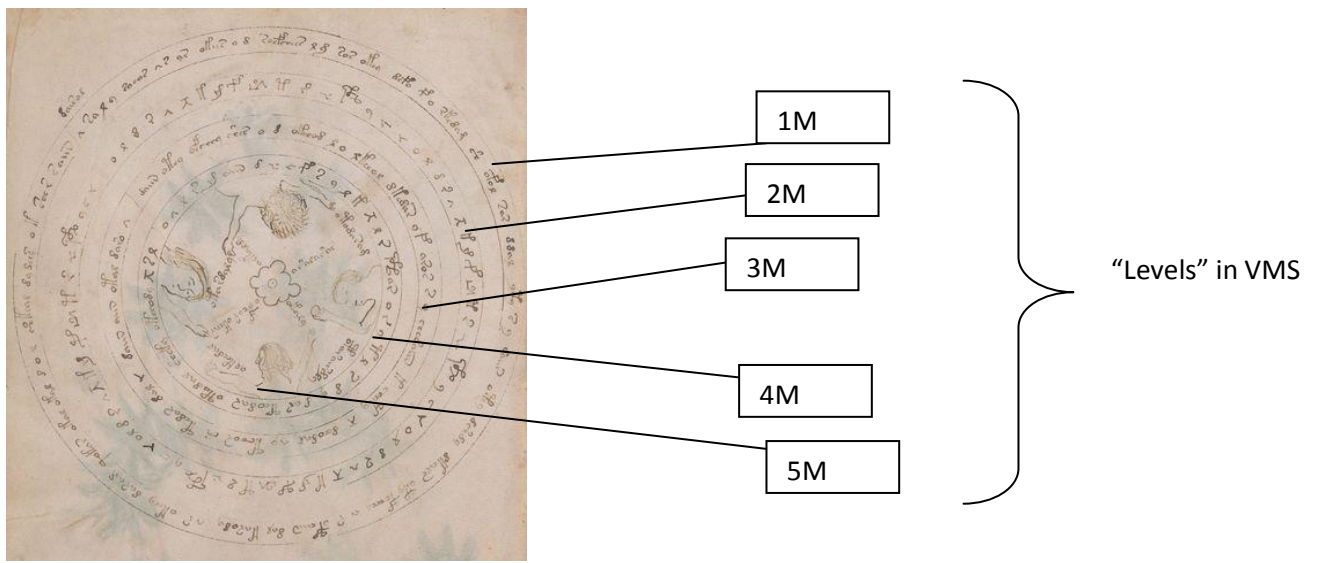

Fig. 77 The image from VMS page 57v (according to library identification)

These 2 images are roughly the same (according its meaning) the only differences are - the location of levels in Trithemius book and levels in VMS. As well as VMS image consist one level more.

The relationship between VMS and Tritemius levels is the following:

1T = 5M, 2T = 4M, 3T = 3M, 4T = 2M

The "Level 1M" is not used for coding.

In details: Why 4T = 2M? Of course it will be a good riddle…

The explanation:

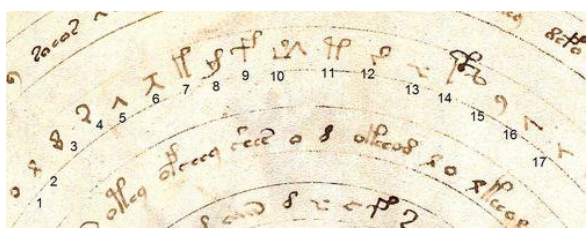

Fig. 78 The image from VMS page 57v with numeral data identified.

You can see the sequence of 17 numerals (from 1 to 17) on that page. It repeated 4 (four) times and also dividing the whole circles for 4 equal parts as in Trithemius book (he also used 17 digits – in the center part of the image). In Trithemius book all numerals have a special values, which are different to the numerals in VMS. The reason is in the coding principles used by Trithemij and VMS author who understood the weakness of classical method and modified it, using not letters, but words for coding. With such modified method the values of numerals were not important and VMS author used only to allow to readers to identify the numerals.

The diagram from "Steganographia" is under protection of magic power of 4 Great Dukes (used from «Magic of Salomon»)– Dukes of North, East, West and South. Their names are listed in the text of level 1T. In VMS author used not names but images of that Dukes on the level 5M. The proportions of the images are fully corresponds with power of each Duke from the Trithemius book.

Using the diagram in VMS you can simply can find ad read some secrets how VMS was coded. You will find the following instruction in the diagram:

«On *(pages – A.U.)* 1-2-3-4-… use 1-2-3-4-5-…  N use, as M …» and so on. The sequence  "1-2-3…" I used just an example – in the diagram you will find other numerals used in the sequence from  1 to 17.

If you will follow to the instruction from the diagram – you will  find the pages which contains Latin letters in different fonts, colors and sizes. The diagram sets the rule how to construct words from these letters. That is another coding level of VMS. The good example will be the page 4r – you will find here at least 5 letters. If you will try to read it starting from the top you will find "AFTOR". According my opinion that word is a snag or random set of letters – the diagram instruction tells us to collect letters from different pages.

Unfortunately some pages like page 12 (which letters according diagram we should use) are not exists.

Another problem is that some letters are written by minore fonts  and current images resolution is not enough to recognize  that is it – the letter or the parchment defect and gives us erroneous data for analysis.

The image in VMS is a modified coding principles offered by Trithemius.

That again links us to the date analysis.

The "STEGANOGRAPHIA" was written in 1500…

Means VMS was definitely written later. Means VMS was written in 16th century.

Now still a question to confirm the authorship of Kelly, because somebody else can left the images and the his signature on the last VMS page.

## 6.2   Additional search and conclusions concern VMS authorship.

1) As we already found that manuscript (VMS)  is written in English  - Tritemius most probably is not wrote it
2) VMS Author used the Trithemius "Steganographia" ( so - VMS written in 16th century)

3) VMS Author is used a lot of specific scientific data from different areas of sciences.
4) VMS Author was interested in development of hard and reliable ciphering methods.
5) VMS Author was linked to Alchemy (bird symbol on the VMS page 1r – means Phoenix, the VMS text starts from alchemical marriage etc.)

Again – we've found very good marker confirming that VMS was written in 16$^{th}$ century – modified Trithemius coding diagram (VMS page 57v) containing description how and where to find some codes and keys.

The Trithemius "Steganographia" with his original diagram ([4] p.55) was finished as handwritten manuscript only in year 1500.

The first it official printed version [4] was published only in year 1608.

In the same time we definitely knew that John Dee used Trithemius manuscript to his own investigation in a mid of 16$^{th}$ century.

According John Dee's archives we definitely knew that Dee found the handwritten copy of "Steganographia" during his diplomatic mission. He was so impressed by Trithemius ideas that he finally copied the exemplar of "Steganographia" by himself [5].  We even knew the date when that work was completed – 15$^{th}$ of  February 1562, and also place where that was done:

[…] Antwerpia apud Gulielmum Silinum in Angelo aureo : in platea, vulgariter, Den Camer straet, vocata […]

The handwritten Trithemius diagram performed by Dee was found also in Dee's archives [6]. That one is dated as a year 1591.

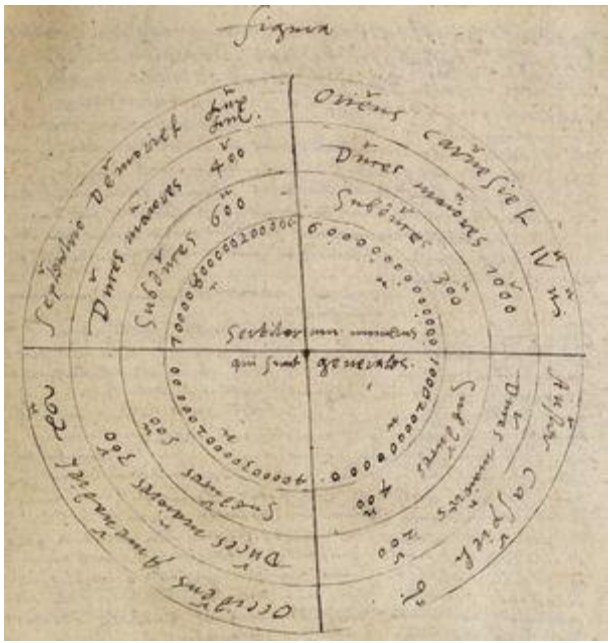

Fig. 79. Image with diagram of coding principles with handwriting of John Dee dated as 1591 [6]

That means that Dee for many years continuously worked with "Steganographia" starting from 1562 (or maybe from the end of 1561 – it depends how much time he needed to copy the manuscript) and he absolutely understood the principles of Trithemius coding methods as well as it weaknesses.

Again in short: **we have the manuscript written in English and coded by principles of "Steganographia".**

As we see also - VMS is readable and look like a real book.

Now let's try to answer the following important questions:

- The purpose of creating the encrypted manuscript
- The manuscript pre-story
- The manuscript future fate

Let's start our investigation from simple question coming from our results and conclusions:

Do we know any other encrypted manuscripts written possibly in English (or in England) in the same period of time and which are linked to the same persons?

Definitely- YES!

It is a **"Book of Dunstan"** – the book which was in hands of Edward Kelly during his co-operation with John Dee…

According the existing legend:

- it was found by Edward Kelly. He brought it to John Dee in 1583 before their journey to the continent ([7] p.lxi)
- The manuscript called as "Book of Dunstan" has been found in Glastonbury Abbey ([7] p.lxii). That place was associated with St.Dunstan service.
- Another version tells what Kelly purchased the manuscript during his journey to Wels in one tavern.

But what is more important - the only source which tells us something about the "Book of Dunstan" and where we have at least it scarce description – is the notes of John Dee… Only! I am especially point your attention to that word.

Anyhow – even according both versions re how the "Book of Dunstan" was found there were no any solid reasons to call the book as a "Book of Dunstan" (may be excluding very precarious link to Glastonbury Abbey) and both - Dee and Kelly as an educated and smart people - were in the known about that.

Otherwise (if it will be a real St. Dunstan manuscript) we can assume that here will be much more different independent  sources linked to the newly discovered "Book of Dunstan", because even in that time the discovery of real  unknown work (manuscript) of legendary St. Dunstan linked to "sacred alchemy" – that will be a great sensation. Many scientists and clergies will be happy to copy it immediately. But nothing happened.

Let's compare  the story of the "Book of Dunstan" and VMS:

- Nothing known about "Book of Dunstan" before 1583
- Nothing known about VMS before end of 1590-es or even till 1612
- The "Book of Dunstan" was taken to the joint journey of Dee and Kelly
- The ownership of the "Book of Dunstan" linked to Kelly
- After the deaf of Kelly (most probably in October 1595 - for that investigation will be dedicated another special article) nothing known about the future fate of the "Book of Dunstan".
- The mysterious manuscript called now VMS was acquired by Rudolf II  (1552 - 1612) definitely after 1586, but in

- the same period of time when Kelly (and probably Dee) were closely related to Rudolf's II court
- The "Book of Dunstan" was in hands of Edward Kelly who joined Rudolf's court at least in 1586 -1587 and who was dependant from Rudolf's reign till October 1595.
- The "Book of Dunstan" motherland is England and there is a big probability that it was written in English.
- VMS is written in English.

To many matches…

One more important note.

There is very interesting Dee's note related to 12$^{th}$ of December 1587 ([8] p. 25;[9]).

It tells the following:

Dec. 12th [1587], afternone somwhat, Mr. Ed. Keley his lamp overthrow, the spirit of wyne long spent to nere, and the glas being not stayed with buks abowt it, as it was wont to be ; and the same glas so flit.ting on one side, the spirit was spilled out, and burnt all that was on the table where it stode, lynnen and written bokes, as the bok of Zacharius with the Alkanor that I translated out of French for som by spirituall could not ; Rowlaschy his thrid boke of waters philosophicall; the boke called Angelicum Opus, all in pictures of the work from the beginning to the end ; the copy of the man of Badwise Conclusions for the Transmution of metals ; and 40 leaves in 4°(*), intitled, Extractiones Dunstani, which he himself extracted and noted out of Dunstan his boke, and the very boke of Dunstan was but cast on the bed hard by from the table."

*((*) 4° – 4 stacks with 10 extracted pages in each – A.U.)*

It is a last note related to the "Book of Dunstan".

From it we knew that about 40 pages were extracted by Edward Kelly from the "Book of Dunstan". And we also knew that similar quantity of pages lost in VMS…

By the way – it was necessary to know the codes and keys of the manuscript, otherwise the preparation of any "Extractiones Dunstani" will be uselessly… So. That Dee's note contradicts with his initial assumption that Kelly brought him the "Book of Dunstan" for help with it deciphering…

Because of that my conclusion will be:

**VMS and manuscript called "Book of Dunstan" – is same book and it was created in only one exemplar especially for the specific mission.**

There were some important reasons which makes the possibility to own the encrypted book called "Book of Dunstan" very useful for Dee and Kelly:

- The "Book of Dunstan" was directly associated with St. Dunstan (as mentioned above). Means it associated with the man who dedicated all his life to the God and Christian service. The man who had unblemished Christian reputation and who was well known everywhere in Europe. That was very important message for Catholic Rudolf II.

- If so – the encrypted "Book of Dunstan" or better say now the "Book of St. Dunstan" - consist of some visible signs linked to alchemy and can be a very good bait to some powerful people as Rudolf II was… Because it subconsciously should tell to him that the manuscript is a good or even blessed book and consist of some initially proved text aloud to read for Christians.

As we see from the history – after the "Book of St. Dunstan" was found - Dee and Kelly starts their journey to the continent (21st of September 1583). It looks like a private mission, organized by Albert Łaski.

Using Łaski relationships network they created their own private relationships network with many powerful people in Europe.

Finally they reached Rudolf's II court…

According to my assumption that was Dee's and Kelly's main destination. The general purpose of their mission were to set a network of spy sources to receive the fast and important information re Rudolf's II planes and actions directly from Rudolf's II court.

The "Book of Dunstan" with it story and all surrounding things were especially created to convince Rudolf II that the journey of Dee and Kelly with their families were absolutely their private deal linked to alchemy sciences and blessed by angels. There were no any links to Queens order.

How Rudolf II will insure that the "Book of Dunstan" is a real valuable treasure?

It was necessary to convince him by some visible features of the manuscript:

- Old ancient expensive parchment
- Mysterious pictures and diagrams
- Strange encrypted fonts
- Alchemy signs on the first page (Phoenix and possibly Uroboros)
- Legend re authorship of St.Dunstan confirmed by authoritative well known scientist (Dee)

And… One of very important issues  - the language. Especially English language was selected to convince Rudolf that the "Book of Dunstan"  is a real one, because English was Dunstan's native language.

St. Dunstan lived in England in $10^{th}$ century and his English was different to modern one in $16^{th}$ century.

That was the reason why Dee and may be Kelly found the idea to use  semantic coding – which simply will hide the incompetence of Dee and Kelly in English of $10^{th}$ century, but sound as an simplified old English.

In addition – it is not a first attempt when Dee tried to attract the Emperor by alchemy manuscript. We knew the fact that he tried to use the copy of Trithemius "Steganograpia" as well as his own  "Monas Hieroglyphica" ("The Hieroglyphic Monad") to attract Maximilian II, Holy Roman Emperor  [5] in an effort to gain patronage.

Now let's try to resolve another riddle – the castle image

## 6.3    Riddle with image of the castle in VMS

The one of the solid arguments for apologists of VMS creation in early $15^{th}$ century is the image of castle on the VMS pages 85r and 86v.

The main feature of the castle which is confusing people - is the dovetail teethes of the castle walls. Such design was initially developed and used in Italy, especially before 15th century (which is nicely fits to the parchment analysis – dates and possible manuscript initial location). That type of castles fortification design came to Europe much later.

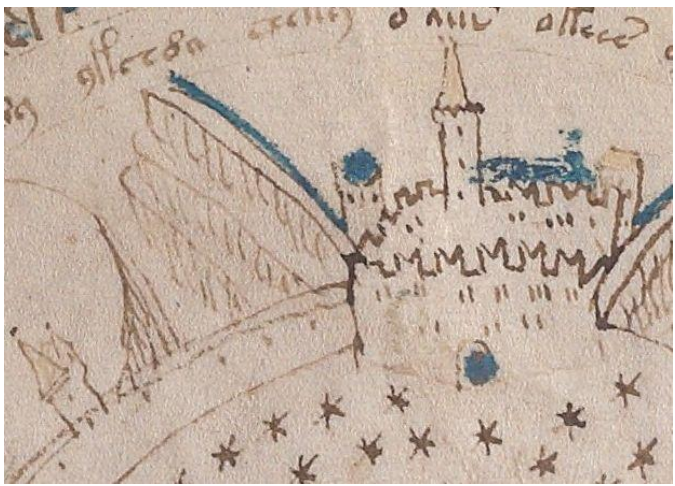

Fig. 80 The part of the image of the castle from VMS page 85r-86v with dovetail teethes on the castle's walls.

Let's analyze that image first.

As always in VMS - we see the allegory images. The castle is a part of specific VMS scheme and it is also allegoric image linked to specific human organs functionality.

First of all – the castle on the image is a very simplified castle, consisting no any visual info where the real castle can be located (if such exists), or any other info re castle specific construction.

If you will pay your attention to the symbolic images of the windows located on the castle walls as well as to the internal castle structure – you will see unreal information:

- To many windows located on the walls and also it location is looks unusual for fortification building
- No internal infrastructure which can tell us how these windows can be achievable from the castle yard.

That tells us that there is also probability that VMS author never seen the real castle, but seeing somewhere it simplified image.

It was my statement based on very good example – see the image below:

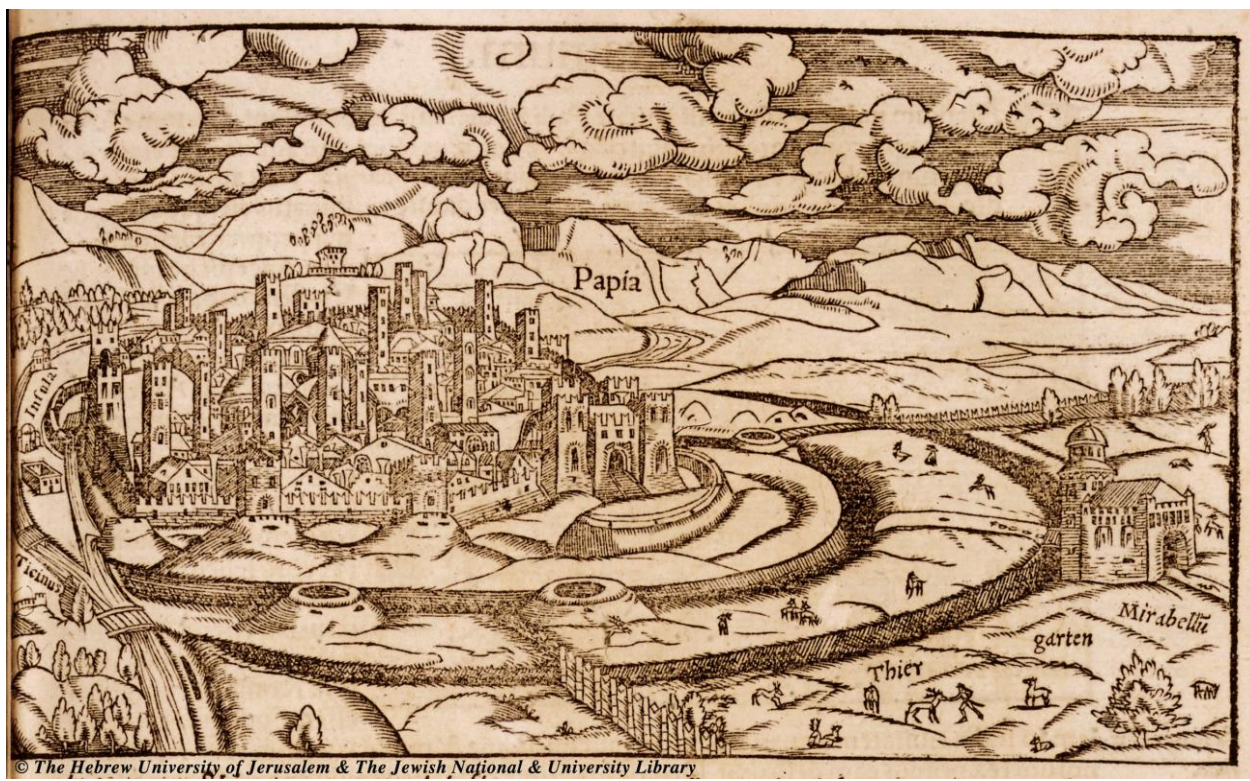

Fig. 81 The image of the castle of Italian city Papia (Pavia) taken from Sebastian's Münster (1489—1552) «Cosmographia» [10,11] with dovetail teethes of the castle's walls.

The «Cosmographia» of Sebastian's Münster was firstly published in Basel in 1544.

It was very popular book and it was translated to several languages – German (origin), Latin, English, Italian, French and even Czech

In total it was re-published about 40 times during 1544 - 1628 years.[12]

The image on fig.81 was initially added to the Latin edition in 1550 year.

It is only the image with Italian castle consisting dovetail teethes of the castle's walls in whole book.

The most important thing – Sebastian Münster never been in Papia (Pavia)… He created that image based on descriptions of travelers and using images of other authors.

It is easy to verify – just find the corresponding articles, pictures and paintings related to medieval Pavia.

The second important thing – the Sebastian Münster's «Cosmographia» edition in French you can find in John's Dee library catalogue [13]...

As I mentioned initially – to many matches.

In addition - same technology of the " dovetail teethes" were used for building of Prague Castle You can easily find it on the South wall. This is "production" of early or maximum middle 16${}^{th}$ century.

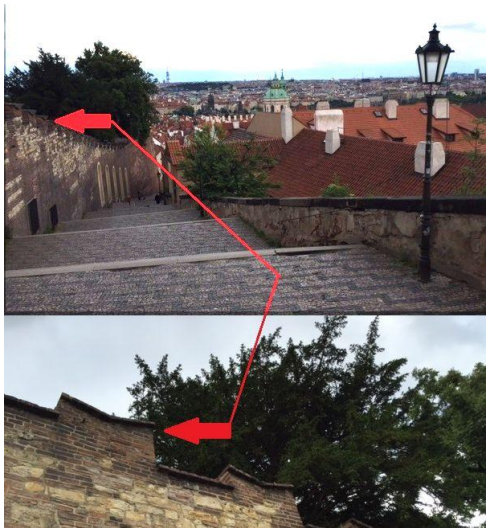

Fig. 82 South wall of Prague Castle (photo - A.U.).

If we can look further to the same VMS pages - we can find more information.

There are another fortification buildings images:

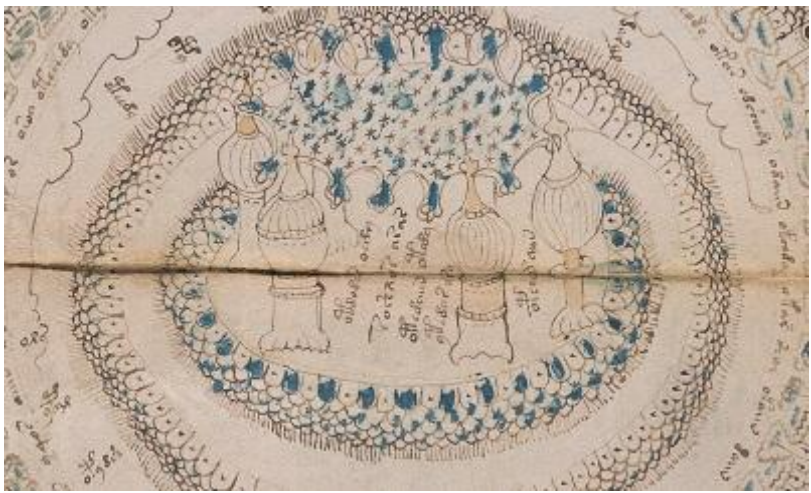

Fig. 83 The part of the fortification images from VMS page 85r-86v.

Let's try to find something similar in Sebastian's Münster «Cosmographia».

What about Venice?

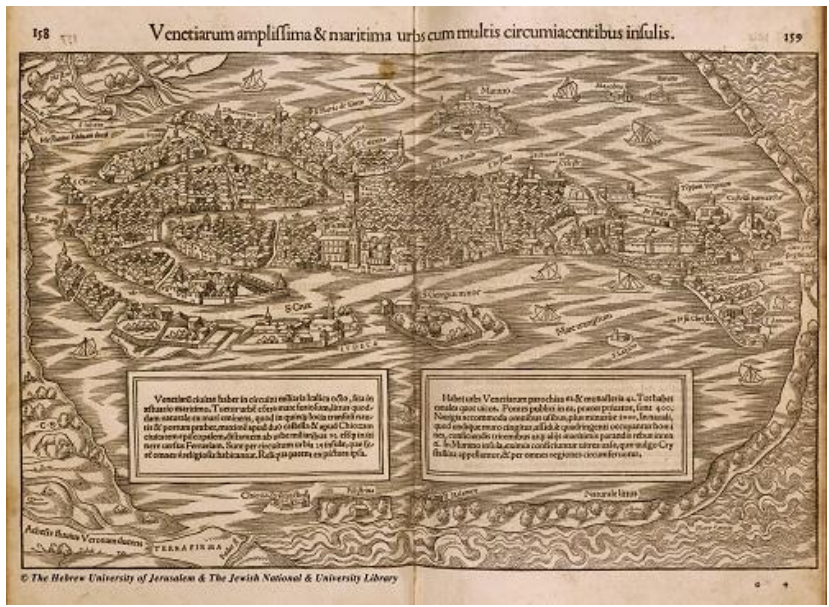

Fig.84 The image of the Venice taken from Sebastian's Münster (1489—1552) «Cosmographia» [10,14].

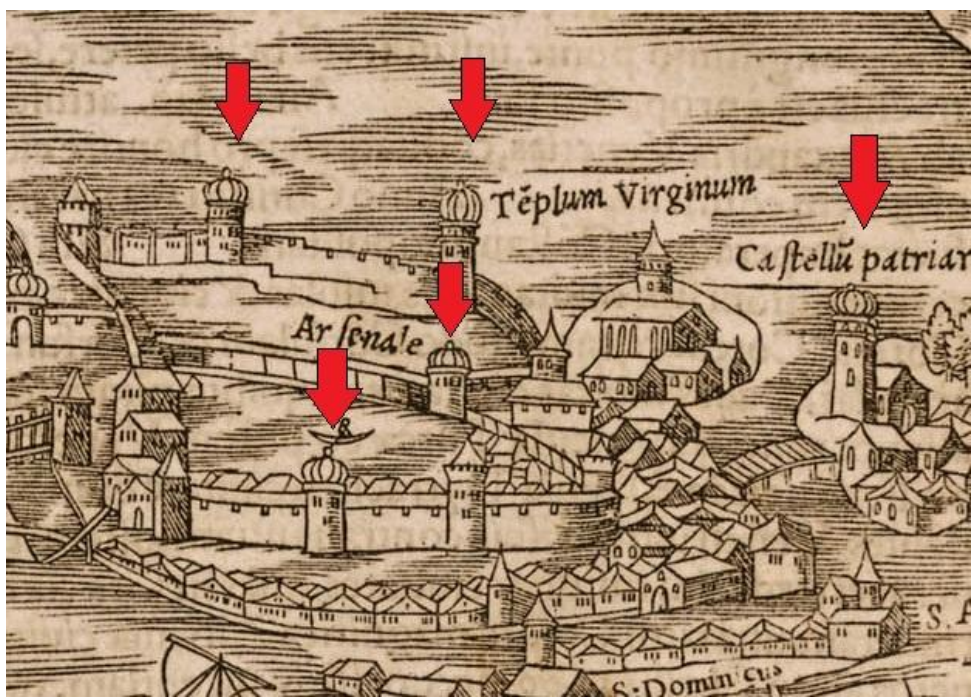

Fig.85 The image of possible candidates of the prototypes of fortification images from VMS page 85r-86v found in Venice image.

As we knew - Sebastian Münster never been in Venice… He created that image based on descriptions of travelers and using images of other authors.

Again - it is only the image in the «Cosmographia» containing such strange round towers…

The other circle images on VMS page 85r-86v are also interesting.
Something similar I've found in Sebastian's Serlio (1475 – 1554) "I sette libri dell'architettura"("Seven Books of Architecture") in the books/chapters #3 and #5.  As we knew John Dee was very interested in architecture studies.

John's Dee library is a brilliant source for answers of some VMS riddles.

### 6.4  Additional conclusions re VMS authorship and dating.

- It is a manuscript most possibly written by John Dee. At least under his direct control. There is a probability that Edward Kelly was really involved in it writing.
- It is written possibly around 1585, may be finally completed in late 1587 or early 1588.
- The real name of VMS is – the "Book of Dunstan"

### 7.  Samples of deciphering.

The codes and keys identification is just a small part for future and long term investigation of professionals.  Even if you knew codes and keys you should also know how to read the text as well as  how to interpret it.  Below I'll show how many visible "traps" were left by VMS author.

At first I would like to demonstrate few examples how the keys which were identified are used for decoding.

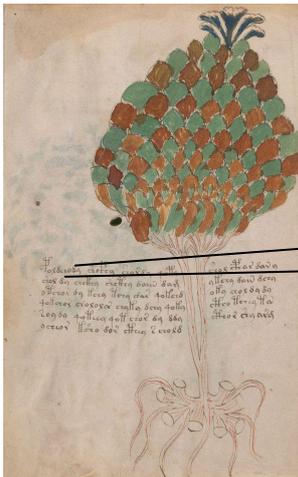

Let's use the image of VMS page 11v (according library identification) [1]

It were used the first 2 strings of the text

Fig. 86  The image of VMS page 11v

Below you will see the reconstruction of the text.

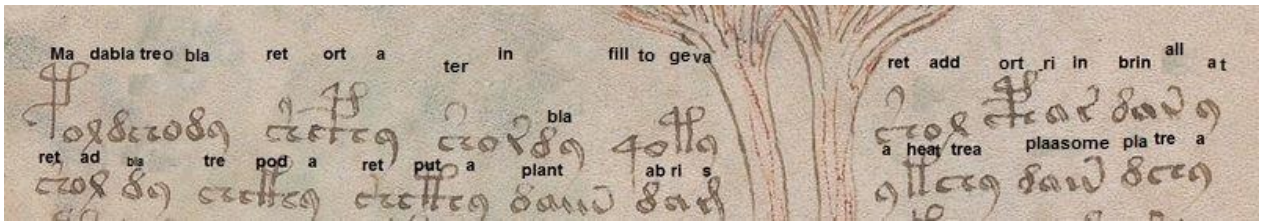

Fig. 87 Reconstruction of first 2 text strings of VMS page 11v

That tell us possibly the following:"**Ma to dabla tre to plat ret ort at ter in plat fil gevat ret ad ortri in brin all, ret ad pla at tre potat ret putat plant pla riis at heat tre at plasom pla tre at**"

The phrase and words reconstruction can be the following:
"**Ma to dabla tre two plat, retortat, tern in, plat fil, gevat ret ad ortri in, brin all, ret ad, pla at trepotat, ret putat plant, pla riis at heat treat, plasom pla treat**"

The meaning can be the following:
"**Find the tree with double trunk** *(as shown on the picture – A.U.)* **with plates and retorts. Turn in.. Plates fill, given red ad once in. Bring all. Add red. Place all (plates) on the the tripods** *(probably "tripod" means "a basement for heating device" or "heatre" – A.U.)*, **put red plant, place rise for heat treatment, place blossoms for treatment.**"

Here is an exercise for you (I'll give the only how it sounds – the interpretation I'll leave for you).

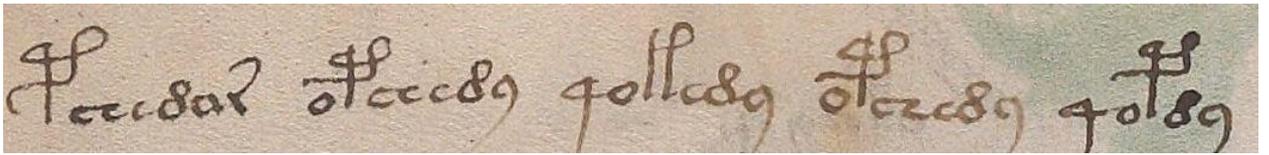

Fig. 88 The image of first string (partly) from VMS page 75v

That is anatomical part. VMS page 75v consist of image similar to human trachea.

**"Ma tre hear pla riin  to ma tre hear plat fil heat hear plat to ma tre hear  fil ma plat"**

**"Ma trehear pla riin, to ma trehear plat fil, heathear plat to ma trehear,  fil ma plat"**

The meaning can be the following: **"Main trachea place in the fresh running water, after put it to the plate, place trachea to heather plate,  fill main  plate"**

One important remark: In that particular case (above) the symbol 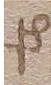 ("MA") is used in most probably in 2 meanings – "take" and "major (mature, main)". Because of the symbol of rule "o"…

Let's come back to the VMS page 100r from pharmaceutical part. Let's try to identify some plants. Important remark – not all of the plants on the page are present with it names, for some there are just the instructions what to do with it(cut, grind etc.). That means that not titled plants should be easily identified by specialists (I'll will mark my assumptions in red).

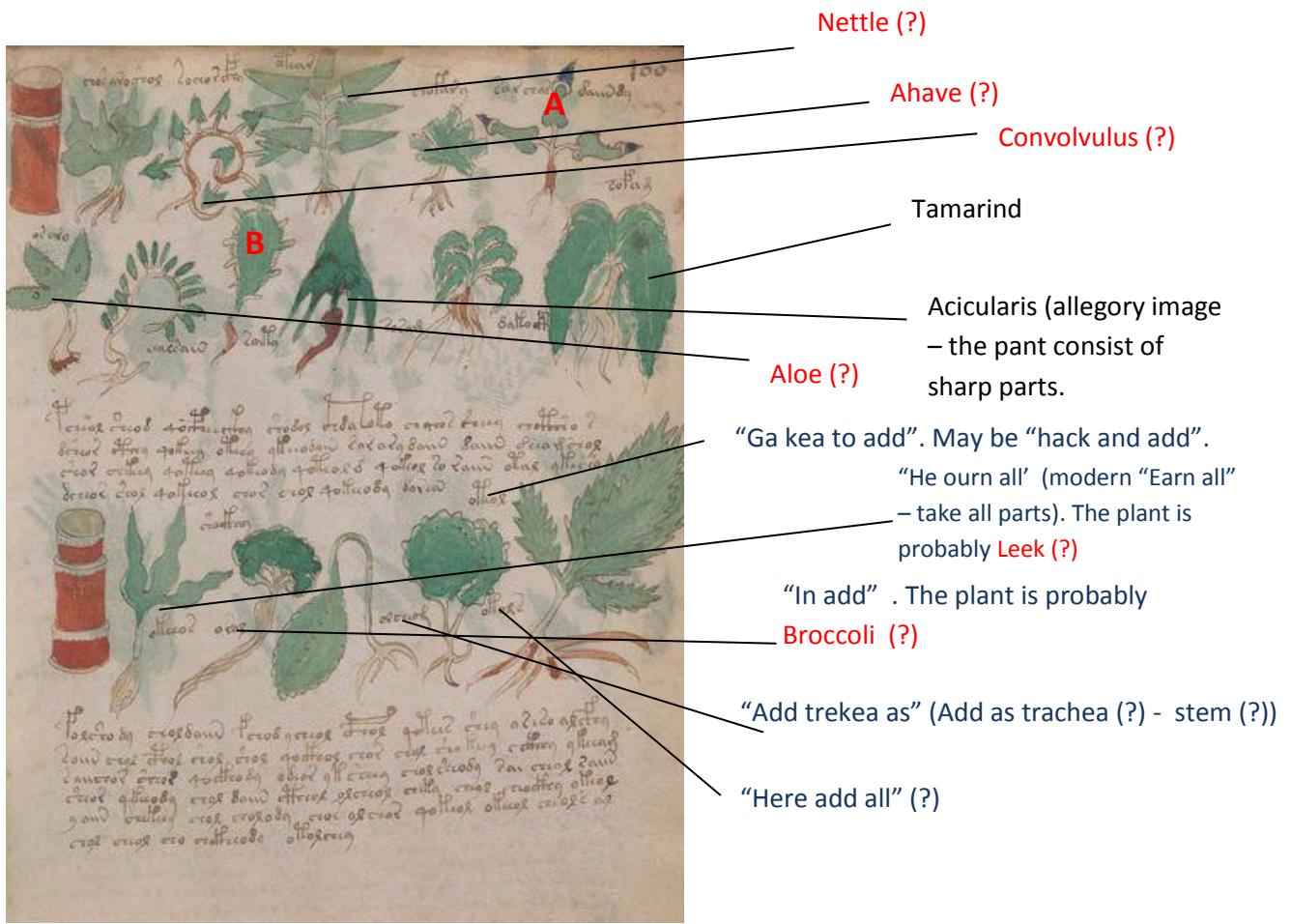

Fig. 89 Identified plants from VMS page 100r

The two of most interesting for me specimens I marked as A and B. Let's pay our attention to Specimen A.

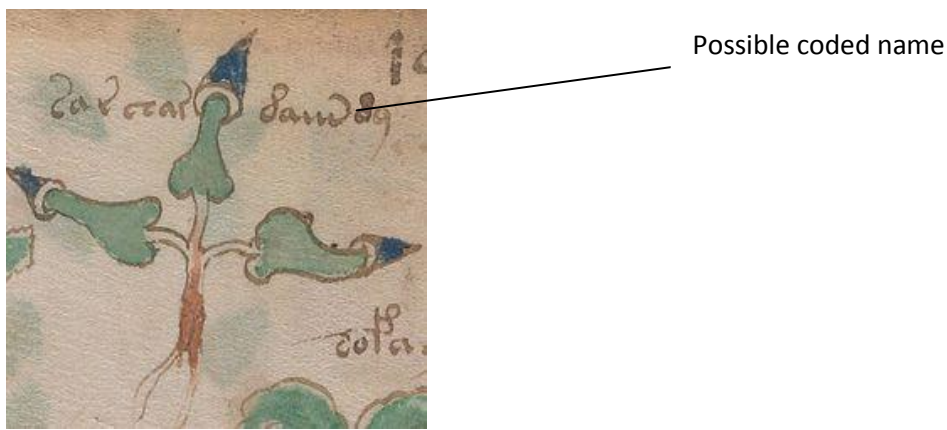

Fig. 90 Specimen A

That specimen can not be visually and logically linked to any real plant. Because of that VMS author should sign it with coded real name. The strings contains the word "PLANT" and located on the top left corner. That sounds like "PLANTBLA" or "PLANTAPLA".

We knew one very nice and interesting plant called in Latin – "Plantago" or, in modern English – "Plantain".
Let's check – what will be possible solution…

It can be for example - Plantago lanceolata [15]

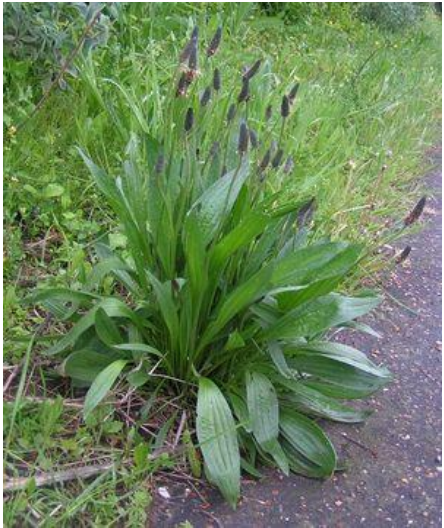

May be not all of it parts – "peaks" only. But also possible leaves and "heads of peaks" because of image in VMS is allegorical.

Fig. 91 Image of Plantaga lanceolata.

Let's pay our attention to Specimen B. I used it because found that in 2014 some specialists [16] identified it as an Opuntia from American continent. And even recognized it name in Aztecs language as Nochtll or Noshtll…

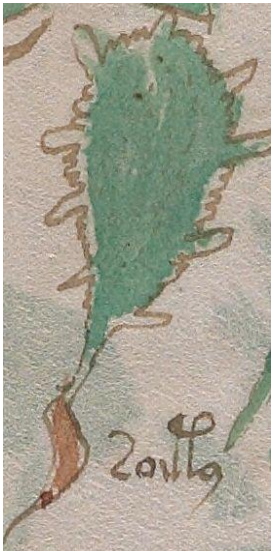

First of all that plant can be compared with many other plants with similar form factor. Even on the same VMS page below the image of specimen B there are 2 more candidates to be recognized as Opuntia…

That means – VMS author left us the name.

Fig. 92 Image of specimen B.

The caption for that plants tells us "ALL TO GEV AT"… In modern English – means we have to utilize "ALL TOGETHER" – whole parts of that plant. The image capture is far away from Opuntia. But according author idea that plant should be easily recognizable.

Next 2 examples I used from the work [17] which consists a lot of ideas re possible identification of plants and words.

I would like to start from the page 46 of the mentioned source [17].
It is linked to analysis of the VMS page 31r – especially for the analysis of the first word on the page which can be possibly read as "KOOTON" (Cotton):

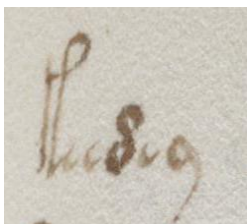

Fig. 93 Image of first symbols of the first string of VMS page 31r.

The first phrase sounds (according our keys) like the following: "Heat ern pla here at" – in modern English it most probably corresponds with "Heather place here"…

VMS author used several word to describe the heating devices - "heathere", "heatern", "heatereh" and so on, with different meaning – roaster, boiler, dryer…

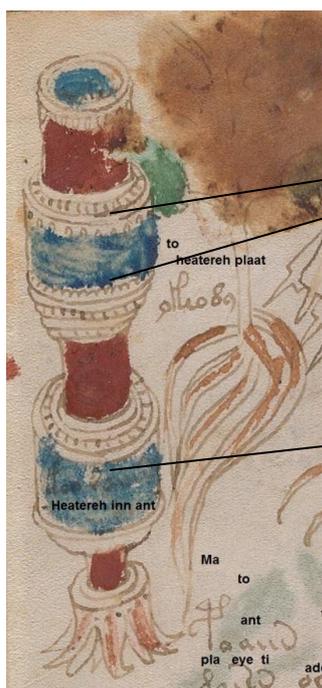

«Two plates of heather»

«Heating chamber» («Heatereh inant» - «heather inside»)

Fig. 94 Example of the heating device from the VMS

Let's come now to the astronomical and astrological VMS parts.

We will point our attention to the same work [17] first, to it page 19 where VMS page 68r analysis was proceed.

The author [17] identified at first the Pleiades constellation and as a result – TAURUS on the diagram.
Small remark – it will be really difficult or even impossible to identify the TAURUS without something significantly and easily recognizable as Pleiades, left by VMS autor.

As author wrote [17] in the chapter "TAURUS":

*It was noted at the beginning of this article that no word of the VM has been convincingly translated or glossed, but in fact there is one word which has received a degree of consensus. On page 68r, in a dramatic diagram apparently showing the moon in the heavens, a set of seven stars has been suggested to show the 'Pleiades' sometimes known as the Seven Sisters, in the constellation of Taurus (……) and the accompanying word has sometimes be interpreted to indicate TAURUS (Zandbergen 2004-2013)   […]*

Truly sad – unexpected conclusion according to my opinion.
There is an astronomical part with TAURUS described twice by VMS author. Here Taurus directly mentioned as Taurus.
And what does it means – "sometimes"? And how it can be interpreted in other cases?

The symbolic caption, mentioned by author [17]  is visually (at least to me) linked to Pleiades, as well as other 2 captions to the star and to the "curve line".
It is unclear - why only one symbolic caption was selected to analysis of word "Taurus".

The Pleiades sign  – in my initial assumption – was used just to allow the navigation in that calendar diagram as a marker for understanding how to calculate the right months/days. If VMS author left us sign of Pleiades there is no need to add the "Taurus" name…

Let's have a look to the mentioned page and to the captions on it.

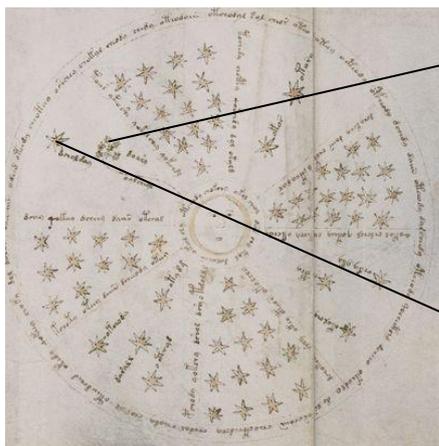

That is the Pleiades and some of symbols compilation slightly above.

In my opinion if Pleiades/Taurus is identified, that means that the star following Pleiades at slightly left should be also recognized as Aldebaran (α Tau, α Tauri, Alpha Tauri)

Fig. 95 Image of diagram on the VMS page 68r with Pleiades.

It is clear that at least the rules which were used by VMS author for naming of "Taurus" or "Pleiades" should be also applicable to naming of at least "Aldebaran" or/and also for "curve line". We are even not talking about the language VMS author used. We are talking just about some possible identification methods which should allow us to find correlation between symbols and captions, which finally can give us the idea for understanding of rules how VMS was written.

If we cannot find the rule linked to astronomical/astrological names – we have to use another way.

Of course now we can read the words on this diagram, but the most important thing that VMS author left us here (in astronomical and astrological parts) some ideas and even explanation to which direction we should go first.

Especially there are 12 constellation on VMS pages.
Look like all Zodiac constellation excluding Capricorn and Aquarius (am I right?) are presented.

The next question will be – Why author is not used Capricorn and Aquarius?

My significant assumption (as I noticed in the beginning – I start the analysis from zero – to try to avoid any stereotypes) was – because it is winter months for the area where VMS author lived.

If so – why he excluded these two constellation from his calendar?

May be this winter months are useless for plants grow…

Again – not to many variants exists – most probably it is a lunar calendar important for plants grow and usage…

How to confirm that? Let's try to understood why Aries (part of VMS page 70v and 71r) and Taurus (part of VMS page 71v and 72r) are presented twice.

My assumptions regarding pages with Aries the following:

- at that case there are the Capricorn images and not Aries
- There is a Latin caption above the Capricorn images telling us – "Abril"(clear – April)
- Capricorn on both images eating something (on the first page smaller piece, on the next bigger).
- If we will check the lunar calendars we will find that that means - the required and important period is starts or ends in a case if the Moon in April moves to Capricorn constellation. The most important is to fit in the described Moon phases and probably in specific days which probably corresponds with ground "hills" where hoofs of Capricorn are located.

I am not analyzed the 2 x Taurus pages (no doubts – there is definitely Taurus, but principles are the same).

One more interesting page – it is an image of Cancer constellation. It consist of two cancers – red and white. I don't know that it exactly means but it may help also with further VMS date analysis – the red cancer can symbolize the full lunar Eclipse with the specific feature – red color of Moon.

The white cancer can also be counted as some events with Sun… It is just an idea for further investigations.

Now let's come back to the VMS page 68r and read what us telling captions for Pleiades and for Aldebaran.

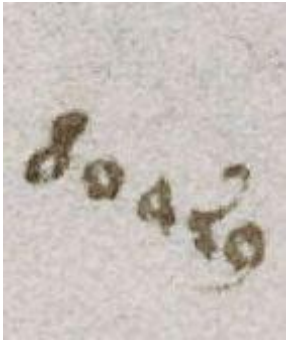

"pla to riin at" - most probably – means "you have to pour (to add water to) your plants twice".

Fig. 96    Image of Pleiades caption from the VMS page 68r.

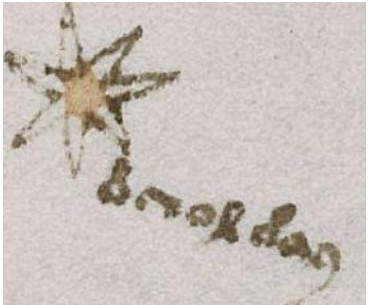

"Pla tre to end pla rin at"
- one of possible translations:

"Plant tree. In the end pour it"

Fig. 97    Image and caption of Aldebaran from the VMS page 68r

I guess it will be also interesting to know that the caption for the "curve line" tells.

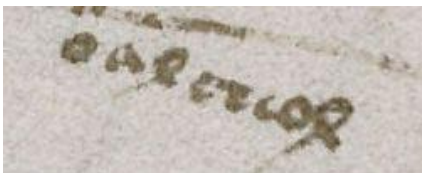

Fig. 98    Image and caption of "curve line" from the VMS page 68r

Let's read it: "to rind tre here ad".
That probably means that the 2 parts of bark of the tree should be added...

You will find here a lot of simple singleroot words related to herbs and trees, it growing and it usage. Like: "plant, grind, rin, rina, herba, grid, grin, brid, brind, pla, apla, aplant, ortea, cut" and so on…

## 8. Last VMS page text analysis.

We already identified some keys for the main text of VMS

The last it page coded by different keys which were used mostly once or twice in the manuscript.
That means that we should find the logics how we can decipher the text.

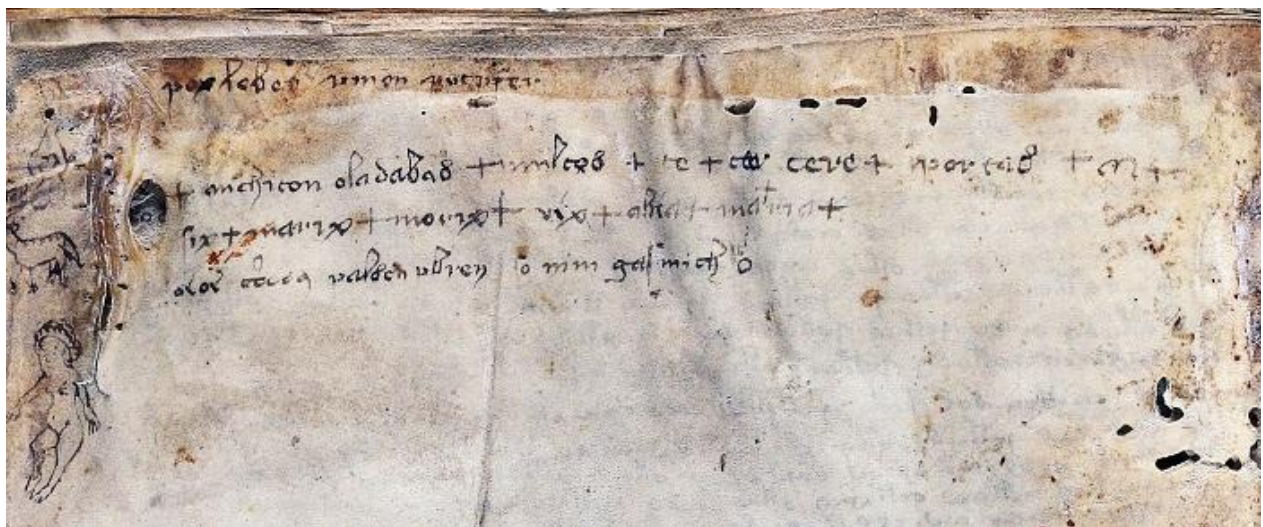
Fig.99   The top part of the last VMS page.

## 9. Basic tasks for codes & keys investigation.

Let's first come back to the whole structure of VMS. It consist of  the following sections:

 - Botanical
 - Astronomical
 - Biological
 - Cosmological
 - Pharmaceutical
 - Prescription

That means that the VMS author used many specific sources for such fundamental development – at least he had the long term access to the specific library.

From that point of view John Dee – is a good candidate for real VMS authorship.

Moreover he had his library catalogue, started in September 1583 and  we can compare some VMS data to the content of that catalogue.

One of my initial assumptions re sequence of coded strings separated by sign "+" was – it is a list of authors who were involved in VMS writing. It was not far away from the final conclusion – it is the list of books or it authors which were used for VMS writing. Looks, that was a right direction for further research.

Again – the general assumption regarding last VMS page text – it contains list of books from John Dee library.

## 10. Last VMS page text decoding.

First of all the last page consist of some earlier identified symbols.

According my opinion the handwriting on the last VMS page is different to other VMS parts and performed by another person and performed sloppy.

That creates some difficulties for symbols identification, especially because of that some symbols similar to already identified were not used for decoding.

There is also VMS page 66r where we can find some symbols also used on the last VMS page.

At the moment we are not able to identify that symbols, but probably it gives us the chance to verify identified symbols.

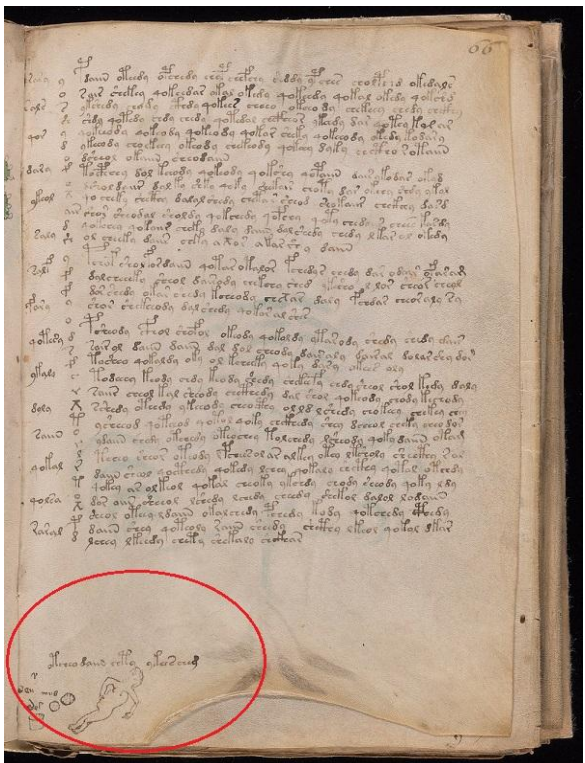

Fig.100 VMS page 66r with location of codes similar to last VMS page.

Let's start our search.

Fist candidate to the identification will be the following symbols string:

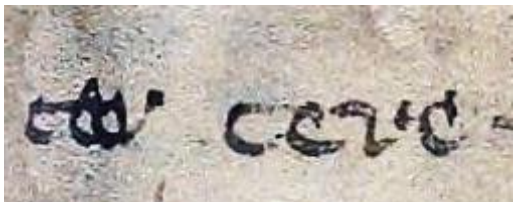

Fig. 101 – the candidate to deciphering

We knew that the used coding method initially developed by Trithemius or Trithemi. The selected string consist of symbol "TRE" which spelling can be also [tri:]. At least a chance to catch something.

Just a small reminder – for each identified key I shown how it can be spelled in a case of separate location and in combinations with other symbols.

Let's associate the longest part of symbols strings starting from "TRE" with possible deciphering as "TRITHEMI".
The shortest part also look like starting from "TRE" or modified "TRE"

Please see below the possible solution.

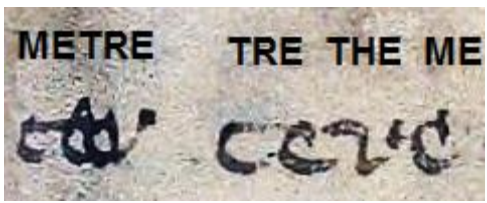

Fig. 102  Possible identified text "Metre Trithemi"

So. We probably find new keys.

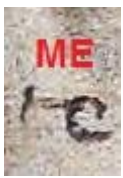

Fig. 103 -   Symbol "ME" ([mi:] and [mə])

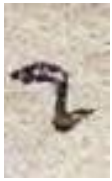

Fig. 104 -   Symbol "THE"

Regarding "METRE" – According to my assumption here were used the symbols "TRE" and "ME" with the following rule – if the part of the last symbol (in that case "ME") overlays on the symbol staying in prior (in our case "TRE") – the order of reading should be
  - To read last symbol at first – "ME"
  - To read the symbol, located in prior, next – "TRE"

We will see same trick later.

Anyhow Trithemius books and manuscripts are present in Dee catalogue.

The next assumption was that the text should contain the word "BOOK" or, to be correct, in the time of John Dee it can be also written as "BOK" (you can simply find that many times in his notes).

How to identify the location of that symbol.

At least some times it should be used in combination with symbol "THE" in prior (as many times in Dee notes).

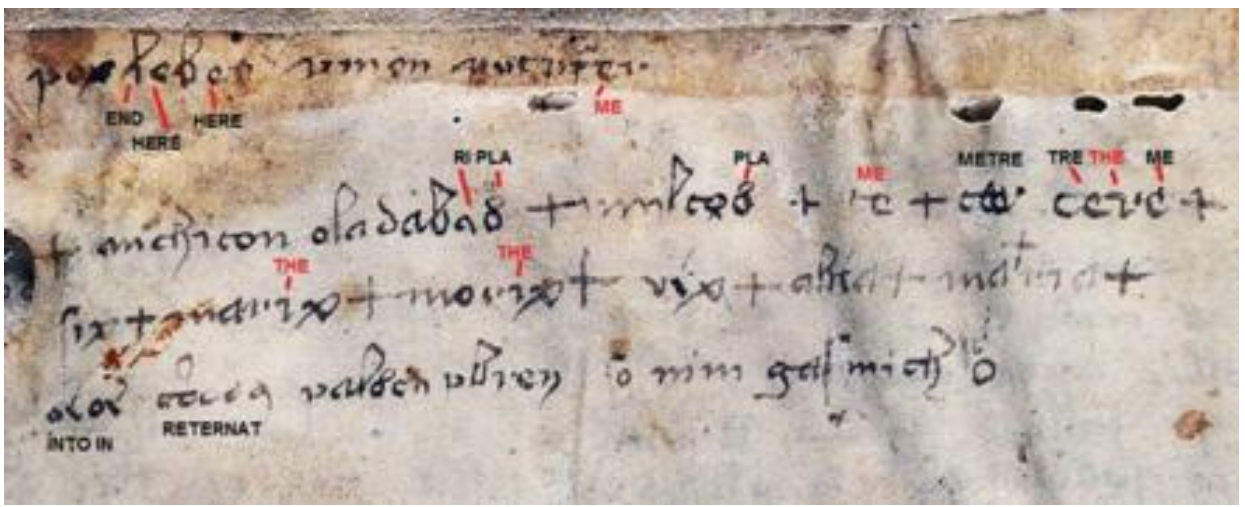

Fig. 105 Identified symbols on the part of last VMS page (in black – used for main text, in red – newly identified) .

Now we've find locations of symbol "THE". Let's follow our logics that the next symbol should mean "BOK" or "BOOK"

The candidate to be verified is:

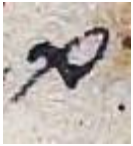

Fig. 106 Possible symbol "BOK" or "BOOK"

Now we have:

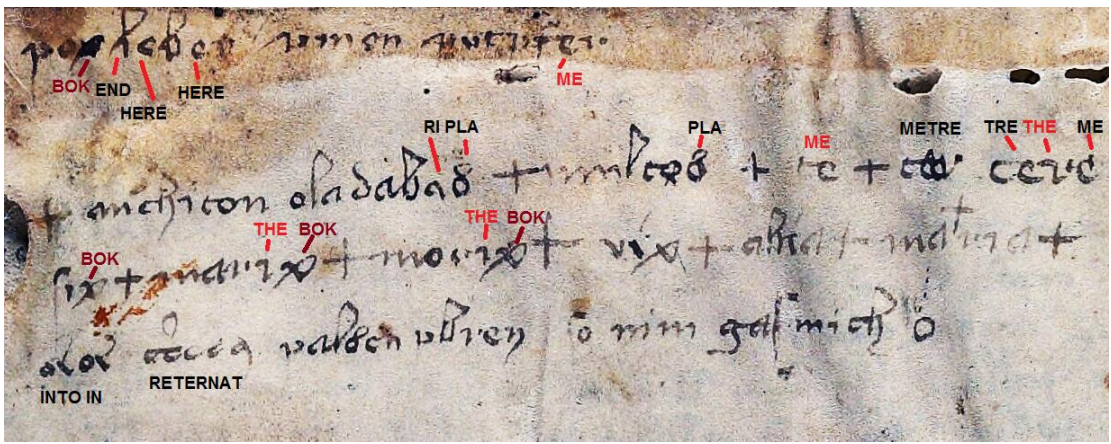

Fig. 107  Identified symbols on the part of last page.
Next my assumptions based on VMS content was to try to find the location for the following books or it authors authors – Avicenna and Villanova.

At first the good candidate for Villanova was

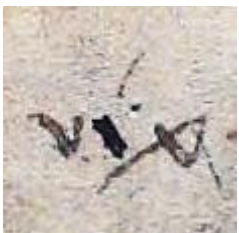

Fig. 108 Possible candidate to VILANOVA

So. If assumption above is right let's fix new keys:

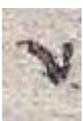

Fig. 109 Symbol VI [like modern eng. "we"]

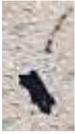

Fig. 110   Symbol "IL" or "ILA" [ila:]

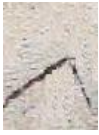

Fig.111   Symbol "NO"

In that case we have the same  rule as with "METRE" – symbol "NO" overlays symbol "VI".

Finally we can read "VIILANOVI".

According some collected statistics we can also fix one more rule regarding how to read the text – if symbol located in prior ends to the same letter from which starts the symbol located next – that letter should be written once.
In a case with "VIILANOVI" – the right it description will be "VILANOVI"

Let's add identified symbols to the last page

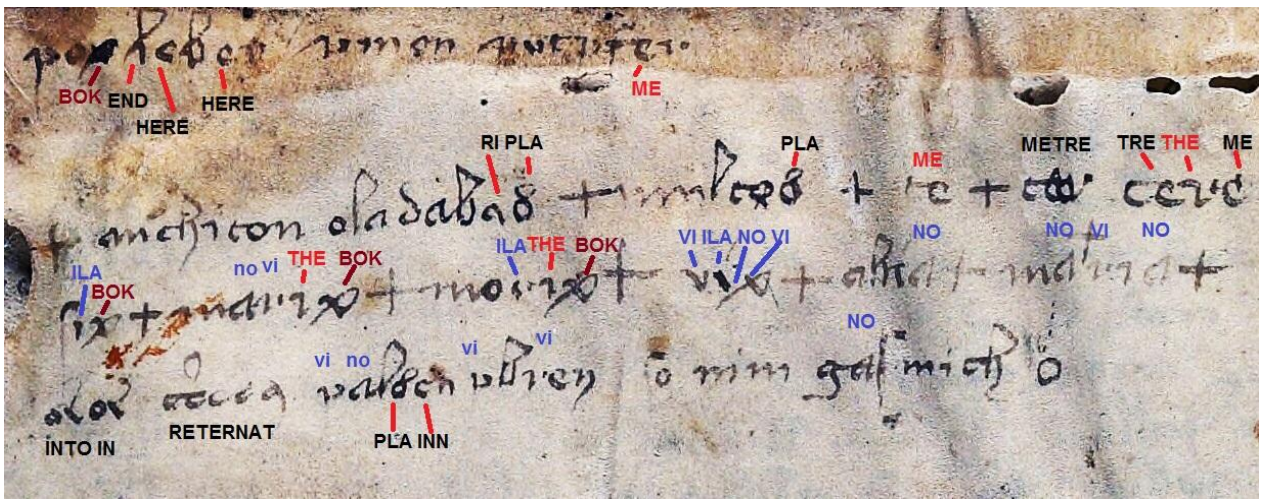

Fig. 112 Last VMS page with new identified symbols added.

At the moment we need some more hints for further research.

Let's come to the VMS page 66r and try to identify  possible keys.

Maybe the following image can help us, because it shows something like bones description:

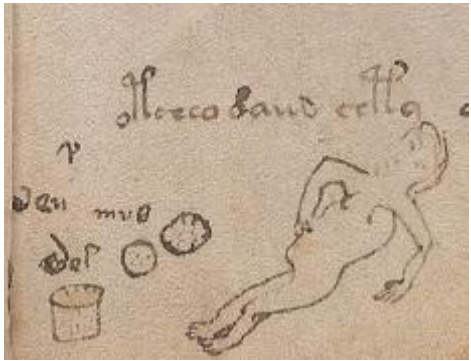
Fig.113 Image from the left bottom side of the VMS page 66r

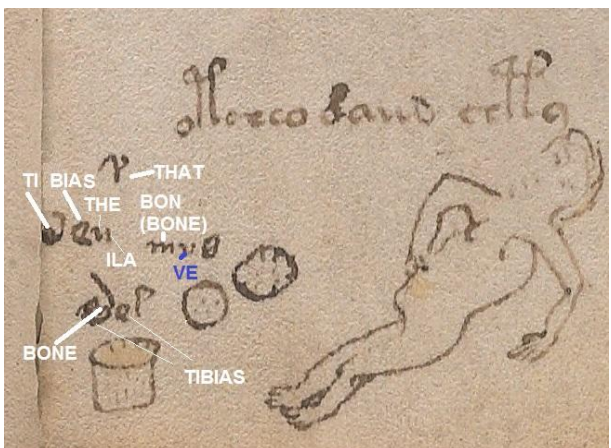
Fig. 114  VMS page 66r with possible solution.

So…
We have the following symbols:

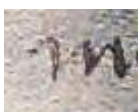

Fig. 115   Symbol "BONE"

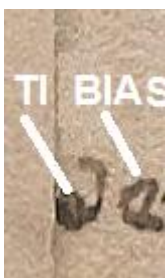

Fig. 116 Symbols "TI"[ti:] and "BIAS" (needs verification). Combination of both meaning "TIBIA" bone.

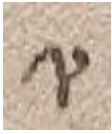

Fig. 117 Symbol "THAT"

Next one which requires more precise verification:

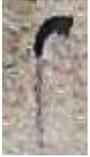

Fig. 118  Symbol "SIC"

Why? As I mentioned – all symbols are equal to unique single root words, which were commonly used in that time.

Also – the general assumption regarding last VMS page text – it contains list of books from John Dee library.

Let's have a look to the string #3 of the last VMS page.

It starts from symbol on the fig.118 and follows by symbol "ILA" and next by symbol "BOK".

Let's right now count the symbol on the fig. 118 as "SIC", we will verify it later.

Now we have the following combination: SICILA BOK.

Let's come to the Dee library catalogue and try to search similar combination.

The result can be the following:
"Joh: de Sicilia in canones Arzachelis de tabulis Toletanis.— Quaestiones mathematicales."

May be John of Sicily is mentioned person, but that can also be another mention – Sicilian book with no special links to the specific author. At least in the medieval history we can find some links to the translations of scientific books (i.e. Almagest) performed by Sicilians.

That needs more investigation.

Finally we have the following possible decoding:

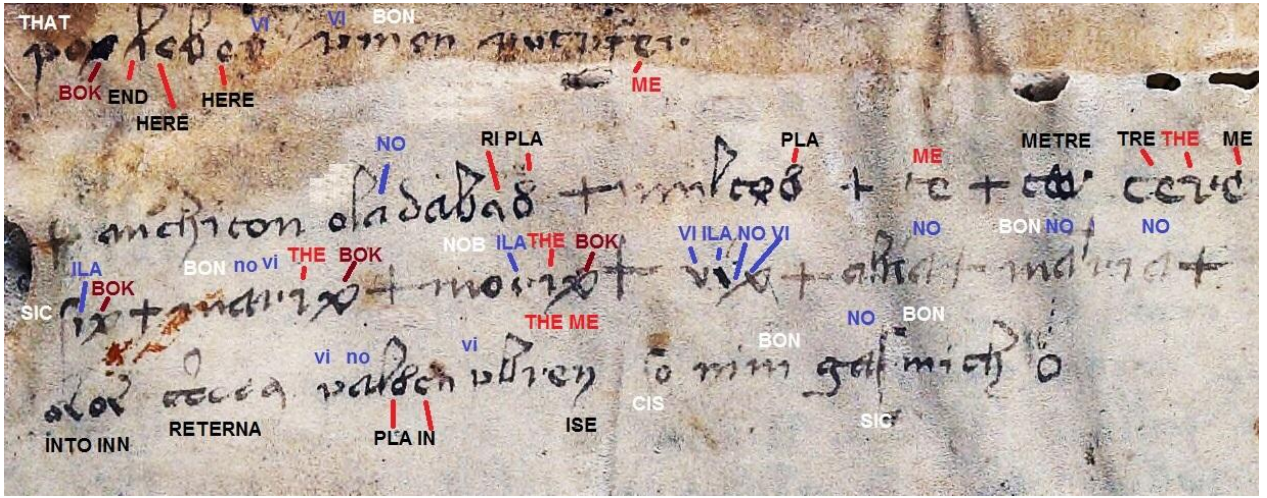

Fig. 119 Last page with possible decoding.

1) First symbols string:
   THAT BOOK END HERE HEAVY. WE BON ……………… ME….

2) Second symbols string:
   ………… …NO (*symbol of rule is used*) … RI PLA + (*may be in the meaning R.I.P*)…. ANY …. PLA + ME + METRE TRITHEMY + ……**...PLA + … + (these two last string sets are not shown on the Fig.119, but they exists on the last VMS page)**

3) Third symbols string:
SICILA BOOK + BONNOVI THE BOOK + NOBILA (*symbol of rule is used*)  THE BOOK + VILANOVA + …NO + BONNO + VI..NO

4) Fourth symbols string:

INTO INN RETERNA WE NO … PLA IN (*may be PLAIN or PLAN*)  … WE…… THEM ISE… CIS (*symbol of rule is used*)… BON …NO SIC BON …..
For further analysis we need more quality high resolution images of the last VMS page.

## 11.     Now it is a time to say "Goodbye" to VMS! Let's say "Hello!" to the "Book of Dunston".

That was just my initial assumption. I even not expected such quick results…

Now the confirmation was found. During verification of the last VMS codes, because as I already mentioned in the chapter

10 – the handwriting on the last VMS page is different to other VMS parts and performed by another person and performed sloppy… So, some symbols like "NO" or "RI" were not so easy to recognize correctly. Now result is fixed.

The other uniqueness of the last page – the most of keys are used only once and it only verification possible in a case if the deciphering using that keys gives us a logical text.

I am especially left chapter 10 unchanged to keep it as an example of logics of search. Not all results in the chapter 10 are correct, but even with such basis as a start point – finally became to the success.

At first let's come back to the following Dee's notes related to 12$^{th}$ of December 1587 ([8] p. 25;[9]).

**"Dec. 12th [1587], afternone somwhat, Mr. Ed. Keley his lamp overthrow, the spirit of wyne long spent to nere, and the glas being not stayed with buks abowt it, as it was wont to be ; and the same glas so flit.ting on one side, the spirit was spilled out, and burnt all that was on the table where it stode, lynnen and written bokes, as the bok of Zacharius with the Alkanor that I translated out of French for som by spirituall could not ; Rowlaschy his thrid boke of waters philosophicall; the boke called Angelicum Opus, all in pictures of the work from the beginning to the end ; the copy of the man of Badwise Conclusions for the Transmution of metalls ; <span style="color:red;text-decoration:underline">and 40 leaves in 4°(\*), intitled, Extractiones Dunstani, which he himself extracted and noted out of Dunstan his boke</span>, and the very boke of Dunstan was but cast on the bed hard by from the table."**

*((\*) 4° – probably means folder of 40 extracted pages or 4 folders or stacks with 10 extracted pages in each – A.U.)*

Let's also pay our attention to the previous note, dated 11$^{th}$ of December 1587:

*"Dec. 1st to 11$^{th}$, my Lord lay at Trebon and my Lady all this tyme[…]"*

Any tiny detail here is very important.

Truly sad – best trick I ever seen…

Let's start!

What we knew from these 2 notes?

(1)   It is a **winter** time – mid of **December** 1587 (probably snow and/or ice is present outside)

(2)   **Dee** and **Kelly** staying in **Trebon.**

(3)   **Dee** and his wife or Kelly and his wife or all of them are **ill.** It is difficult to understood whom Dee mentioned as "my Lord" and "my Lady". My assumption – he is telling about himself and his wife.

(4)   **Big fire (or flame)** caused in the room where Dee and Kelly worked with **books.**

(5)   That was the fault of Kelly.

(6)   One of the book – is the "**Book of Dunstan**"

**(7)**   There are also $4^0$ stacks called **Extractiones Dunstani**…

(8)   Kelly noted in the book of Dunston that he extracted 40 pages.

The **handwrite** of last VMS page is different to other parts of book and looks like performed by **another hand.**

**Again:**
- **according my initial assumption – our VMS = "Book of Dunstan"**
- **authorship: idea – John Dee, writing –Dee and Kelly, or one of them**

Now I'll show you the real text of last VMS page. String by string (see Fig. 120)

As you remember – I've already focused your attention that for understanding the text you should point you attention to spelling of the words. May be it is more easy for me because English is not my native language and my mind more flexible to understood the mentions on medieval English.

As an example in my case: a)"ve" = [vi:] = modern "we"  and also "vy"(as the ending of the word) - we are talking about primary key spelling mentioned by VMS author in some cases it should be spelled as [və]

b) "me" = [mi:] = modern "me"

in the same time it can be spelled as [m] or as [mæ] – it depends of the symbol location.

c) "ni" = "ny" = [ni:]

d) "ri" = "ry" = [ri:]

e) more complicated:
"AND" – cosound also as "END" = [ənd].
In the beginning or in the middle of sencence it means "AND", in the end means "END"

That is a rule for any keys of VMS. The authors plays not with grammar, but with spelling or better say – sounds of keys he used and what is interesting – we always see significant solution.

But let's start.

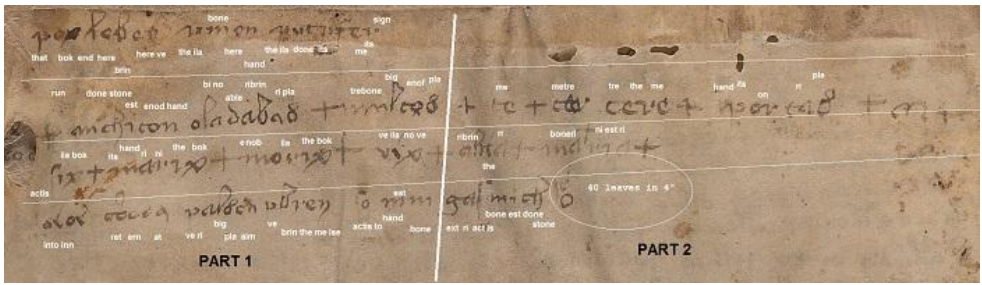

Fig. 120 Last VMS page

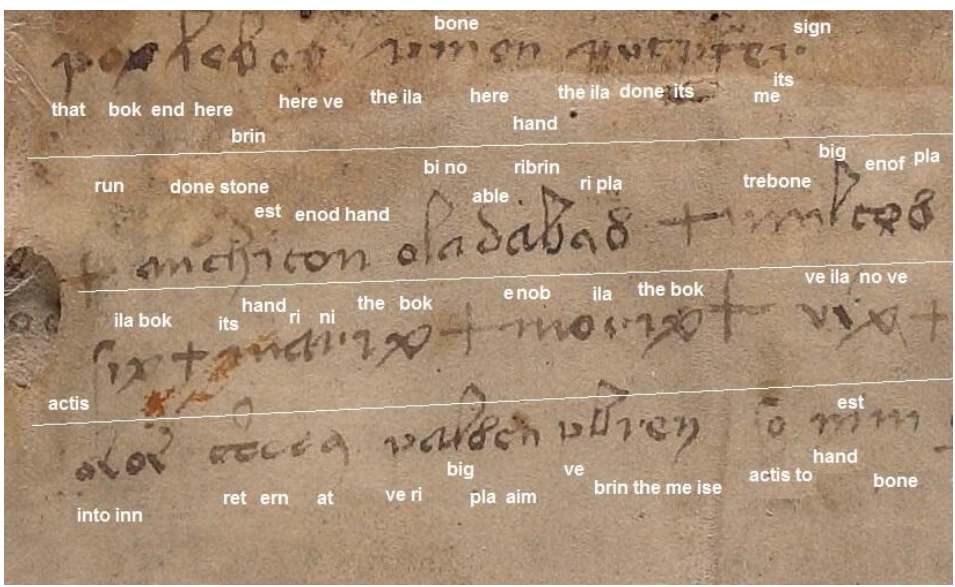

Fig.121 last VMS page PART 1

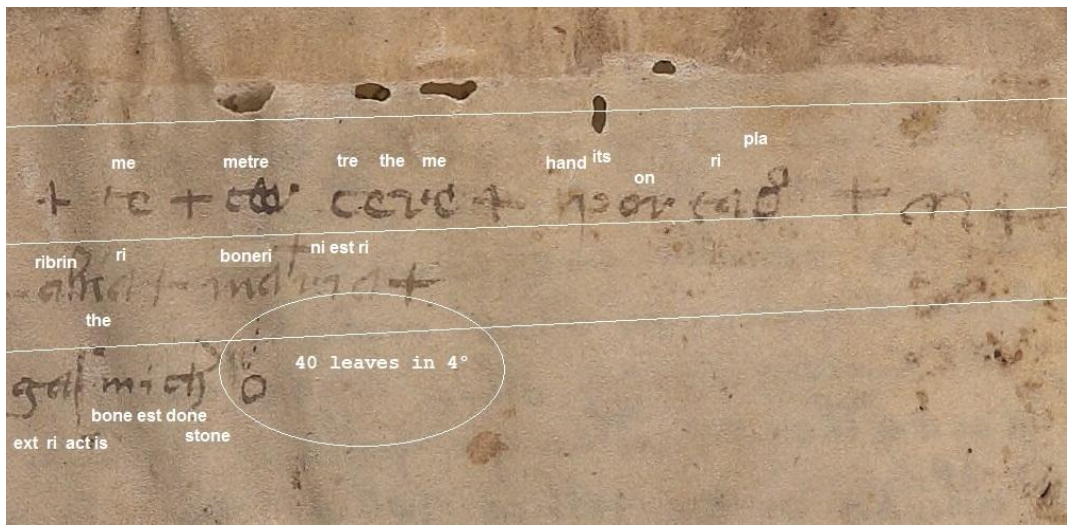

Fig. 122 Last VMS page PART 2

**String #1.**

**"That bok end here. Brin here ve the ila bone here hand. The ila done its me its sign"**

So. That string says:

**"That book ends here. Bring heavy illness of the bone of the hand. The illness donates me its sign(s)"**

So. We knew from Dee's notes – he was ill that time (see (3)) above.

**String #2.**

**"Run done stone est enod** (symbol of rule "o" is used after the word "done") **hand bi** (symbol of rule "o" is used before the word "big") **no able ribrin ripla + trebone big enof pla + me + metre tretheme + hand its on + ripla + ….."**

So. That string says:

(1) **"Run Dunston asthenic (est enod - can be also translated as "another") (hand. Be no able to re-bring or replace. Lay in Trebon big enough. Me. Metre Trithemij. His hand works. Replace** *(AU: probably – replaced)….. "*

**OR**

(2) **"Run Dunston using  enother hand.  Be no able to re-bring or replace. Lay in Trebon big enough. Me. Metre**

> **Trithemij. His hand works. Replace**(AU: probably – replaced)….. "

So. Our author tells us that he is writing Dunstone (6) by asthenic hand (!) – that's why we see that last page performed by another hand (8)

### String #3.

**"actis ila bok + its hand ri ni the bok + ebon ila the bok + velanove + ribrin the ri + its hand ri + ni est ri + "**

So. That string says:

**"Working ill with book. His hand running the book. A nobili the book. Vilanova. Re-bring very. Boneri**(probably – Edmund Bonner – Bishop of London is mentioned – AU) **. Niestri** (probably – "nostri" - AU) **"**

So. Our author still sic. "His" – most probably – Kelly is mentioned.

As you can see in all Dee notes - he is very often using mix of languages – English and Latin. In our case we see the same situation. As well as Dee often used shorthand writing of the words. Here we have the same.

### String #4.

**"into inn reternat ve ri big pla aim we brin the me ise actis to handest bone  ext ri actis bone est done stone . . . . o"**

So. That string says:

**"After coming back - very big flame.  We bring them** (probably – they used the snow and ice to extinguish the fire - AU) **ice. Working on hand's bone. Extractis  bonest Dunston. 40 leave in 4°** (probably extracted stacks of book of Dunstan are gone in the fire – AU) **"**

So. Now we see the part of story described in Dee's notes of 12th of December 1587. We knew that it was Kelly's fault.

Do you remember Dee's note **– "…which he himself extracted and noted out of Dunstan his boke…"**

Dee wrote that Kelly noted that fact in the book of Dunstan.
**Now we have found that note…**

I'm not 100% sure, but I am still stay on the position that the manuscript author is John Dee.

So… That's the end of the story.

**Verified keys for last manuscript page:**

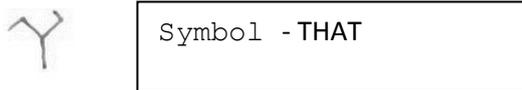

Fig. 123   Symbol # 31 "THAT"

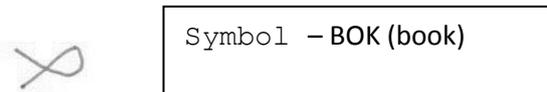

Fig. 124   Symbol # 32 "BOK"

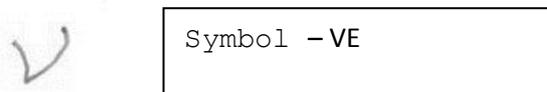

Fig. 125   Symbol # 33 "VE"

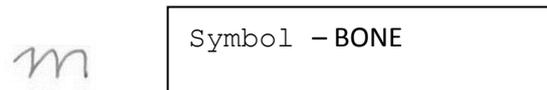

Fig. 126   Symbol # 34 "BONE"

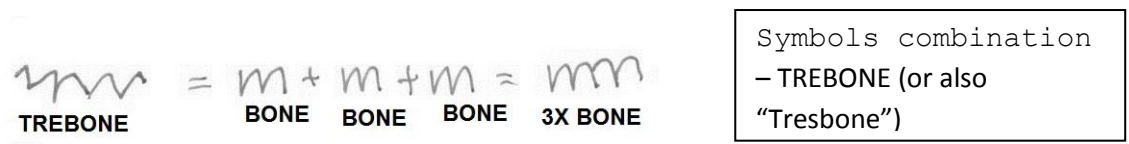

Fig. 127   Symbols combination "TREBONE"

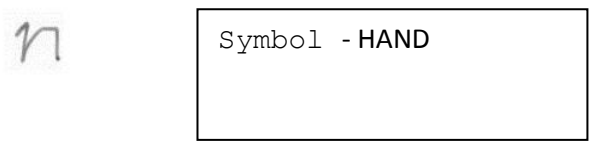

Fig. 128   Symbol # 35 "HAND"

Symbol – BRIN (bring)

Fig. 129   Symbol # 36 "BRIN"

Symbol – NI [ni:]

Fig. 130   Symbol # 37 "BRIN"

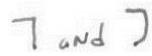  Symbol – ITS

Fig. 131   Symbol # 38 "ITS"

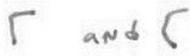  Symbol – DONE

Fig. 132   Symbol # 39 "DONE"

Symbol – EST

Fig. 133   Symbol # 40 "EST"

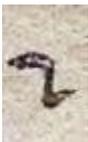  Symbol – THE

Fig. 134   Symbol # 41 "THE"

Symbol – ME

Fig. 135   Symbol # 42 "ME"

Symbol – STONE

Fig. 136   Symbol # 43 "STONE"

```
Symbol - NO
```

Fig. 137   Symbol # 44 "NO"

```
Symbol - ILA
```

Fig. 138   Symbol # 45 "ILA"

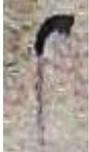
```
Symbol – ACTIS or ACTUS
or ACT
```

Fig. 139   Symbol # 46 "ACTIS".
That symbol initially was not correctly identified in the chapter 10 (see Fig.118). Now we've corrected the mistake.

```
Symbol - EXT
```

Fig. 140   Symbol # 47 "EXT"

```
Symbol – SIGN
```

Fig. 141   Symbol # 48 "SIGN"

```
Symbol – ENOF (enough)
```

Fig. 142   Symbol # 49 "ENOF"

```
Symbol – ABLE
```

Fig. 143   Symbol # 50 "ABLE"

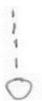
```
Symbol – 40   or/and   4⁰
```

Fig. 144   Symbol # 51 "40 or/and $4^0$"

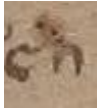
```
Symbol – AIM
```

Fig. 145   Symbol # 52 "AIM"

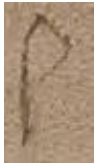
```
Symbol – BEE or BI or BIG
```

Fig.146   Symbol # 53 "BEE"

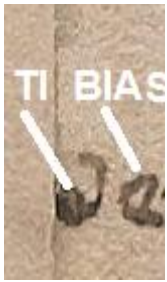
```
Symbol – "TI"

Symbol – "BIAS"
```

Fig. 147 Symbol #54 "TI"[ti:] and  symbol # 55 "BIAS", identified in the chapter 10.

So. Only on the last page manuscript author used 23 (one symbol was not been identified) additional keys for coding… It is comparable with quantity of keys used for all other  parts of the manuscript.

Let's come back to the verify symbol#25

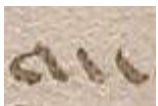
```
Symbol – RUN [ran]  or RAN [ran]
```

Fig. 148 Symbol #25 (RUN)

It verification was found on the manuscript page #65r. It contains only one symbol string.

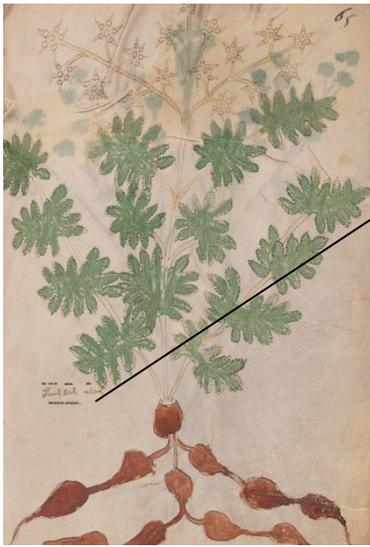
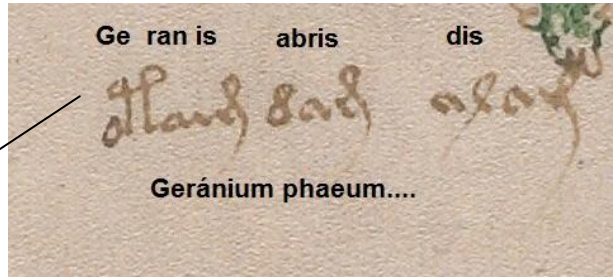

Fig. 149   Page #65 with "GE RAN IS"

So, it is finally confimed now that:

- **Voynich manuscript  initial title was "Book of Dunstan".**

- **It is was written in the period  1583 – 12<sup>th</sup> of December 1587**

## Interesting riddle with possible solution .

Let's come back to the page 66r.

Let's have a detailed look to the image of naked woman.

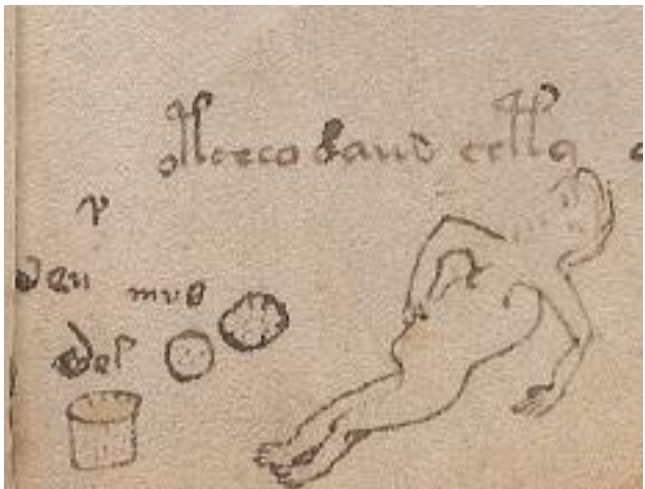

Fig.150 image of naked woman  from the page 66r

There is a lot of  women images in VMS (nymphs)
But that one is more interesting.

What we can tell about her?

- at first, it look like we can't see the bed (or couch) on which she lays. If we will mentally add the bad we will receive something like on the image below. [18].

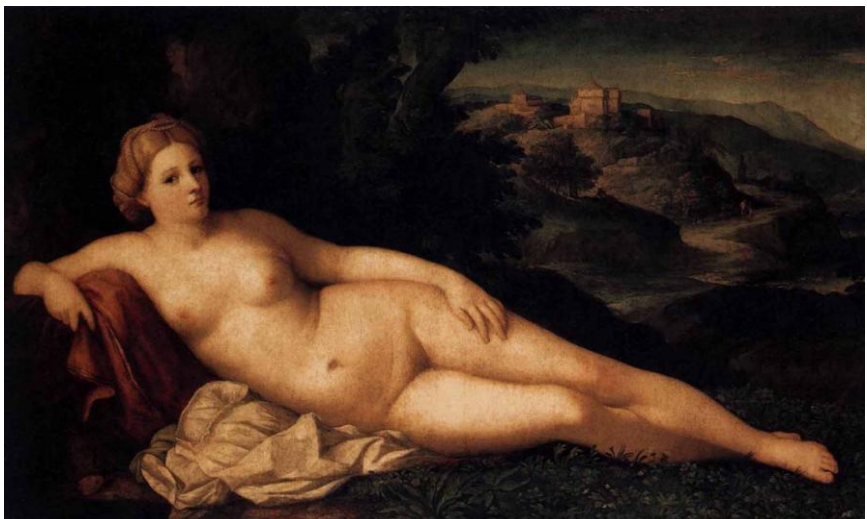

Fig. 151 Palma il Vecchio "Venus" [18]

- at second, the girl on the Fig. 150 looks pregnant…
- at third, the face of that women (even drawn very primitively) - looks like face of live woman, with light sadness. Like manuscript author draw it from real woman.

Let's try to identify possible person.

We need to use John Dee notes again [8].

The note of 28$^{th}$ of February 1588 tells us the following:

«Feb. 28th, mane paulo ante ortum solis natus est Theodorus Trebonianus Dee […] (*Trebonianus – who was born in Trebone - - A.U.*) »
So, John Dee son – Theodorus Trebonianus Dee - was born in 2,5 months after the "Book of Dunstan" was finished... So in December 1587 Mrs. Jane Dee was Pregnant already seven months.

So – on the image on Fig.150 most probably the "rough" portrait of Jane Dee – wife of John Dee.

I tried to find the original portrait of Jane, but unfortunately I am not find any in internet.

But I found the modern one, which was performed by illustrator, who probably have seen her original portrait.

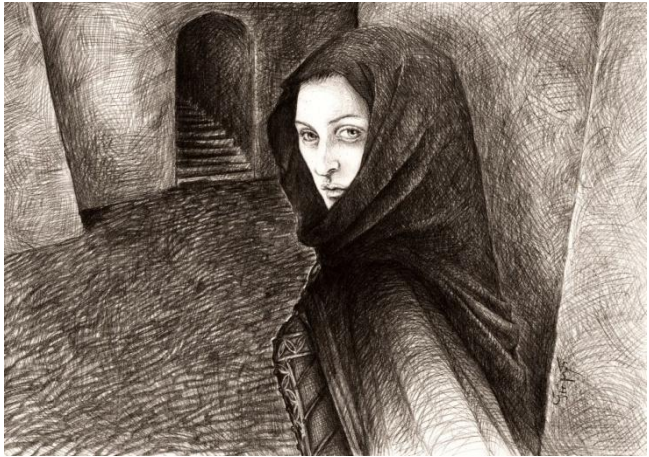

Fig. 152 Jane Dee by Meisiluosi [19] – the illustration for movie by novel of Lucie Lukačovičová «Green dragon, red lion» (Zelený drak, karmínový lev)  - historical fantasy about alchemists in Prague in  Rudolf II period.

Here I want to point out your attention to the semantic code on the page  66r. The cheat is not direct and not that simple:

- the images of cylinder and circles we identified as – «bone».

- the pregnant woman should (logically) birth a child, So – child will born.

"Bone" and  "Born" – different grammar, different meaning, but spelling is close to each other.

## 12. Final conclusion re VMS coding methods.

I only would like to point you attention to the quantity of coding methods the VMS author are used and also I would like to give you some ideas for further research (hopefully it will be useful):

1) "simple" steganography  - using images with masked letters (for example for last VMS page).

2) "modified" Trithemius steganography for main text. Modified means: that author used not a single letters but commonly used singleroot words ("ret", "ort", "as", "key", "ant" etc), special symbol "o" which sets the rules how to read neighboring symbols; arbitrary method for symbols combinations for phrases constructions when even long phrases look like a simple small words; author not used the punctuation. Also semantic level which allow to construct words even with incorrect grammar but with significant meaning.

3) Symbol "a" (RI, RY - [ri] & [rri]) – another symbol of rule. If there are the symbols string in the following sequence "a" "other symbol" "a"  - symbols "a" are not readable, other symbol cutted till last consonant letter, and that last consonant letter should be readable with letter "i" ([i:]) following next. I.e. if we have symbol "INN" in between symbols "RI" the final spelling will be "NI" ([ni:]), if the middle symbol is "ALL" – final spelling is "LI" etc.

4) Last VMS page most probably performed by the person different to one who wrote main VMS text. That page consist of unique – single used keys which are coding also the list of books (or/and it authors names) which were used for VMS preparation.

5) Additional rule how to read the text – if spelling of symbol located in prior ends onto the same letter from which starts the spelling of the symbol located next – that letter should be written once.
I.e. In a case with "VI ILA NOVI" – the right it spelling will be "VILANOVI"

6) Allegory images like on Fig.1 where main feature of the plant required for the right choice is -  the tree with double trunk,  but not it strange leaves or other parts. The second sign can be red color or tree blood as VMS author meaning.

7) "Inverted" colors – like in a case with sunflower (see VMS page 33v). The middle part should be black or dark blue, petals – yellow. It is a very nice trick to refocus your attention from keys location to useless features and decisions.

8) Latin letters scattered on the pages, like on the page 4r. Let's try to find all the letters on that page which gives you word "AFTOR". It is also allegory and for reading other hidden text you need to follow the instructions which are present on the image on VMS page 57v. Unfortunately some pages to which numbers links us the instruction are lost.

9) Same letters but in different colors (and scripts). You should know which letter with which color should be used for words construction. Same - you need the instructions from VMS page 57v.

10) Special symbols like on the page 8v – in the right bottom corner of the page.
It follows in some order  - pages marked by special symbols: 8,16,24,32, 42, 58 etc.  - you can find it by yourself.
My assumption that it also divides the calendar year for certain periods important for growing and usage of plants.

11) Probably symbolic signs like on the page 11v  - in the middle of the left side of the page – which look like "88"

12) Images of constellations for coding roman numerals as on the last page of VMS

13) Images of nymphs  - allegory of diseases.

14) Allegory images of human organs

15) Allegory images in astronomical and astrological parts

16) Allegory images in pharmaceutical part

17) Self developed numerals like on the page 57v in combination of Arabic numerals.

18) Special traps or tricks for refocus your attention from right decoding direction or may be some additional hints for right interpretation of text like presented on the page 49v (Arabic numerals vs VMS author created symbolic numerals).

**Main:**
1) **VMS = The "BOOK OF DUNSTAN",**
2) **Manuscript is not a sham because of it readable and structured content.**
3) **In the same time it is a nice fake of ancient manuscript, performed in one significant exwmplar.**
4) **It performed by Dee (and somhow Kelly), started in England in 1583 and finished in Trebon in December 1587.**

I would like to add that John Dee most probably never sell that manuscript to Rudolf, as well as never was a middleman in the deal  - most probably manuscript was donated to Rudolf by Kelly, or Rudolf confiscated it and added to his library after Kelly's death.

## 13. Addendum.

### IMPORTANT DATA

The high resolution images available on VMS web-site gave us the possibility to investigate very tiny details on the VMS images. One of the most interesting openings you can find below.

### 3 QUEENS

In the astrological part (Moon calendar) we can see the 12 circle diagrams consisting of zodiac symbols and we can find 3 specific diagrams with "nymphs" which has the crowns on their heads.

1) CANCER

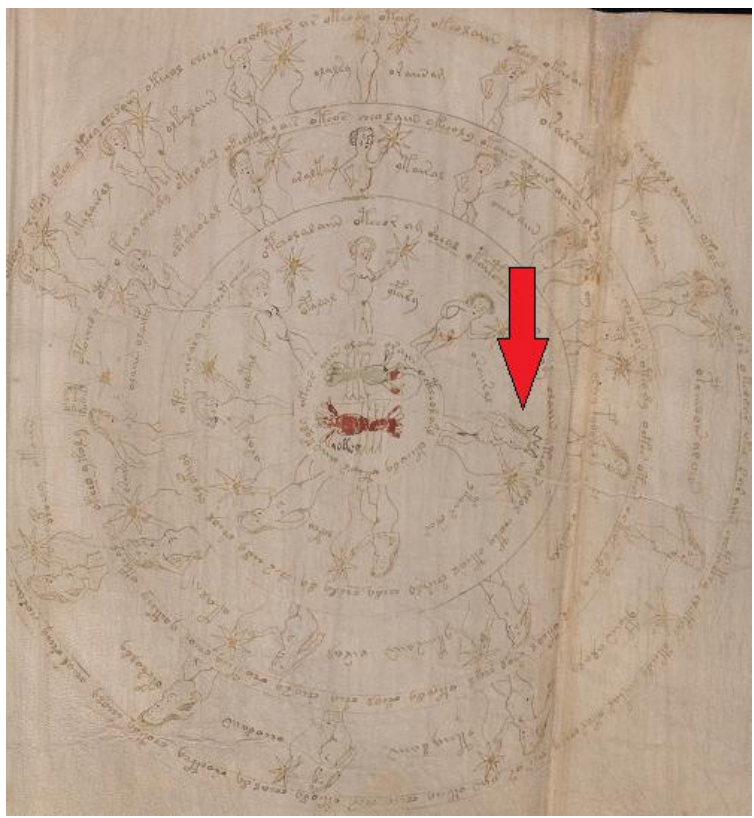

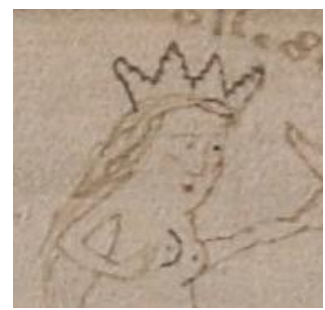

Fig. 154 Cancer's "queen" (№1)

Fig. 153 Zodiacal diagram of Cancer with "queen" location.

2) LION

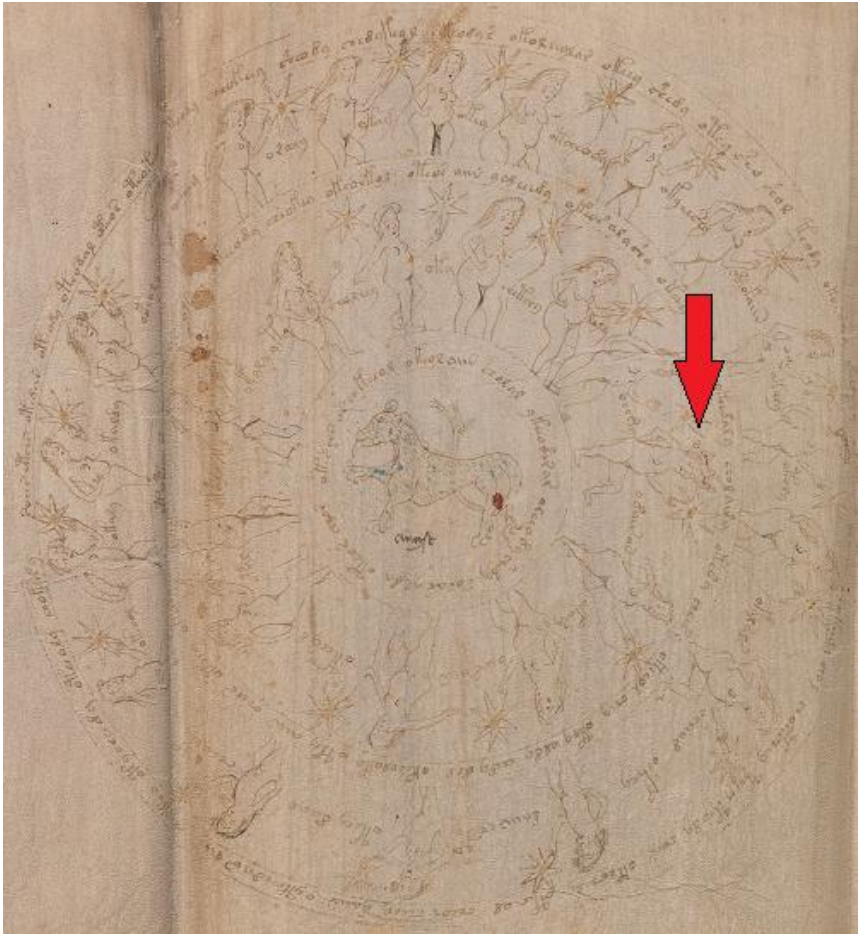

Fig. 155 Zodiacal diagram of Lion with "queen" location.

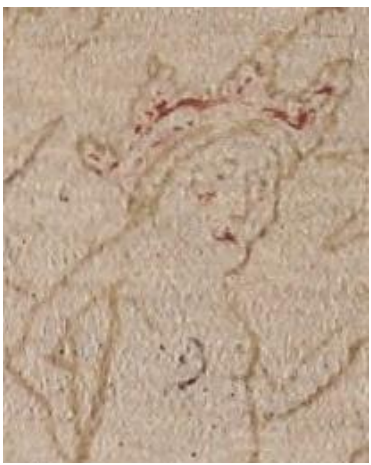

Fig. 156 Lion's "queen" (№2)

And finally most impressing "queen"…

3) LIBRA

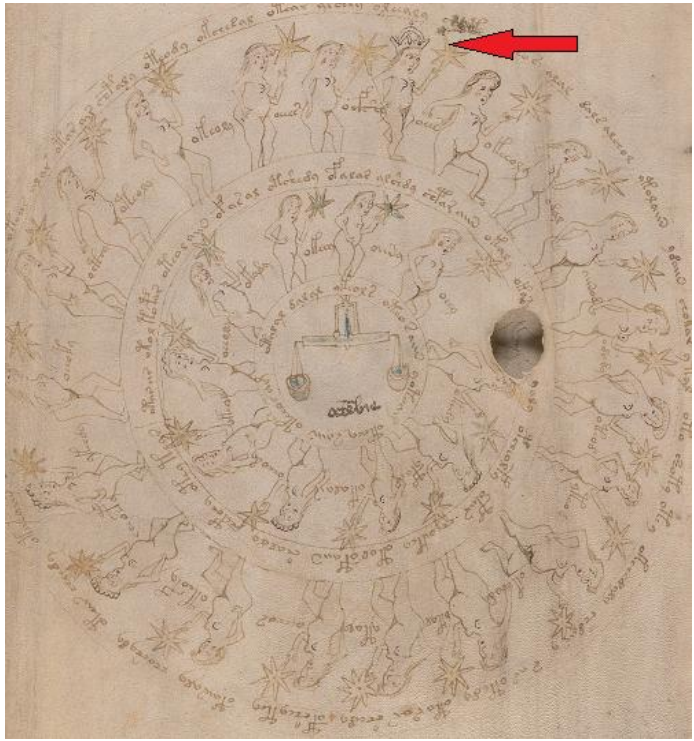

Fig. 157 Zodiacal diagram of Libra with "queen" location.

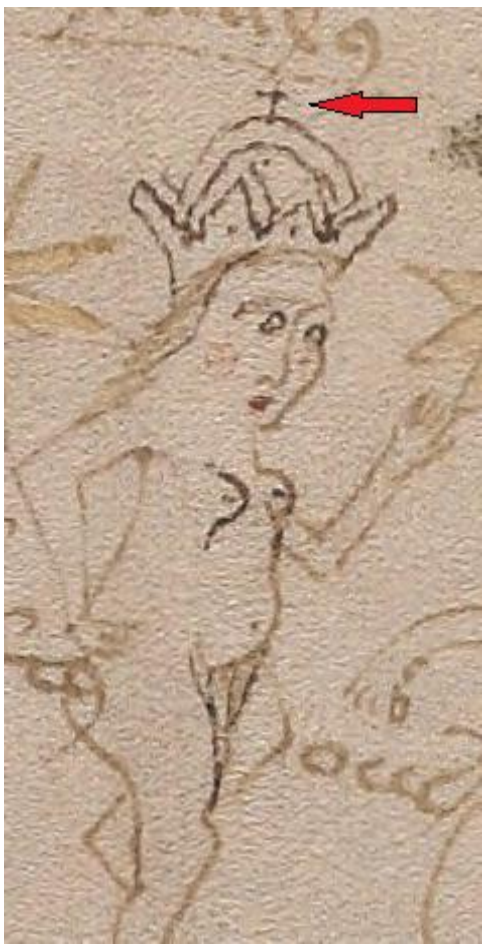

Fig. 158 Libra's "queen" (№3)

We see the Christian crown! That confirms that the author of manuscript was a Christian!

That also tells us that all versions linked to Hebrew or Jewish, Arabic, Islamic, Asia or America origin of manuscript are highly likely wrong…

What can be the origin of that crown?

So. It can be copied from another image, it can be seen by authors of VMS – Dee and Kelly during their continental journey in Poland (crown of Batory) or in Prague (crown of Rudolph II).

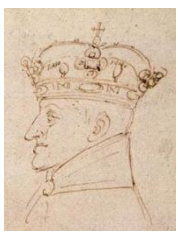

Fig. 159 Crown of Rudolph II from one of his portraits done by Giuseppe Arcimboldo (1575, from the collection of national gallery in Prague). [20]

**MORE INTERESTING - WHO ARE THESE 3 QUEENS and why only 3 queens?**

1) CANCER's QUEEN is the… **Lady Jane Grey** from the House of Tudor (c. 1537 – 12 February 1554)! Known also as Lady Jane Dudley (after her marriage) and as "the **Nine Days' Queen**" - d*e facto* Queen of England and Ireland from 10 July until 19 July 1553.
2) LION's QUEEN in the **RED CROWN,** red mouth, red spot on ugly face is… the Queen **Mary I Tudor** (18 February 1516 – 17 November 1558) was the Queen of England and Ireland from July 1553 till her deaf in 1558. She is known as "**Bloody Mary**" (that why her crown is red and she has red spots on her ugly face). John Dee was badly hurt because of her actions.
3) LIBRA's QUEEN is the Queen **Elizabeth I** (7 September 1533 – 24 March 1603), **The Virgin Queen** or **Good Queen Bess -** the last monarch of the House of Tudor. Truly Christian Queen of England who was loyal to all the confessions of Christianity.

Let's go deeper… It looks more and more interesting. Let's use the astronomy for help.

So, John Dee was the astronomer at first. For that coding he used real (not astrological Zodiak sign) data of the sun position for the moment when each queen has started to be queen. So…
**Lady Grey** – started when sun physically was in Cancer constellation… On the diagram we see 2 cancers – red and white. That means most probably - eclipse. Lady Gray started as a queen on 10[th] of July 1553… That Exactly the day of the annular solar eclipse when Sun (red cancer) was partially covered by MOON (white cancer).
**Maria I** – 9 days after when sun physically was in Lion constellation…
**Elizabeth I** - 17 of November 1558 when sun physically was in Libra constellation…
Exactly all that we see on the diagrams...

16-th century was a unique one because 3 queens from the House of Tudor successively changed one another. Only the case in the English history.

That is additional proof confirming Dee authorship of manuscript.

## 14. Important remarks.

i) First of all I would like to thank all the readers (especially many thanks to Mr. Eugene Tsyganov) who was so kind to send me the comments and info re some grammar and stylistic errors in the text, existing because of my poor (and basically technical) English.

ii) All the assumptions, results and conclusions - are the private opinion of author and it is the matter for further detailed research and discussions.

iii) Regarding links between Tritemius "Steganographia" and VMS (especially with of assumption of similarity of circle diagrams on the Fig.52, 53, 55) – I would like to ask you to do comparison by yourself using the assumption provided and original sources if necessary. Note: that can turn you to another direction like Magic of Salomon and so on which is not corresponds with a tasks for current article.

iv) Some of samples of deciphering in were especially taken in relation to different earlier publications of other authors who claimed codes identification. I'll try to add corresponding links if possible.

v) I would like to apologize for any mistakes in the text because of my poor English. I'm doing my best to improve it.

I would like to ask the VMS specialists and fan's community to co-operate for further research for codes, keys and rules verification.

The following operations were performed for contrast enhancement – autotone, autocontrast, automatic color contrast& manually set contrast level 200.

# Манускрипт Войнича («Книга Дунстана») - методы кодирования и декодирования.

**Аннотация**: Манускрипт Войнича (VMS) – книга, датируемая 15 или 16 веками, написана с использованием загадочного кода. Данная статья посвящена расшифровке манускрипта, установлению авторства и даты его написания, а также оригинального названия – «Книга Дунстана». Главы 1 – 7 посвящены поискам кодов к основному тексту и сбору косвенных «улик», подтверждающих первоначальные предположения. Непосредственные доказательства представлены  в главе 11 с расшифровкой текста последней страницы.

Ключевые слова: Voynich manuscript, манускрипт Войнича, коды, codes, кодирование, coding, декодирование, расшифровка, decoding, decipher, VMS, Эдвард Келли, Джон Ди, Kelley, Kelly, Dee, Тритемий, Дунстан, книга Дунстана, Trithemij, Trithemius, стеганография, steganographia, steganography.

## 1. Введение

Существует средневековый манускрипт, знакомый многим исследователям под названием «Манускрипт Войнича» или «Рукопись Войнича» (для его обозначения часто используется аббревиатура – **VMS**, которой я также буду пользоваться в данной статье). Этот манускрипт также называют – "Книга, которую никто не может прочесть", поскольку ни язык, на котором она написана, ни ее автор, ни место написания, ни дата написания – неизвестны.

Свое современное название манускрипт получил от имени последнего известного владельца  - Михаэля Войнича (1865 - 1930) – польского революционера («Вилфрид» - его псевдоним), библиофила и антиквара, мужа известной писательницы Этель Лилиан Войнич.

Войнич приобрел манускрипт для свой коллекции в 1912. Увы, это никак не помогло расшифровке рукописи, но, в конечном итоге, повлияло на ее дальнейшую судьбу – в 1959 наследники семьи Войнич продали манускрипт известному книготорговцу антикварными книгами – Хансу Крайсу, который в 1969 подарил манускрипт библиотеке редких книг Бенеке Йельского университета (Yale University Benecke rare book library), где он и хранится до настоящего времени.

На официальном вэб-сайте библиотеки редких книг Бенеке Йельского университета [1] приводится следующее описание манускрипта:

Позывной: Beinecke MS 408 (позывной – это специальный библиотечный номер для рукописей с оригиналами которых можно поработать в специальном помещении)

Альтернативное название: Манускрипт Войнича

Дата написания: [расчетная 1401-1599?]

Жанры: Манускрипт, Ботанические иллюстрации, Астрономические диаграммы, Рисунки, Ручная роспись, Иллюстрации

Материал: Различные материалы

Описание: Пергамент. 102 листа (современная фолиация, арабская нумерация; не все листы подшиты) + i (бумага), включает 5 двойных листов, 3 тройных листа, 1 четверной лист и 1 шестерной раскладывающиеся листы. Размер 225 х 160 мм.

Краткое содержание: Научный или магический текст на неизвестном языке, с шифром, видимо, основанным на малых Романских символах.

Физическое описание: 1 том.

Цветные иллюстрации

23 х 16 см. (в переплете)

Основной текст манускрипта и иллюстрации выглядят аналогично приведенным на Рис.1 (все изображения страниц манускрипта взяты с официального вэб-сайта библиотеки редких книг Бенеке Йельского университета [1].

Большинство растений, приведенных в манускрипте, в том виде, как они изображены, в природе не существует.

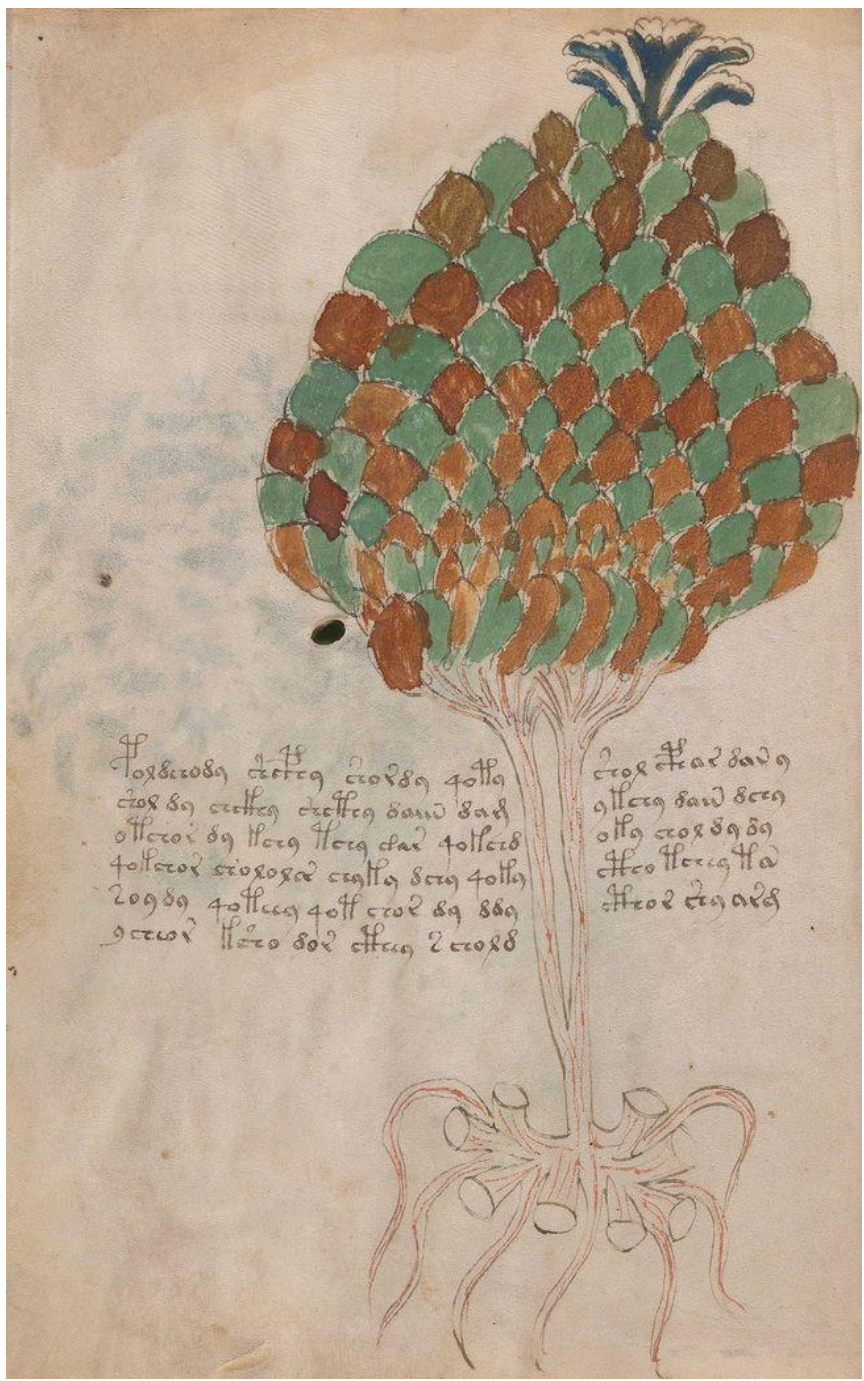

Рис. 1 Типичное изображение растения из манускрипта (страница VMS 11v в соответствии с нумерацией библиотеки).

## 2. Базовые предпосылки и установки, использованные для анализа манускрипта.

1) **Никаких стереотипов** – После ознакомления с рядом посвященных манускрипту статей, основным подходом был выбран «начать с нуля» – в попытке избежать влияния стереотипов, и тем самым избежать критических ошибок. Это, в частности, объясняет отсутствие длинного списка референций к данной статье.

2) **Верь своим глазам**  – давайте, доверяя собственным наблюдениям, попробуем получить некую квинтэссенцию наблюдений:

- Манускрипт имеет логически упорядоченную структуру(мы наблюдаем ботанические, астрономические, анатомические и фармацевтические данные)-  в некотором роде структура манускрипта напоминает «Канон врачебной науки» Авиценны;
- манускрипт содержит шифрованный и не шифрованный текст;
- манускрипт содержит аллегорические изображения;
- манускрипт содержит записи арабскими цифрами
- манускрипт в основной части написан четким стабильным почерком (за исключением последней страницы и еще ряда мест) с использованием как минимум 4-х различающихся рукописных шрифтов, каждый из которых имеет свое назначение: шрифт стенографического типа (для основного текста), латинский готический (например, в астрономической части  для маркировки зодиакальных созвездий), италик (для текста, отдельных букв латиницей и арабских цифр, используемых для шифрования некоего послания, разбросанного на страницах с основным текстом), «старый» (наиболее вероятно - «черный английский» ("black English"), аналогичный использовавшемуся  при написании т.н. роговых книжек (hornbooks) -  он так же задействован в системе шифрования дополнительного текста)  - все вышеперечисленные рукописные шрифты приведены на странице 4r манускрипта.
    - манускрипт содержит текст разных цветов
    - манускрипт написан на дорогом тонком пергаменте – велене – изготовленном из кож мертворожденных телят и/или козлят
    - на страницах манускрипта с основным текстом есть следы физического воздействия (соскабливания, стирания), следы букв латинского алфавита.

Теперь посмотрим на манускрипт более пристально – книга визуально разделена на следующие разделы:

- Ботанический – наиболее объемный. Он содержит изображения, и, возможно, описания неких растений, в определить которые по

внешнему виду в большей части не представляется возможным даже приблизительно.

- Астрономический – содержит схемы, где распознаются символы солнца, луны, звезд и зодиакальных созвездий. Зодиакальная часть содержит 12 соответствующих изображений созвездий, за исключением Козерога и Водолея, но при этом дважды повторяются диаграммы с Овном и Тельцом. В дополнение ко всему, зодиакальная часть содержит подписи к созвездиям (легко читаемые названия месяцев), выполненные готическими латинскими буквами.

- Биологический – раздел с аллегорическими изображениями человеческих органов – в виде резервуаров (типа бассейнов и т.п.) и каналов, наполненных жидкостями разных цветов. В каналах и бассейнах располагаются некие «дамы», или, как их некоторые называют – нимфы.

- Космологический – содержит изображения и диаграммы, содержание которых не совсем понятно и требует дополнительных данных.

- Фармацевтический – раздел, содержащий уменьшенные изображения растений, а также изображения гипотетических фармацевтических емкостей или сосудов – все вместе предположительно могут являться своеобразной инструкцией по составлению травяных смесей.

- Рецептурный – последний раздел, насыщенный короткими параграфами, начало каждого из которых обозначено символом звезды. Этот раздел практически не сдержит иных изображений.

Итак, благодаря такому рассмотрению манускрипта, первый вывод был, что это не фальшивка – в том понимании, что манускрипт содержит некий осмысленный текст.

Если данное предположение верно – это означает, что автор VMS (у которого не было компьютера под рукой) имел возможность читать текст, без использования каких-либо сложных кодов.

Если так – автор манускрипта использовал более-менее простой подход при кодировании манускрипта.

Если так – автор манускрипта, возможно, оставил ряд подсказок, как читать или где искать ключи и коды к манускрипту – для себя и посвященных читателей.

## 3. Базовые условия для поиска кодов и ключей.

### 1) Какие методы шифрования использовались во времена написания манускрипта?

Для начала, обратимся к датировке. Радиоуглеродный анализ пергамента, проведенный Университетом Аризоны, дал следующий результат: период 1404 – 1438 г.г. с вероятностью не хуже 95% [2].

Это очень точные цифры, которые вызывают сомнение.

С ними можно было бы согласиться, если бы в качестве референтного образца использовался образец, как минимум - из той же местности, где был написан манускрипт. Но, к сожалению, исходя из начальных условий задачи – место написания неизвестно.

Сравнение спектров проводилось с допущением, что распределение изотопов углерода в земной атмосфере однородно в течение некоего короткого промежутка времени, и, соответственно, результат измерений не зависит от того, где географически обитали животные, из шкур которых был выделан пергамент.

Здесь нужно сделать важное замечание, без учета которого можно совершить критическую ошибку – эта ошибка, кстати, свойственна профессиональному комьюнити исследователей манускрипта.

Нельзя путать датировку пергамента (т.е. попросту выделанных кож) с датировкой непосредственно манускрипта, т.е. сброшюрованного пергамента, содержащего текст и рисунки. Датировка пергамента дает нам только лишь период, в который были забиты животные, чьи кожи пошли на его выделку. Манускрипт мог быть написан позже или даже намного позже.

Существовала также и практика перезаписи манускриптов – в тот период весьма распространенная – поэтому следы соскабливания текста, заметные на страницах VMS, вполне могут являться дополнительным подтверждением того, что для его написания мог быть использован пергамент от другого, более старого манускрипта, содержавшего изначально иной текст.

Кроме того, на момент, соответствующий данным радиоуглеродной датировки (т.е. на начало XV века) получила достаточно широкое распространение и набирала популярность на равнее с пергаментом - бумага.

Существует статистика, собранная авторитетными исследователями старинных книг и манускриптов, посвященных в.ч. и методам датировки бумажных носителей. Здесь, в первую очередь, следует назвать имя Шарля Брике (Charles Moïse Briquet, 1839-1918), выдающегося швейцарского историка и филиграноведа, который написал один из основополагающих трудов по данной теме- Les filigranes Dictionnaire historique des marques du papier dès leur apparition vers 1282 jusqu'en 1600 (первое издание на французском в 1907, затем, начиная с 1967г., издание многократно переиздавалось на 3-х языках: французском, английском и немецком).

В соответствии с данными Брике, проверенными, и приведенными также в труде выдающегося российского палеографа В.Н. Щепкина «Руководство по датировке рукописей на основании водяных знаков бумаги и её залежностью», срок «залежности» бумаги, т.е. период времени от ее изготовления до первого использования, в зависимости от формата (стандартный или увеличенный) и сортности, мог составлять до 10-15 лет. Причем, статистика [http://www.raruss.ru/slavonic/slav5/1670-schepkin-water-marks.html ] показывает, что для стандартного формата бумаги в период:

- XIII – XIV веков в средний срок залежности (до 10 лет) попадает до 94% исследованных манускриптов, а 6% приходится на залежность от 11 до 14 лет

- XV – XVI веках в средний срок залежности (до 10 лет) попадает около 87% исследованных манускриптов, а порядка 13% приходится на залежность от 11 до 15 лет

А теперь рассмотрим бумагу увеличенного формата (для нас это означает в первую очередь её стоимость, в сравнении со стандартной – она априори выше, её ценовой класс ближе к пергаменту):

- XIII – XIV века - срок вылежки составляет до 23 года (!)

- XV век - срок вылежки составляет до 29 лет(!)

- XVI век - срок вылежки составляет до 32 года (!)

Т.е. для более дорогой бумаги срок вылежки только увеличивался.

Поскольку и бумага и пергамент использовались фактически для единого назначения – письма, то вполне возможно предположить, что для пергамента залежность тоже имела место. Поскольку пергамент являлся более дорогим в производстве и в продаже, и менее подверженным порче при длительном хранении, то для периода XV - XVI веков его залежность вполне могла составлять 30-50 лет без влияния на качество (в случае если использовался чистый пергамент).

Исходя из вышеизложенного, результаты датировки, полученные в Аризоне, были приняты мной за основу, но с корректировкой в более поздние даты – в нашем случае, как минимум, до 1500г.

Хотя, даже Йельский университет - непосредственный заказчик радиоуглеродного датирования, в случае с VMS гораздо более осторожен – верхняя планка возможного периода написания официально указана, как 1599г.

(*Позднее правильность данного подхода будет подтверждена.*)

В рассматриваемый период (1500 год) использовалось несколько методов шифрования, но наиболее прогрессивным для того времени был метод стеганографии, о котором мы поговорим позже.

### 2) Где искать подсказки, коды и ключи?

Важный вопрос. Однозначно, не в основном тексте манускрипта – так как мы не знаем языка, на котором он написан.

Вполне вероятно, что подсказки связаны с изображениями.

Если так, то существует всего 4 возможных локализации подсказок к кодам:

- a. В изображениях, возможно содержащих замаскированные буквы
- b. В подписях к изображениям
- c. С помощью латинских букв, разбросанных по страницам манускрипта (но, в данном случае, мы должны знать или обнаружить правила конструирования слов и предложений из этих букв)
- d. С использованием идентифицированных числовых данных

Теперь перейдем к непосредственному поиску подсказок.

## 4. Поиск и анализ кодов и ключей манускрипта
### 4.1 Анализ изображений на наличие скрытых букв.

Для данного анализа была использована последняя страница манускрипта (Рис.2.)

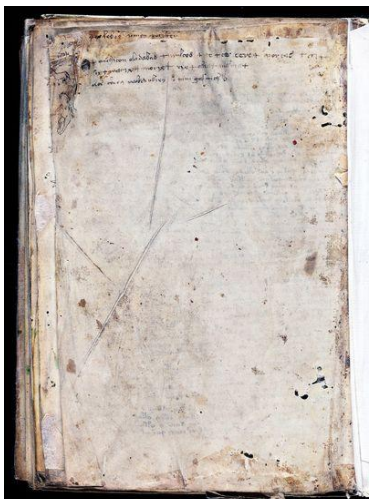

Рис. 2   Изображение последней страницы манускрипта

В верхней части последней страницы можно увидеть изображения и текст.

Причиной, по которой была выбрана именно данная страница – изображения и их компоновка с текстом отличаются от общих правил, использованных автором в остальных разделах манускрипта.

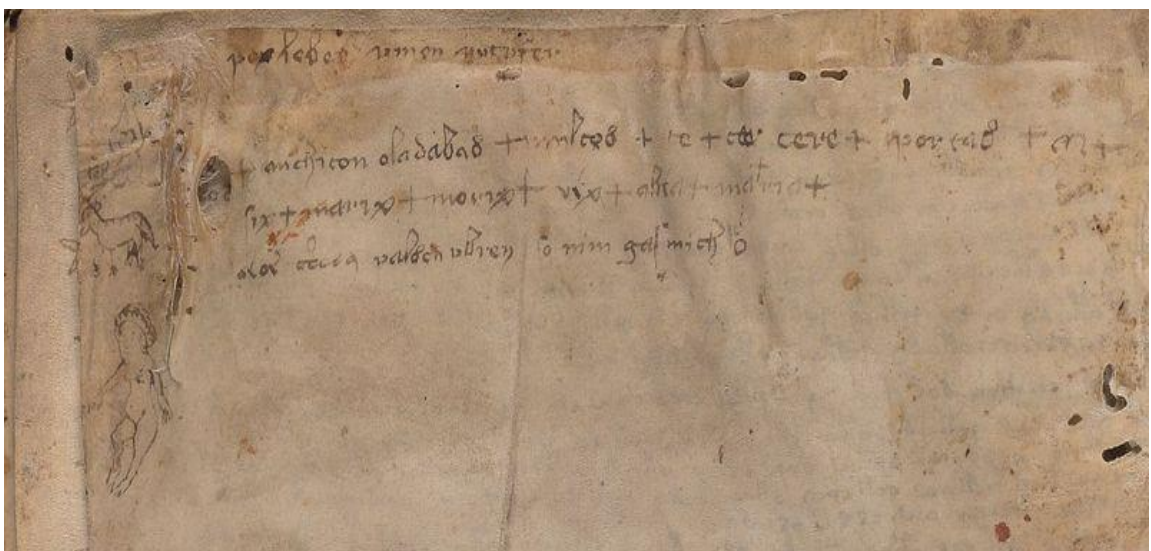

Рис. 3   Рисунки и текст на последней странице манускрипта.

Рассмотрим подробнее изображения на последней странице.

Для начала с помощью программных средств[3] был улучшен контраст.

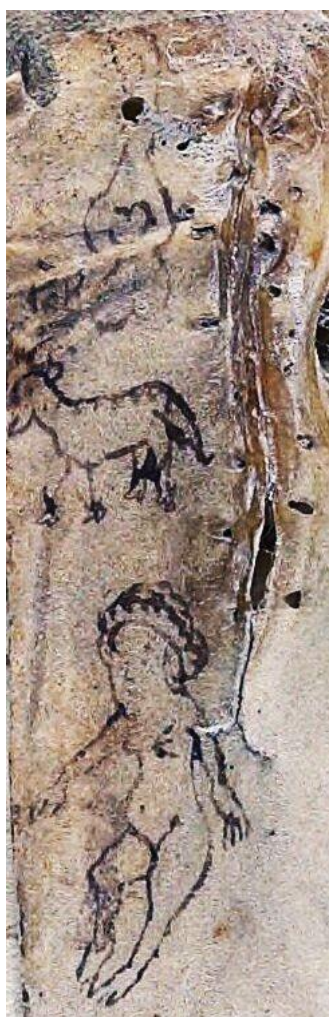

Верхнее изображение выглядит похожим на утку (duck) с опущенным клювом или на дракона (dragon) с раскрытой влево пастью – (D). Тело животного с высокой вероятностью содержит латинские буквы (слева – направо) – возможно F, E и L.

Изображение символьной строки, начинающейся с перевернутой перечеркнутой латинской V (V)

Изображение, похожее на козу (Goat) или Козерога (Capricorn) (C)

Изображение дамы (Madame или Maiden) (M).

В данном случае, не Lady и не Virgo (объяснение будет дано ниже)

Рис. 4   Изображения на последней странице манускрипта.

Итак, почему – Dragon или Duck, Capricorn и Madame (или Maiden) были использованы в качестве возможных названий каждого из рисунков?

Ответ достаточно очевиден – все рисунки подобного типа в остальном манускрипте соответствуют созвездиям. Единственный вопрос остается по поводу "Madame". Дополнительные пояснения касательно Козерога будут даны ниже, в разделе, посвященном разбору примеров дешифровки.

Продолжим. Следуя логике, что изображения последней страницы соответствуют знакам зодиака, можно предположить, что перевернутое и перечеркнутое V также должно обозначать созвездие.

Данное созвездие несложно идентифицировать – это символический знак Змееносца(Ophiuchus), только лежащий на боку. В настоящее время для его обозначения используется перечеркнутая буква "⛎", но в средние века использовалась буква "∀".

Здесь я хотел бы обратить ваше внимание на буквы, которые являются заглавными в названиях созвездий:

D, F, V, C и M.

Подобная последовательность букв может дать дополнительную информацию. Такую, например, как дата – поскольку все эти буквы (за исключением F) могут являться и римскими цифрами. Это частично объясняет причину, почему именно слово "Madame" было использовано для идентификации.

Мы знаем, что радиоуглеродная датировка показала период 1404 – 1438. Таким образом – 15-ый век.

"Madame"(не Lady или Virgo) – как "M" выходит из следующих предположений:

- если автор VMS оставил информацию о дате (даже если мы говорим о 15-ом веке – т.е., в принципе, о любом веке второго тысячелетия) – в римской датировке должна присутствовать как минимум буква "M" – т.е. "Millenium". Единственное изображение, соответствующее "M" – это изображение обнаженной девицы…

Букву, похожую на "F", мы исследуем ниже.

Начнем наш детальный анализ с изображения «D».

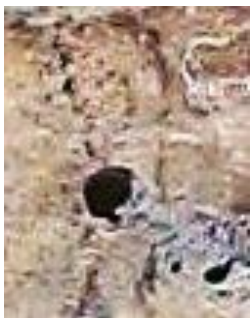

Рис. 5    Голова Дракона или Утки(оригинальное изображение)

Изображение содержит несколько буквенных символов.

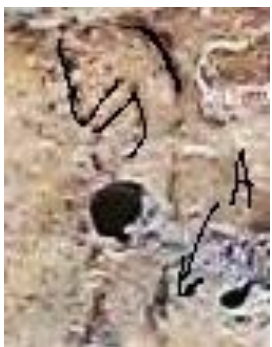
Рис. 6   Голова Дракона или Утки с выделением замаскированных букв.

Несколько слов о том, как читать текст (данное правило будет работать и в остальных случаях): в случае написания в одну строку, текст нужно читать слева - направо, но если буквы разбросаны на разных уровнях - **верхняя (возвышающаяся над соседней) буква (даже если находится правее) имеет приоритет перед стоящей ниже и читается первой.**
В нашем случае самая верхняя буква  -   **V**.

Следующая по высоте расположения - **I**.

Следующая   -   **V**.

Последняя -   **A**

Итак, первое скрытое слово  - <u>**V I V A**</u>

Теперь рассмотрим детально центральную часть изображения...

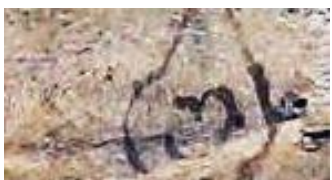
Рис. 7   Центральная часть (оригинальное изображение)

Центральная часть содержит хорошо видимые **L E** и, наиболее вероятно, - **X**, а не F, как предполагалось изначально.

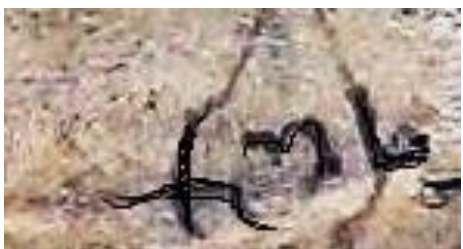
Рис. 8    Скрытый текст в центральной части рисунка.

Набор найденных букв обозначает - **L E X**

Вернемся к римским цифрам: **X** - полностью вписывается в последовательность. Это дает нам данные для анализа даты – **D X V C M = 500 10 5 100 1000**. Выглядит, как некорректное написание **1615**...

Этот странный порядок римских цифр может говорить также и о том, что не все цифры нужно складывать, одну (V) или две из них – (X и V) возможно нужно вычесть... Вычитание двух дает нам такой результат - **1 5 8 5**

Почему я считаю, что **X** и **V** нужно вычесть? **D, C, M** - имеют графическую интерпретацию в виде соответствующих изображений созвездий. **X** и **V** - записаны буквами.

**500 10 5 100 1000** – с другой стороны последовательность содержит 4 цифры с нулями и 1 без, так что возможен вариант и с цифрой **V**. В любом случае, все остальные результаты укладываются в интервал **1585 – 1615**, и пока для нас этого достаточно.

Наиболее интересно – что означает и к чему относится **VIVA LEX**?
Попробуем разобраться со строкой странных символов.

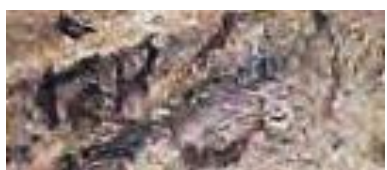

Рис. 9   Строка символов с маскированными буквами.

Мой вариант декодирования (разрешения электронных изображений из Йеля, увы, недостаточно) был таким:

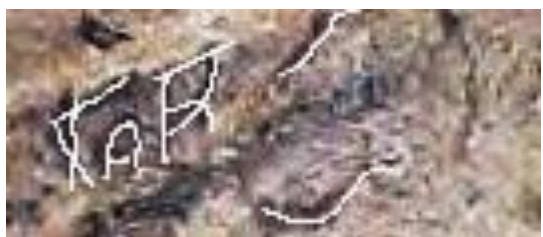

Рис. 10  Символьная строка с указанием маскированных букв. Вариант «А»

или таким

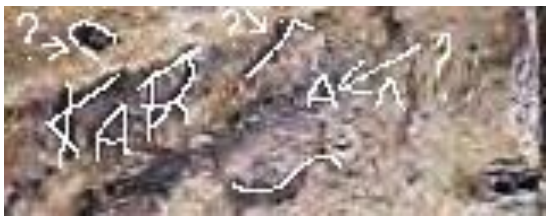

Рис. 11  Символьная строка с указанием маскированных букв. Вариант «В».

Предположительно она читается, как

**V I A R I U S**   (вариант А)

или **V I A R I A S**   (вариант B)

Есть и другие кандидаты – VICARIU(A)S и, даже, ESQUIRIU(A)S

На мой взгляд, наиболее вероятным все же является первый вариант

"**V I V A   L E X   V I A R I U(A)S**" – «Слава закону пути!».

Этот вариант соответствует алхимическому духу манускрипта. Кроме того, написание данной фразы именно на последней странице, соответствует логике успешного завершения алхимического труда.

Рассмотрим теперь один из наиболее интересных рисунков – Козерог (Capricorn).

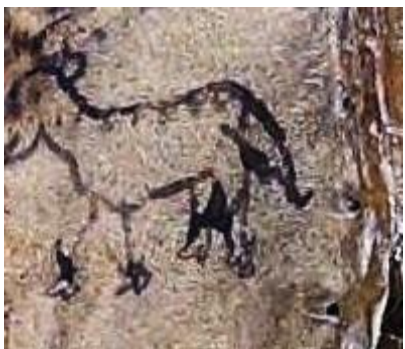

Рис. 12  Оригинальное изображение Козерога (Capricorn)

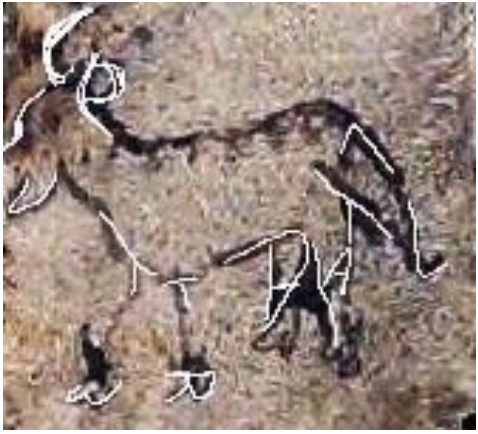

Рис. 13  Изображение Козерога с маскированными буквами

Что-то похожее на имя автора – **C A L L Y A D V A R T E Y U S** .

Все просто – Edward Kelly…

Но, давайте дойдем до конца.

Последнее изображение  - Maiden.

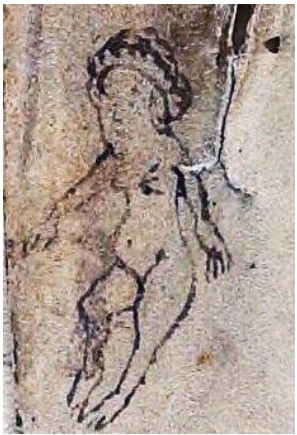

Рис. 14  Оригинальное изображение Дамы

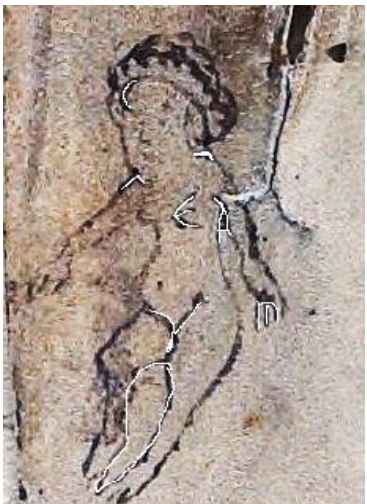

Рис. 15 Изображение  Madame или Maiden с маскированными буквами.

Прочтем...

**C R E A T Y O R** или

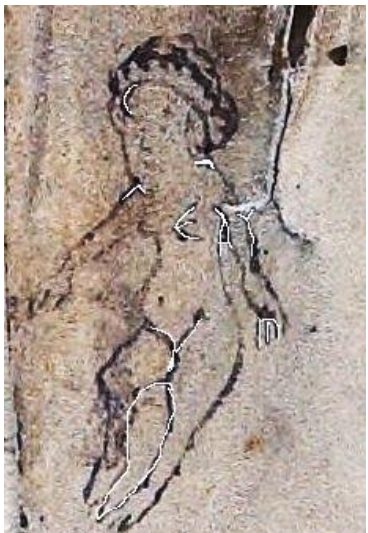

Рис. 16 Изображение Madame или Maiden с маскированными буквами (2-й вариант).

**C R E A Y T Y O R**

Из официальной истории мы знаем, что Эдвард Келли умер предположительно в конце 1597г.- начале 1598г.

Таким образом, скрытый текст последней страницы и дата, получаемая из комбинации заглавных букв названий созвездий весьма точно совпадают.

Итак, наша первая находка:

**VIVA LEX VIARIUS**
**EDWADRD KELLEY**
**CREATOR**
**1585**

Несомненно, прочтение может иметь определенные вариации, но, полагаю, без существенных отклонений от вышеуказанного текста...

Для уточнения нечетких деталей рисунков последней страницы требуются дополнительные исследования оригинала страницы методами световой микроскопии – УФ, ИК, Темное поле, поляризация и т.д.

В любом случае, сокрытие текста в рисунках – один из приемов стеганографии, нередко использовавшийся в средние века.

## 4.2 Первые результаты.

1) Итак, возможный автор манускрипта - Сэр Эдвард Келли(Sir Edward Kelly) (1555 – 1597) – известный английский алхимик.
2) Манускрипт, наиболее вероятно, написан/завершен в конце 16-го века (не ранее 1585г).
3) Возраст пергамента более, чем на 150 лет, превышает период написания манускрипта и, таким образом, при датировке подобных артефактов нельзя опираться лишь на радиоуглеродный метод.
4) Манускрипт, наиболее вероятно, начинал создаваться в Англии.
5) Языки срытого (маскированного) текста – латынь.
6) Язык зашифрованного текста манускрипта с высокой вероятностью должен быть английским («creator», «madame» или «maiden», «duck» или «dragon») с элементами латыни.

Предварительные данные по установлению автора, языка написания, периода написания и места написания нами получены. В дальнейших исследованиях нам предстоит либо их подтвердить, либо опровергнуть.

## 4.3 Рисунки первой страницы.

На первой странице манускрипта есть несколько

## 4.4 Анализ подписей к изображениям на отдельных страницах.

Для данного анализа было выбрано несколько страниц, представляющих различные разделы:

- Анатомический

- Ботанический

- Астрономический

- Фармацевтический

Из астрономической части была выбрана страница с диаграммой, содержащей двойную звезду – хорошо известную в период средневековья.

Впоследствии, в качестве проверки найденных ключей, были использованы некоторые другие страницы манускрипта.

### A) Отобранная страница #1 (78r)

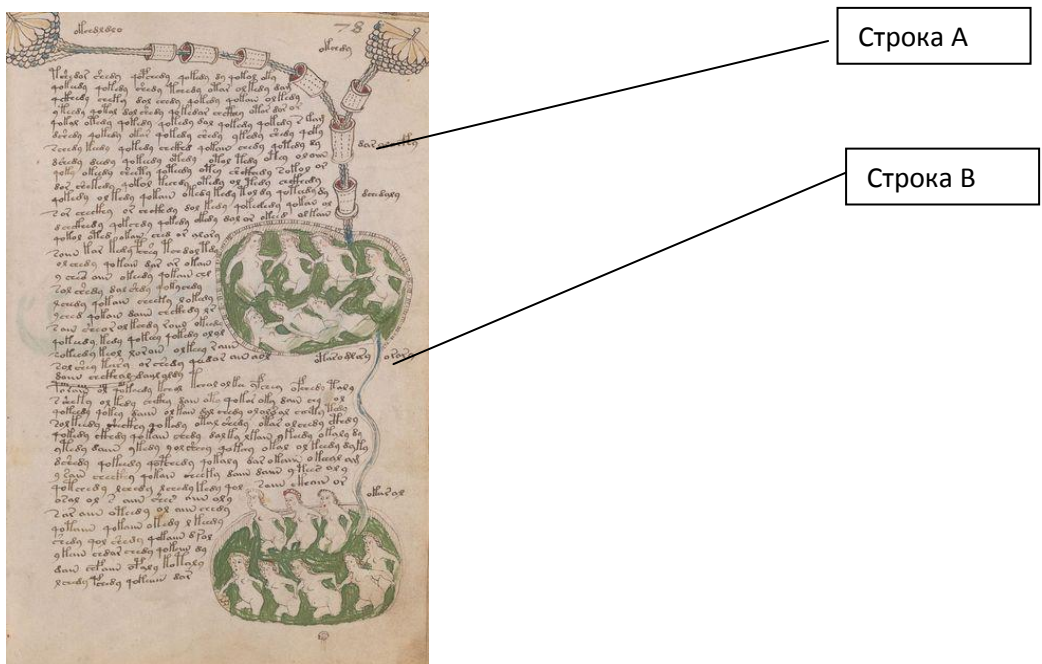

Рис. 17 Отобранная страница #1 с указанием анализируемых символов

### B) Отобранная страница # 2 (33v)

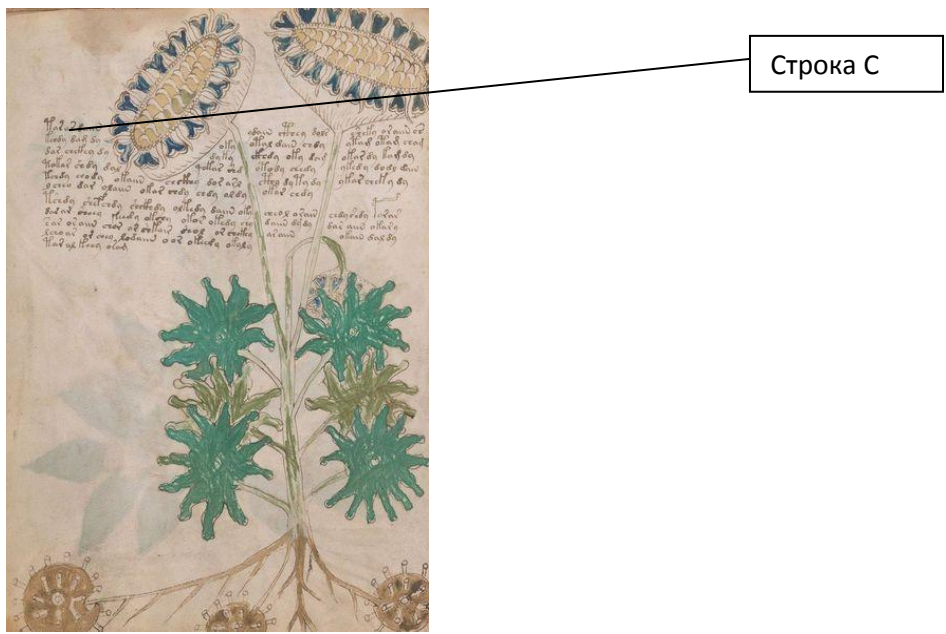

Рис. 18   Отобранная страница #2 с указанием анализируемых символов

**C) Отобранная страница # 3**(70v)

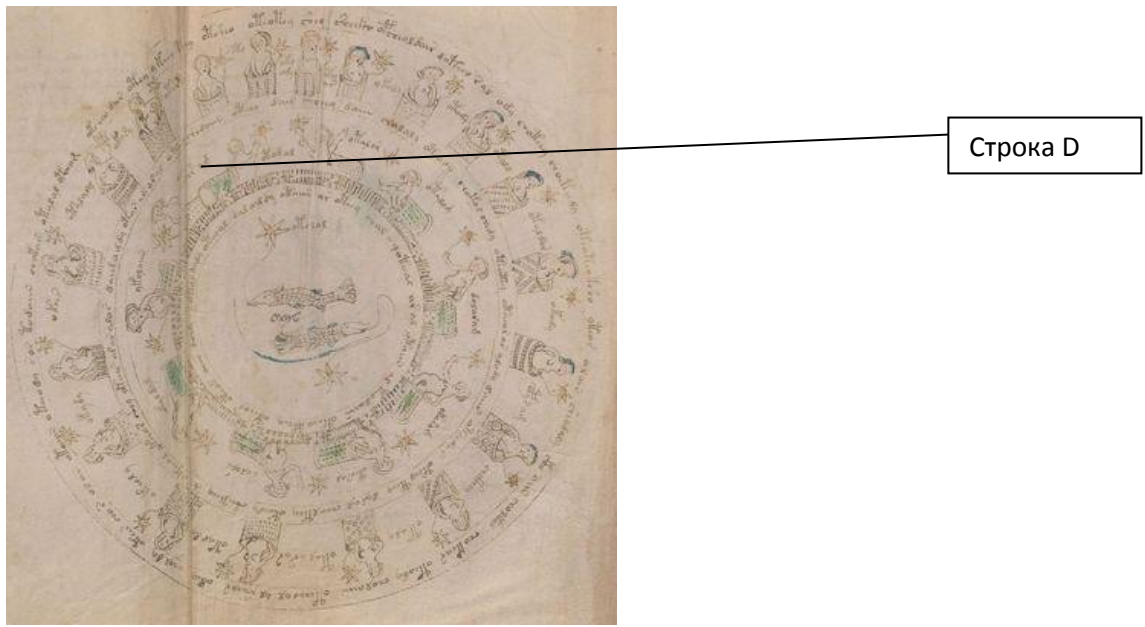

Рис. 19   Отобранная страница #3 с указанием анализируемых символов

**D) Отобранная страница #4**(100r)

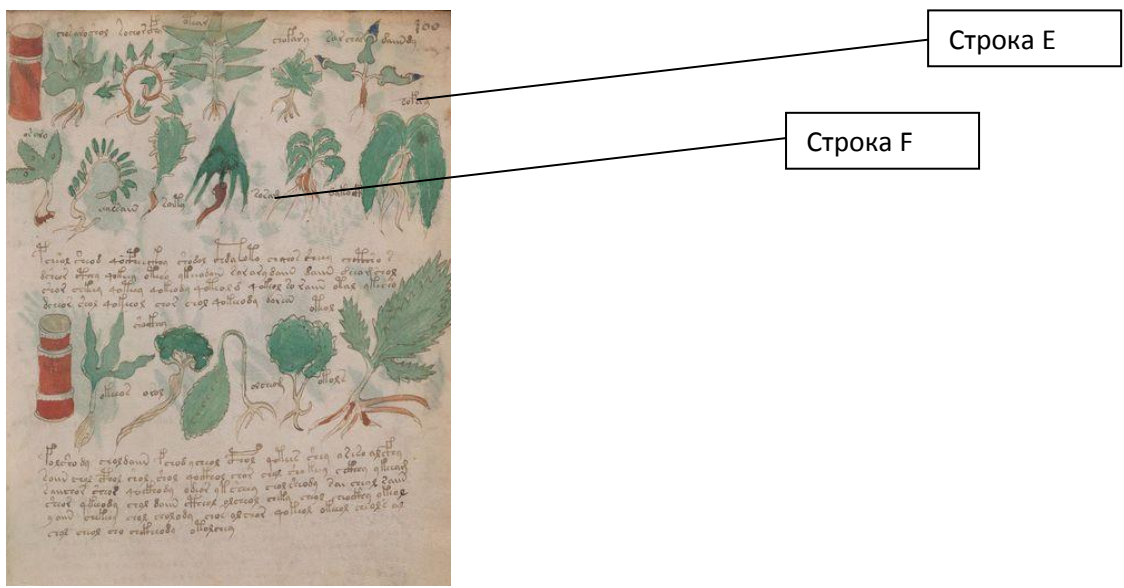

Рис. 20   Отобранная страница #4 с указанием анализируемых символов

Почему были отобраны именно данные страницы?

Главным критерием отбора явились рисунки на этих страницах – их вполне можно отожествить с реальными объектами.

**ОЧЕНЬ ВАЖНОЕ ДОПОЛНЕНИЕ О ТОМ, ЧТО И КАК МЫ БУДЕМ ИСКАТЬ:**

- **МЫ БУДЕМ СОПОСТАВЛЯТЬ ГРУППЫ СИМВОЛОВ ИЗ ПОДПИСИ К КОНКРЕТНОМУ ОБЪЕКТУ С ЕГО ВОЗМОЖНЫМИ НАЗВАНИЯМИ НА ЯЗЫКАХ, ИСПОЛЬЗУЮЩИХ ЛАИНСКИЙ АЛФАВИТ** (несмотря на то, что за основной язык манускрипта мы принимаем английский, в подписях к рисунков могут использоваться термины из других языков, и этот фактор должен учитываться при исследовании).

- **МЫ БУДЕМ ИСХОДИТЬ ИЗ ПРЕДПОЛОЖЕНИЯ, ЧТО КАЖДЫЙ СИМВОЛ ПРЕДСТАВЛЯЕТ СОБОЙ ПРОСТОЕ РАСХОЖЕЕ СЛОВО ИЗ АНГЛИЙСКОГО ЯЗЫКА**. (Это предположение является главным отличием от всех известных ранее предпринимавшихся попыток дешифровки и именно оно, в конечном итоге, явилось ключом к разгадке тайны манускрипта).

- **МЫ БУДЕМ ИСХОДИТЬ ИЗ ТОГО, ЧТО СРЕДНЕВЕКОВОЕ НАЗВАНИЕ МОГЛО НЕ СООТВЕТСТВОВАТЬ СОВРЕМЕННОМУ, ИЛИ МОГЛО БЫТЬ НАМЕРЕННО ИСКАЖЕНО**. (Это предположение подтвердилось при дальнейшей работе с манускриптом).

- **КАК СЛЕДСТВИЕ ИЗ ВЫШЕИЗЛОЖЕННОГО, МЫ БУДЕМ ИСХОДИТЬ ИЗ ТОГО, ЧТО ВСЕ БЕЗ ИСКЛЮЧЕНИЯ ГРУППЫ СИМВОЛОВ НА СТРАНИЦАХ МАНУСКРИПТА, РАЗДЕЛЕННЫЕ ПРОБЕЛАМИ, ПРЕДСТАВЛЯЮТ СОБОЙ СЛОЖНОСОСТАВНЫЕ СЛОВА ИЛИ ЦЕЛЫЕ ФРАЗЫ**. (Это предположение полностью подтвердилось и помогло установить некоторые правила написания текста манускрипта).

### 4.4.1 Результаты
#### 4.4.1.1 Анализ строки А.

На данной странице одним из главных кандидатов на определение ключей выбрано следующее символьное сочетание:

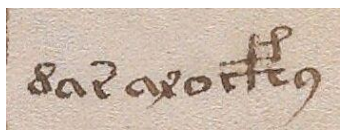

Рис. 21  Изображение строки А.

Изображение, подпись к которому мы изучаем, ассоциируется с реалистичным изображением человеческой аорты (**AORTA).**

**Данная группа символов была выбрана не случайно. Ключ, в виде слова «AORTA», мог оказаться весьма удачной находкой, потому, что для большинства европейских языков, в т.ч. средневековых, написание слова «AORTA» было идентичным и именно благодаря данному факту существенно сокращалось количество итераций при поиске. Сам поиск производился простой подстановкой.**

**Символьная группа на рис.21 существенно длинней по количеству символов, чем количество букв слово «AORTA». Тем не менее, основным предположением являлось то, что символьная группа должна содержать это слово.**

**Локализация слова «AORTA» в выбранной символьной группе могла быть различной, но я решил начать поиски с конца символьной группы, предположив, что слово «AORTA» завершает некую фразу или более длинное зашифрованное название. Получилось, что удача сопутствовала поискам – слово «AORTA» оказалась зашифрованным двумя последними символами группы!**

Окончательные результаты анализа позволили предварительно идентифицировать ряд символов ниже.

**Еще раз хочу обратить внимание на основной постулат - главная особенность манускрипта – каждый символ соответствует простому расхожему английскому слову, а группы символов могут быть фразами или сложносоставными словами.**

Теперь еще одно важное замечание:

 - все указанные ниже значения кодов и ключей, приведены к их латинскому спеллингу и, соответственно, произношению – т.е. «А» читается как латинское «А», а не как английское «Эй». Т.е. как если бы английские слова на слух записывал бы латинянин следуя своим правилам написания. Например, английское «RUN» трансформировалось бы в «RAN».

Ряд ключей в зависимости от определенных сочетаний символов и их расположения в символьной группе (начало/середина/конец) имеют созвучные значения (звонкое и глухое), например: "B"-"P", "D"-"T", "S"-"SH" и т.д.;

- значение и прочтение каждого символа зависят не только от его расположения в группе, но и наличия рядом, и обязательно в связке (*часто, но всегда, такая связка имеет графическое отображение в виде соединительной черточки между символами*) с т.н. главным символом правила – символом "**O**".

Как было установлено - автором манускрипта для ограничения количества используемых символов был применен фонетический код – это касается слов, произношение которых схоже, например «AND» и «END», «INN» и «IN» (таких слов немного, но они являются часто повторяющимися).

В дальнейшей расшифровке, при наличии нескольких значений, жирным шрифтом на первом месте указано наиболее частое значение символа.

Итак, эмпирическим путем был установлен первый символ.

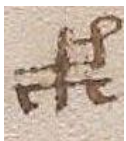
Символ – **ORT** [ɔ:rt]

Рис. 22 Идентифицированный символ #1 (ORT)

Его соответствие указанному значению было впоследствии многократно подтверждено подстановкой в содержащие данный символ произвольно выбранные символьные группы манускрипта.

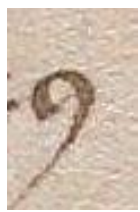
Символ – **AT** [ɑt] or [ət], Используется для обозначения цифры «15» на диаграмме со стр. 57v (см. раздел 6)
**A** [ɑ:] – возможно, если располагается в конце символьной группы/строки

Также, возможно, обозначает "**,**" в зависимости от расположения.

Рис. 23 Идентифицированный символ #2 (AT)

Соответственно, 2 последних символа в символьной группе на рис.21 могут быть прочтены как «ORTAT» или «ORTA».
В процессе изучения символьных групп, содержащих символ #2 в конце группы, появилось предположение, что символ #2 может являться индикатором множественного числа описываемого объекта и в таким случаях всегда читается как «AT».

Возможно, что символ #2 выполняет роль запятой, при перечислении нескольких предметов подряд – здесь примером подобного значения может послужить первая символьная строка на странице 11v.

Оба предположения требуют дополнительного изучения.

Перейдем к следующему, весьма интересному символу.

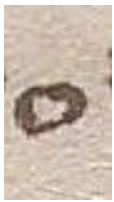
Символ – **TO** [to:] в понимании – to (указатель места или направления) Используется для обозначения цифры «1» на диаграмме со стр. 57v (см. раздел 6)

*Является главным символом правила, определяющим как читать соседний символ.*

Рис. 24 Идентифицированный символ #3 ("TO" или «символ правила»).

Касательно интерпретации данного символа, в качестве буквы «о». Примером может послужить символьная группа, содержащая слово «AORTA». Символ правила расположен на третьем месте от конца группы и вполне может давать прочтение «OORTA», что, в принципе, не изменяет значения найденного слова.

Почему этот символ назван символом правила? Это один из самых распространенных символов манускрипта. Роль символа #3 была обнаружена и многократно подтверждена при подстановке в дешифруемый текст – он обретал смысл только в том случае, когда учитывалось влияние символа #3 на связанный с ним символ. Символ #3 задает правило прочтения связанного с ним символа, сам символ #3 при этом не читается – ниже будет представлено подробное описание работы данного символа на конкретных примерах.

А пока перейдем к идентификации других символов.

### 4.4.1.2 Анализ строки B.

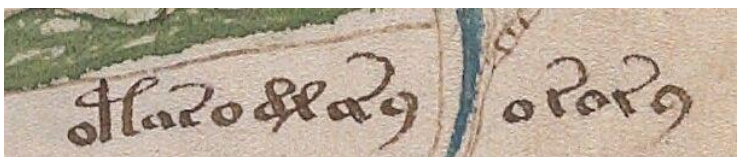

Рис. 25 Изображение строки B.

Эта подпись к изображению, которое очень похоже на изображение желчного пузыря.

Поиск среди европейских названий показал и в первую очередь в английском, что это вероятнее всего название желчного пузыря на староанглийском языке – **GALLBLADER** (это касается лишь первой – самой длинной части символьной группы).

Здесь были идентифицированы следующие ключи:

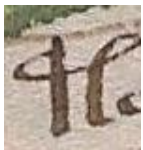 Символ – **GE** [g ə], **GEV** [g ə v], соответствующий современному <give> в зависимости от того, с какой буквы – гласной или согласной – читается следующий символ. Используется для обозначения цифры «11» на диаграмме со стр. 57v (см. раздел 6)

Рис. 26 Идентифицированный символ #4 (GEV)

Работа с текстом проверочных страниц, выбранных случайным образом, показала, что автором манускрипта используются оба варианта – **GE** [g ə] и **GEV** [g ə v]:
- первый – в случае, если после данного символа стоит символ, означающий слово, начинающееся с согласной буквы;
- второй – в случае, если после данного символа стоит символ, означающий слово, начинающееся с гласной буквы.

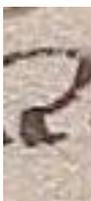 Символ – **ALL** [ɑ:l],

Может иметь спеллинг **LI** [lɪ], если находится между символами 2 похожими на "а" (это вспомогательный символ правила)

Рис. 27 Идентифицированный символ #5 (ALL)

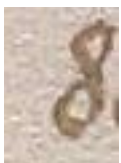 Символ – **PLA** [plɑ:] , **BLA** [blɑ:], Используется для обозначения цифры «3» на диаграмме со стр. 57v (см. раздел 6)

Рис. 28 Идентифицированный символ #6 (PLA)

Анализ символьных групп содержащих символ #6 показал высокую вероятность использования всех приведенных значений символа, в зависимости от положения в группе и соседних символов.
В особенности, в сочетании с символом #7. Примеры будут приведенных ниже.

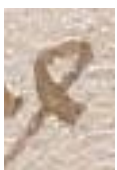 Символ – **AND** [ænd] и **END** [ənd] , **в зависимости от расположения в строке** Используется для обозначения цифры «2» на диаграмме со стр. 57v (см. раздел 6)

Рис. 29 Идентифицированный символ #7 (AND)

Установлено, что отдельно встречающаяся группа символов #6+#7 читается, скорее всего, как «BLADA» или «BLAD»,

соответствующее современному английскому «blood» (кровь). И именно это сочетание мы видим в символьной группе на рис. 25 с прочтением «BLAD».

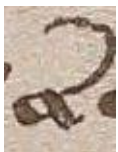
Символ – **ER** [ər], **IR** [ir]

Рис. 30 Идентифицированный символ #8 (ER)

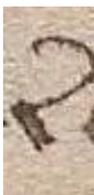
Символ – **INN** [inn], Используется для обозначения цифры «4» на диаграмме со стр. 57v (см. раздел 6)

Рис. 31 Идентифицированный символ #9 (INN)

Продолжим. Вторая, коротная часть символьной строки, содержащей слово «**GALLBLADER**» (*точнее с учетом прямой подстановки значений идентифицированных символов, но без учета неустановленных пока еще правил – на рис. 25 это слово, звучит как «GEVALLBLAANDERA» или «GEVALLBLAANDERAT»*) может быть прочтена как **IN** или **INN** или **INNA** или **ININ** (в смысле внутри чего-либо) или как **NINA**, поскольку изображение желчного пузыря (**GALLBLADER**) содержит изображения нимф – маленьких женщин (или «малышек») – а в испанском «NINA» как раз и означает малышку. Однако, эта часть символьной строки содержит целых два главных символа правила "о" и в деталях мы разберемся с этим случаем несколько позже и результат будет весьма интересным.

Замечу, что благодаря полученным здесь и далее результатам и сравнением их с английским языком 16-го века, весьма близким современному английскому, мы выявим ряд устойчивых правил конструирования слов манускрипта.

### 4.4.1.3 Анализ строки C.

Для анализа была отобрана следующая символьная комбинация:

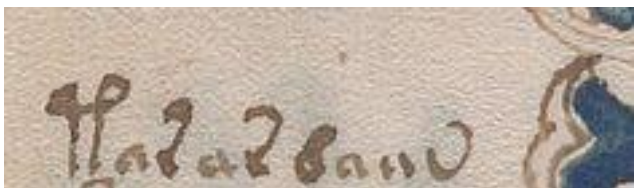

Рис. 32 Изображение строки C.

Эта подпись к изображению, которое вполне можно было бы идентифицировать, как подсолнечник (он же Heliantus или

Sunflower). Данное изображение цветка подсолнечника выполнено в инверсии – лепестки темно-синие, а середина – желтая. Если цвета поменять местами, то получится классическое изображение цветка подсолнечника.

Разделим слово "SUNFLOWER" на 2 простых и получим "SUN FLOWER", что, в свою очередь, можно записать еще и как «HELIA PLANT». У нас уже есть символы, идентифицированные ранее. Применим их к данной строке.

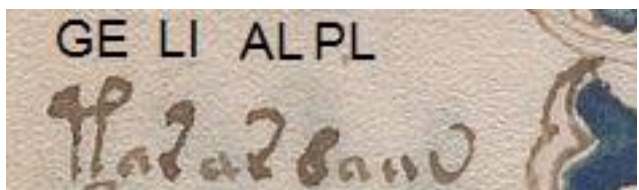

Рис. 33   Изображение строки C с ранее идентифицированными ключами.

Заметим роль вспомогательного символа правила, схожего с современной буквой «**a**» *(этот символ, пишется и произносится, как «RI» – идентификация его будет показана ниже)*:

- в данном случае, нами возможно установлено новое правило прочтения: если первый из символов ALL находится между двумя символами «**a**», то сочетание символов RI + ALL + RI читается, как "LI". Примеры, приведенные ниже показывают постоянство работы данного правила.

Если вернуться к рис. 21 с символьной группой, включающей слово «AORTA», мы можем увидеть в группе сочетание символов RI + INN + RI, однако заметим, что второй символ RI отделен от символа INN пробелом и связан с символом AND, рядом с которым в свою очередь расположен главный символ правила. Рассмотрим все варианты прочтения символьной группы:

1) PLA NI AND/END TO/O ORTAT (AORTA) – возможный вариант реконструкции символьной строки PLAIN END OORTAT («простой конец аорты»)

2) PLA RIINN RIAND TO/O ORTAT (AORTA) – для такой интерпретации написания получается следующий вариант реконструкции:

- PLA (apply, place)

- RIIN (древнеанглийское обозначение проточной воды)

- RIAND = RIND (современное: кожура, корка, шкура)

Результат можно перевести так: «Промой проточной водой поверхность аорты».

На текущий момент оба варианта равноправны.

Следующий пример. Рассмотрим единственную символьную строку со страницы 65r из трёх групп символов, которая содержит (здесь я, немного забегая вперед, использую ключи, идентифицированные при дальнейшем анализе) три составных слова:

**«GE RAN IS», «AB RI IS» и «DI IS»**

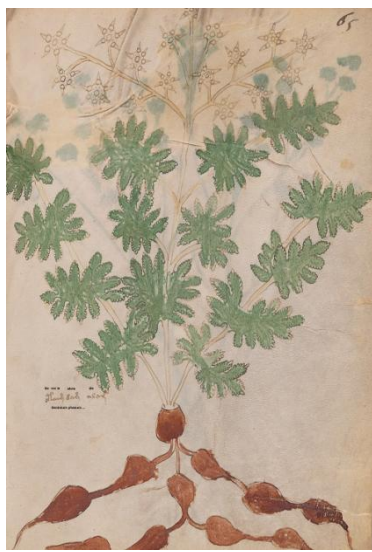

Рис.??? Страница 65r с надписью

«GE RAN IS»

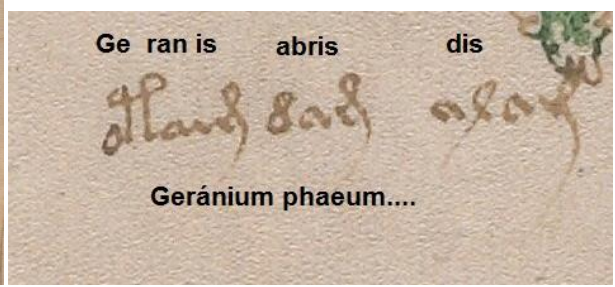

**«DI»**, в последней символьной группе, образовано согласно вышеуказанному правилу из сочетания RI + AND + RI.

Сформулируем данное правило: **если символ в одной символьной группе расположен между двумя символами RI, то этот символ теряет две первых буквы своего написания латиницей, а оба символа RI преобразуются в окончание I [i:] к оставшимся буквам срединного символа.**

Еще одно правило: **в случаях, когда в одной символьной группе впереди стоящий символ, в своем написании латиницей, оканчивается на гласную букву, такую же, с которой начинается латинское написание следующего символа – эти гласные сокращаются до одной.**

Таким образом, символьная строка со страницы 65r читается как: «GERANIS ABRIS DIS» - т.е. «это рисунок герани».

Здесь хотелось бы обратить ваше внимание на слово «DIS». Это еще один прием автора манускрипта для придания языку манускрипта атрибутов древности – упрощение, основанное на созвучии. Сегодня это слово означает "THIS". Возможно «DIS» - взято из одного из английских наречий. В случае, если в написании действительно участвовал Келли – это может быть уэльский вариант.

Зачем такие сложности? Наиболее вероятно – для надежной маскировки ключей. Длина символьной группы сбивает дешифровщика с толку, перефокусируя на поиски заведомо несуществующих ориентиров.

Вернемся к «HELIA PLANT». Прочитав первые символы группы мы обнаружили новый символ, который легко идентифицируется подстановкой недостающей части слова «HELIAPLANT»:

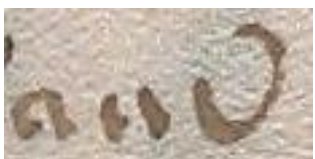

Символ – **ANT** [ɑ:nt]

Рис. 34 Идентифицированный символ #10 (ANT)

### 4.4.1.4 Анализ строки D.

Здесь у нас единственный кандидат на идентификацию ключей – подпись к двойной звезде.

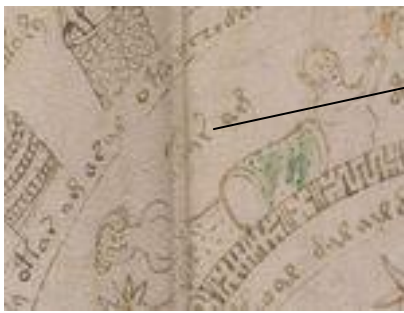

Строка D

Рис. 35   Изображение строки D

Эта подпись к символическому изображению двойной звезды в созвездии Рыб.  Она известна как Alpha Piscium (Alpha Psc, α Piscium, α Psc), но в средние века ее имя предпочитали писать так - Alrescha (Al Rescha, Alrischa, Alrisha, Al Risha)

Как результат:

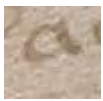

Символ – **RI, RY** - [ri],  Также является вспомогательным символом правила.

Рис. 36 Идентифицированный символ #11 (RI)

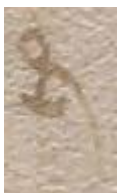
Символ – **AS** [æs] or [æz], **IS** [is] or [iz]

Рис. 37 Идентифицированный символ #12 (AS)

### 4.4.1.5 Анализ строки E.

Выбранный кандидат:

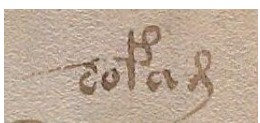

Рис. 38 Изображение строки E

Эта строка содержит главный символ правила "о"(который в случае отсутствия связки (соединительной черточки) читается как – TO или, возможно TA) и символы, читаемые как… RI и AND, которое дает нам современное RIND (кожура, корка, шкурка). Одно из вероятных слов – …TOMARIND или …TAMARIND. После верификации получаем:

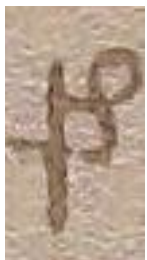
Символ – **MA** [m ɑː], возможно **MAT** [mat] или [mət], в зависимости от того, с какой буквы – согласной или гласной – читается следующий символ. Используется для обозначения цифры «9» на диаграмме со стр. 57v (см. раздел 6)

Рис. 39 Идентифицированный символ #13 (MA)

Если мы обратимся к первой странице манускрипта, к первой символьной группе, мы увидим, что первые два символа – это символы MA и RI, что в конструкции дает MARI, которое при дальнейшем исследовании оказалось известным алхимическим «marry» («пожени», «соедини»). Что, в свою очередь, косвенно подтверждает верное направление поиска и подход.

## 4.4.1.6 Анализ строки F.

Следующий кандидат.

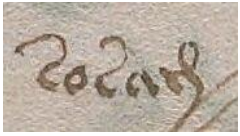

Рис. 40   Изображение строки F

Оно содержит AL или LA, RI, S. После детального анализа был сделан вывод, что это слово ACICULARIS (или, по-русски, ШИПОВНИК), оригинал (выше) может читаться как - CULARIS.

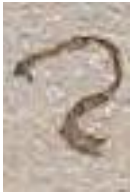

Символ – **CU** [ku:], **CUT** [ku:t]**,** в зависимости от того, с какой буквы – согласной или гласной – читается следующий символ.

Рис. 41 Идентифицированный символ #14 (CUT)

*Если уважаемый читатель обратит внимание на рисунок, к которому сделана подпись, то увидит растение, все части которого выглядят, как шипы. Очередная визуальная аллегория автора манускрипта.*

В итоге, с помощью ранее идентифицированных символов (указаны выше), некоторые другие символы были идентифицированы и проверены.

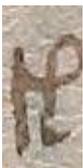

Символ – **HEAT** [hi:t]
Используется для обозначения цифры «7» на диаграмме со стр. 57v (см. раздел 6)

Рис. 42 Идентифицированный символ #15 (HEAT)

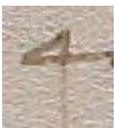

Символ – **LIF** [li:f] в современном значении: **Leaf** (лист)

Рис. 43 Идентифицированный символ #16 (LIF)

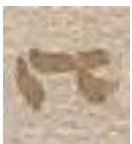

Символ – **TRE** [træ] или [tri:]
При раздельном написании всегда подразумевается современное TREE

Рис. 44 Идентифицированный символ #17 (TRE)

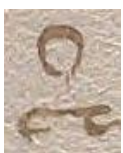
Символ – **RET** [rət]

При раздельном написании всегда подразумевается RED (красный)

Рис. 45 Идентифицированный символ #18 (RET)

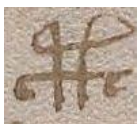
Символ – **PUT** [pu:t]

Рис. 46 Идентифицированный символ #19 (PUT)

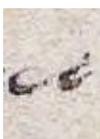
Символ – **ERN** [ɛə rn] или [ɛərnæ]

Рис. 47 Идентифицированный символ #20 (ERN)

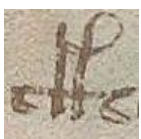
Символ – **POT** [po:t]

Рис. 48 Идентифицированный символ #21 (POT)

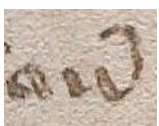
Символ – **SOM** [sa:m] (современное "some")

Рис. 49 Идентифицированный символ #22 (SOM)

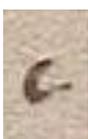
Символ – **HERE** или **HEA** [hə:a:]

Рис. 50  Идентифицированный символ #23 (HERE)

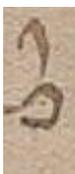
Символ – **ICE** или **ISE** [aɪs ]

Рис. 51  Идентифицированный символ #24 (ICE)

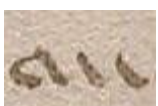
Символ – **RAN** [ran]   в современном значении **RUN** [ran]

Рис. 52  Идентифицированный символ #25 (RAN)

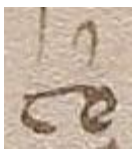

Символ – **REST** [rəst]

Рис. 53   Идентифицированный символ #26 (rest)

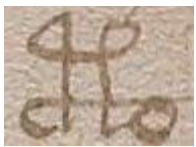

Символ – **TOP** [to:p] в современном эквиваленте также используется как "cover".

Рис. 54   Идентифицированный символ #27 (TOP)

Символ #27, вообще говоря, является сочетанием символа #21 с главным символом правила (#3) и иллюстрирует работу последнего – **если главный символ правила в связке с другим символом расположен сразу за ним – данный символ читается справа налево, а главный символ правила не читается**. Таким образом, образуется новое слово: было «POT» (чайник, сосуд для жидкости) – стало «TOP» (верх, крышка). Это весьма оригинальный способ уменьшить количество используемых для написания текста символов.

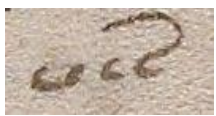

Символ – **EYE** [aɪ] или **AI**

Рис. 55   Идентифицированный символ #28 (EYE)

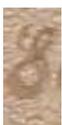

Символ – **EPL** [əpl]
Современное значение - apple

Рис. 56   Идентифицированный символ #29 (EPL)

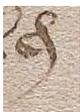

Символ – **SA** [sa:] или [za:]/ или **SALT** [salt]
Используется для обозначения цифры «8» на диаграмме со стр. 57v (см. раздел 6)

Рис. 57   Идентифицированный символ #30 (SA)

На текущем этапе ряд символов остался не идентифицированным. Вернемся к ним позже.

## 4.4.2 Некоторые слова и их спеллинг.

Этот раздел посвящен маленькому словарю идентифицированных слов.

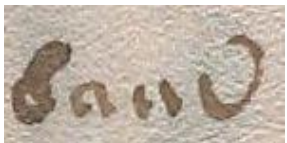 **PLANT** [plant] (всегда означает растение)

Рис. 58 Слово PLANT

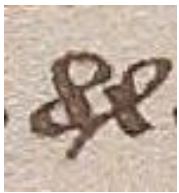 **BLADA** [blɑ:da] или **BLAD** [blɑ:d] (современное – BLOOD (кровь))

Рис. 59 Слово BLAD

Как уже было указано выше, изучение конструкций типа **BLADA** [blɑ:dɑ:] и **BLAD** [blɑ:d] на предмет их корректного написания – дало понимание одного из основных правил получения окончательных вариантов прочтения символов-слов. **Наиболее вероятно, что в случае, если впереди стоящий символ заканчивается на букву, на которую начинается следующий – то эти буквы не удваиваются, остается только одна, как примерах ниже:**

PLA + ANT = PLANT
BLA (PLA) + AND = BLAD
PLA + AT = PLAT (тарелка, блюдо)

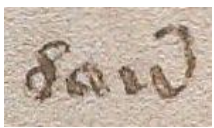 **PLASOM (современное «blossom» - цветок)** –[pla:a:sa:m]

Рис. 60 Слово PLASOM

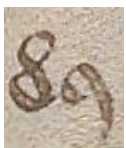 **PLAT** (современное "plate")
Символ PLA, отдельно может означать также глаголы «класть» (place) или применять (apply), в этом случае символ AT выступает в роли запятой. Это достаточно четко определяется контекстом

Рис.61 Слово PLAT

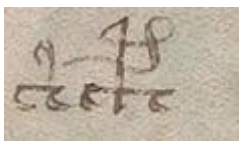
Символьная комбинация, обозначающая одну реторту –
**RETORT** [rətort] (retort)

Fig. 62A Слово RETORT

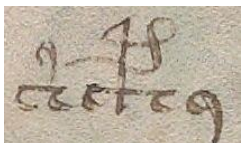
Символьная комбинация, возможно обозначающая множественное число реторт
**RETORTAT** [rətortat] (retorts)

Рис. 62B Слово RETORTAT

Здесь приведен пример того, как символ "AT", использованный в конце слова "retort" может указывать на множественное число этого слова. Возможно также, что символ «AT» может обозначать запятую.

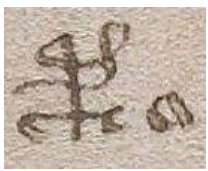
**ORTRI** [ortri:]
В современном значениях скорее всего – "straightforwardly" – «прямо», «непосредственно», возможно встречается также и в значении «единоразово».

Рис. 63 Слово ORTRI

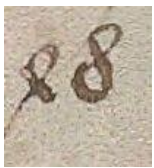
**DABLA** [da:bla:] или **DAPLA** [da:pla:]
(современное - double or twice)

Рис. 64 Слово DABLA

Здесь символы «AND» и «PLA» поменялись местами и вместо значения «кровь», мы получили значение «двойной».

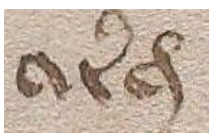
**RINS** [ri:nz] или **RINSA** [ri:nza:]

Рис. 65 Слово RINS

Для сочетания символов RI и INN (и только лишь), я сделал исключение в спеллинге и прочтении, приняв это прочтение с повторяющимся I, т.е. сочетание читается, как RIIN – в старо и древнеанглийском слово именно с таким написанием означало проточную воду. В манускрипте это символьное сочетание по логике текста вполне соответствует проточной воде.

### 4.4.3 Главный символ правила "о".

Одна из интереснейших загадок автора манускрипта.
Символ отдельно (без связки) однозначно читается как "ТО" (или в некоторых случаях, возможно, как "О"), **но в зависимости от расположения данного символа может полностью изменить смысл слова, в состав которого он входит.**
Часть примеров с главным символом правила мы разобрали выше. Рассмотрим еще несколько случаев.

Пример с символом#7 – AND

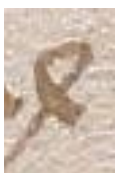
Символ – **AND** [ænd] or [ənd],

Рис. 66 Символ #7 (AND)

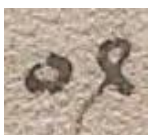
В сочетании с символом правила – **AD** [æd] or [əd], в современном значении ADD

Рис. 67 Слово AD

Т.е. в случае, когда "о" задает правило чтения, сам символ правила не читается.

Этот пример дает нам направление, позволяющее установить новые правила действия символа «о».

Изначальное "AND" трансформировалось в "AD" – похоже, что при расположении символа правила перед другим символом это означает, что мы должны не читать среднюю букву в его написании – "A**N**D". Так "AND" трансформируется в "AD".

Более сложный случай.

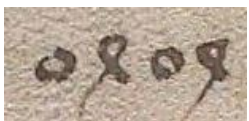

Рис. 68 Комбинация символов, требующая интерпретации.

Фактически мы видим дуплет из символов на Рис.67, который в конечном итоге мог бы звучать как ADAD. Возможно это и вариант, но по смыслу манускрипта этот текст читается "**AD TO**

**END**" (современный вариант "add to finalize" – «добавь, чтобы закончить»)

Проверим, как это правило работает на других примерах.

Вернемся к символу #9 (INN)

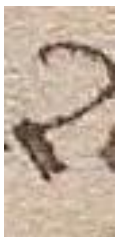

Рис. 69 Символ #9 (INN)

INN – в манускрипте подразумевает некое внутренне пространство.

В случае, с использованием символ правила
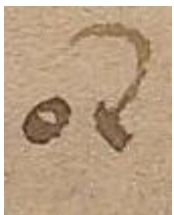

Рис. 70 Символ INN в комбинации с символом правила.

Согласно нашей версии «INN» (пространство) трансформируется в предлог "IN".

Кстати, "AD" и "IN" очень расхожие слова и это отчасти объясняет, почему их символьное обозначение так часто встречается в манускрипте.

Рассмотрим более сложный случай.

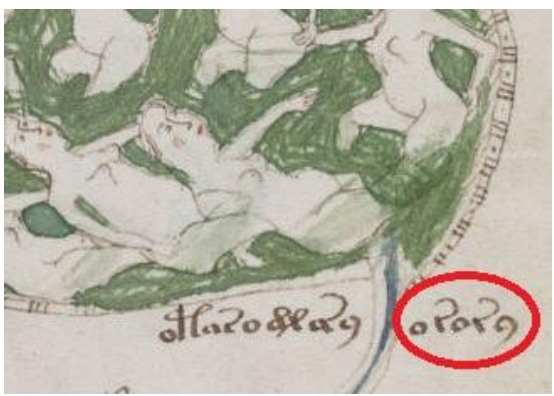

Рис. 71  2х символа "INN" в комбинации с 2х символами правила с последующим символом "AT" (выделено красным на рисунке).

Этот пример нам знаком из раздела с поиском базовых ключей и кодов – это желчный пузырь со страницы 78r.

Выделенная комбинация звучит как "IN TO INN A" (символ «AT» стоит в конце и читается как «A»), окончательный вариант будет "INTO INNA" в смысле "inside" (внутри). Таким образом общая подпись означает "gallbladder inside" – желчный пузырь внутри, что полностью соответствует изображению.

Еще более сложный и интересный пример со страницы 17v…

Здесь рассмотрим следующую строку

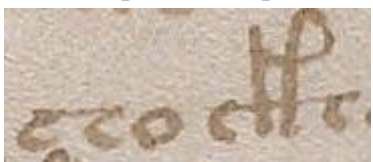

Рис. 72  Символ TRE, символ правила, символ POT

Возможные результаты – "TRE TO POT" или "TREO POT" или "TRE PT". Обратим внимание на связку символа TRE и символа правила. Важно отметить – символ правила расположен после символа TRE.

Из рассмотренного ранее примера «POT»-«TOP» вытекает, что символ «TRE» должен читаться как «ERT».
«ERT» - либо специально упрощенное, либо взятое из одного из диалектов или более древнего языка – в современном языке имеет значение "**EARTH**"

И, в данном случае, искомый результат будет "**ERT POT**", т.е. «СОСУД С ЗЕМЛЕЙ».

Здесь сделаю одну маленькую ремарку. Поскольку анализ последней страницы показал возможную причастность Эдварда Келли к созданию манускрипта, то параллельно поиску кодов я изучал дневники Джона Ди – фактически основной источник информации о Келли. И в этих дневниках есть запись, согласно которой Келли отправлялся в путешествие по 11 английским землям с целью сбора образцов почвы этих земель. Совпадение? Дальнейшее исследование покажет, что таких совпадений слишком много.

Рассмотрим еще один сложный пример:

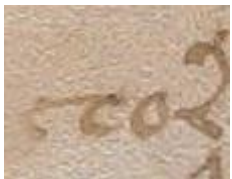

Рис. 73 Символ TRE, символ правила, символ INN.

Здесь важно отметить, что символ правила работает только в отношении одного из символов.

Сложность данного случая объясняется тем, что все символы написаны очень плотно и прочтений может быть несколько:
- TRE TO INN (т.е. «дерево внутрь»)
- TREIN (средневековое значение требует уточнения)
- ERT IN или ERTIN (возможно в значении «землю в…», «на земле», а возможно это означает числительное «18» (eighteen)). Если обратить внимание на числа, оставленные автором в зашифрованном виде (см. раздел 6), то их всего семнадцать – от 1 до 17. Соответственно, для обозначения чисел превышающих семнадцать автор манускрипта вполне мог использовать их буквенную запись.

Позже мы попробуем проверить значение символьной строки исходя из контекста всего предложения. Данный пример был приведен, чтобы показать влияние символьных взаимосвязей на результат прочтения.

Однозначно установлено, что символ правила"о", стоящий за символом "RET" и в связке с ним означает, что вместо слова "RET" нужно читать слово "TER". Иногда отдельные перевертыши кажутся на первый взгляд бессмыслицей, однако не следует забывать, что большинство из них стоят в связке с другими символами и общее сочетание дает однозначное и осмысленное прочтение.

Дополнительно можно утверждать, что символ правила, стоящий в связке с другим символом перед этим символом применяется исключительно для односложных слов типа "AND", "INN", "RET", "TRE" и т.п. и не применяется к более сложным словам. В случае более сложных слов символ правила имеет (чаще всего) значение "TO".

## 5. Основные выводы.

Проведенный выше анализ позволяет сделать следующие выводы:

- Основной метод кодирования – STEGANOGRAPHIA – подтверждается. Говоря «Стеганография» я имею ввиду классическую методику и идеи шифрования, описанные Иоганном Тритемием (1462 - 1516) в его одноименной работе "Steganographia" (1500г).

- Автором манускрипта стеганография модифицирована и использует более мощный криптографический механизм.

- Основной язык манускрипта – староанглийский – подтвержден. В тексте встречаются также отдельные включения специфической терминологии из Латыни (например, для названий некоторых растений, созвездий, органов тела и т.д.) и из Арабского (как, например, в астрономической части – "Al-Risha"). Возможно, при дальнейших изысканиях появится что-то из греческих слов.

- Язык основного текста выглядит как намеренно искаженный или упрощенный английский, или как некий диалект английского или, возможно, древний английский.

- Для кодирования методом стеганографии автор манускрипта использовал не отдельные буквы, а простые, часто однокоренные слова, являвшиеся расхожими в его время.

- Автор манускрипта в качестве дополнительной защиты своего кода использовал символы правила – "o" и "a"

- Даже небольшие комбинации символов представляют собой различные структуры – отдельные сложносоставные слова, фразы и предложения.

- Комбинирование символов, вероятнее всего, осуществлялось случайным образом с единственной целью – ввести непосвященных в заблуждение, что каждая группа символов является одним словом.

Единственное ограничение, примененное автором при комбинировании символов – их однозначная интерпретация при прочтении.

- Автор манускрипта использовал визуальное отсутствие пунктуации для запутывания потенциальных дешифровщиков. Это означает, что автор, вероятнее всего, имитировал более ранние правила написания манускриптов.

После проведения анализа спеллинга идентифицированных символов, обнаружился следующий подуровень кодирования части текста, подчиняющийся 3-м основным принципам:

- Произношение символов соответствует их побуквенному латинскому прочтению (т.е., как минимум, не соответствует спеллингу современного английского языка)

- Автор сфокусировался на фонетике символов, практически проигнорировав грамматику – это позволило ему использовать меньшее количество символов, при этом значение каждого такого символа определяется его локализацией (как пример "AND" и "END") и взаимосвязью с соседними по группе символами или указаниями символов правила.

- Спеллинг слова или фразы играет ключевую роль, даже если ее написание выглядит некорректным.

- Ряд символов в зависимости от их локализации имеют либо звонкое, либо глухое произношение, например: "B"-"P", "D"-"T", "S"-"SH" и т.п.

То, что автор манускрипта столь творчески подошел к модернизации классической стеганографии Тритемия, и использовал простые, чаще – односложные слова в относительно произвольном группировании (под «произвольным группированием» имеется в виду только деление символьной строки на символьные группы различной длины) оказало значительный результат на все попытки дешифровки манускрипта. Все специалисты, использовавшие в своих поисках метод частотного анализа, столь удобный для других случаев, автоматически закладывали искаженные начальные данные. С другой стороны, поскольку каждый из символов представляет собой слово английского языка (возможно модифицированного языка, в основе которого лежит английский), что соответствует реальному языку, то при частотном анализе это соответствие реальному языку должно будет обнаружиться. Собственно, этот факт нашел подтверждение в исследованиях Марчело Монтемурро из Манчестерского университета в соавторстве с Дамианой Занетте из Атомного центра Барилоче […].

## 6. Метод кодирования

Нужно обязательно рассказать об основном методе кодирования – СТЕГАНОГРАФИИ.

Для того времени это был передовой метод кодирования. Его название "Steganographia" впервые было употреблено Иоганном Тритемием (Johannes Trithemius (1462 - 1516)) в его одноименном труде:

"Steganographia" (рукопись закончена в 1500, впервые официально опубликована в 1608, а в 1609 в список запрещенных книг (Index Librorum Prohibitorum))

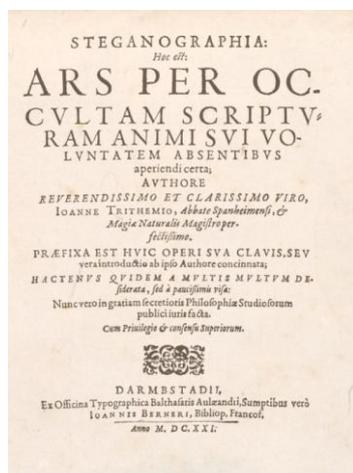

Рис. 75 Титульный лист первого печатного издания "Steganographia", Joh. Tritemius [4].

В своей книге Тритемий подробно описал, как работают искусство сокрытия тайного послания в общедоступном тексте и в дополнение привел весьма громоздкий, но, в итоге, слабоустойчивый к дешифровке, массив таблиц. Существует множество описаний работы методов «Стеганографии». Один из них подробно описан в романе Умберто Эко «Маятник Фуко» в главе 19.

В тоже время Тритемий не ограничивал использования иных приемов сокрытия посланий, таких как использование иностранных языков, неизвестных в данной местности, или использование специальных, выдуманных символов, значение которых известно только адресату. Последний приём несомненно относится к нашему манускрипту.

Здесь я хотел бы обратить ваше внимание на интересную диаграмму из книги Тритемия    ([4], стр.55)

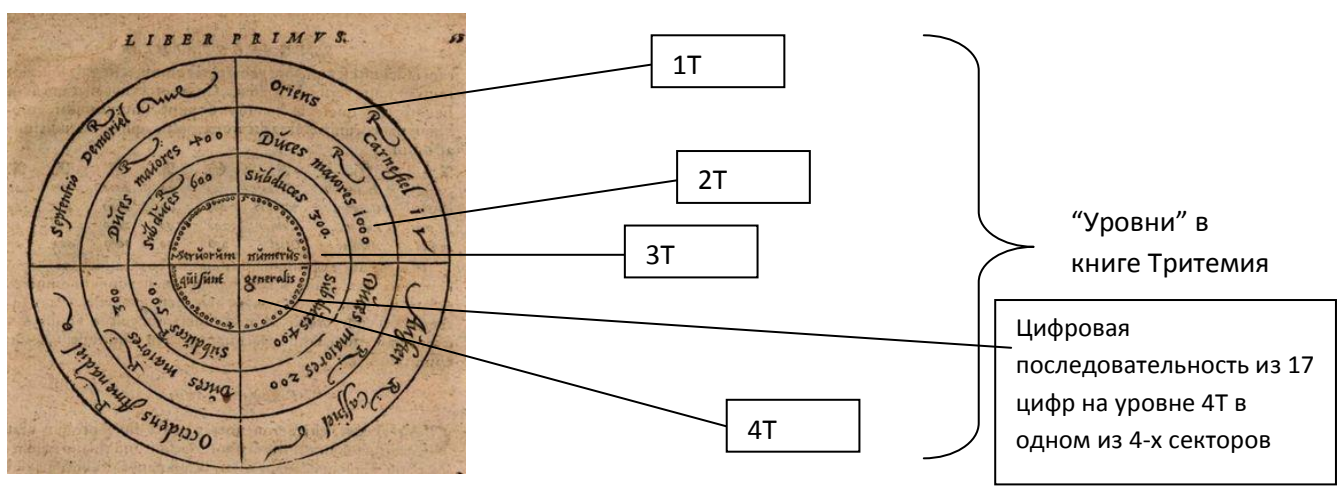

Рис. 76 Диаграмма с принципами кодирования в  "Стеганографии"

Теперь если мы обратимся к странице # 57v *(более подробную информацию о содержании диаграмм см. в разделе 13)* нашего манускрипта, мы тоже увидим некую диаграмму…

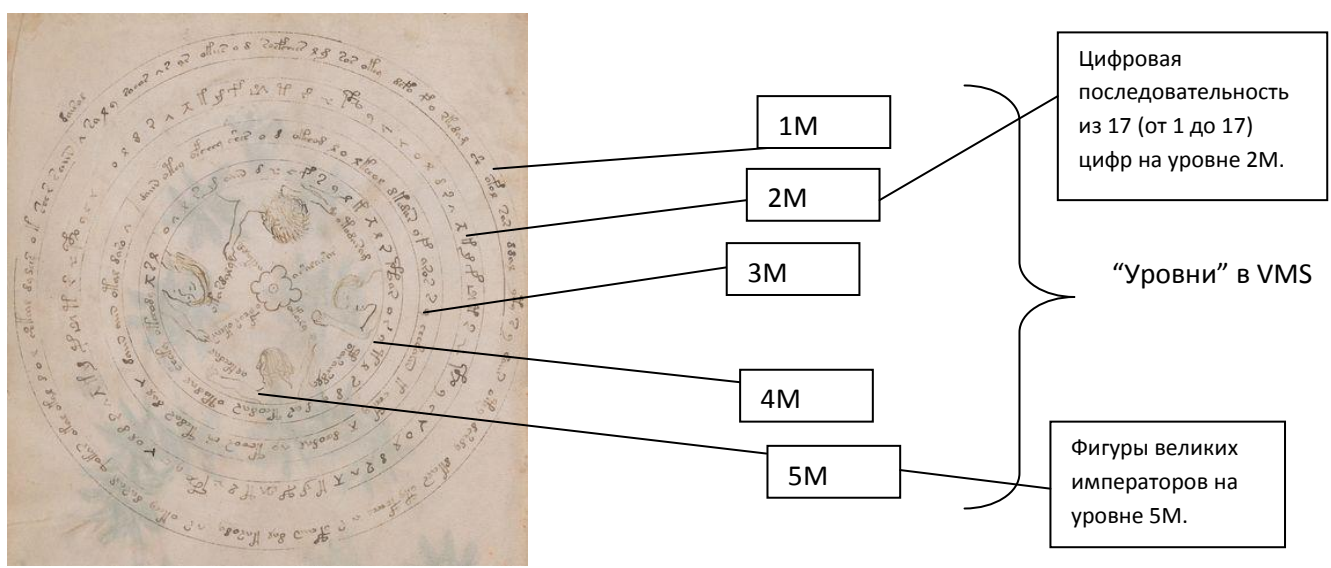

Рис. 77 Диаграмма со страницы #57v манускрипта

Скажу сразу, что обе приведенные выше диаграммы по сути – одно и тоже. Разница лишь в способе изображения и расположении и количестве уровней – в манускрипте на один уровень больше.

Корреляция между уровнями в «Стеганографии» и в манускрипте следующая:

1Т = 5М, 2Т = 4М, 3Т = 3М, 4Т = 2М

При этом "Уровень 1М" не используется для кодирования.

Более детально: Почему 4Т = 2М? Объяснение:

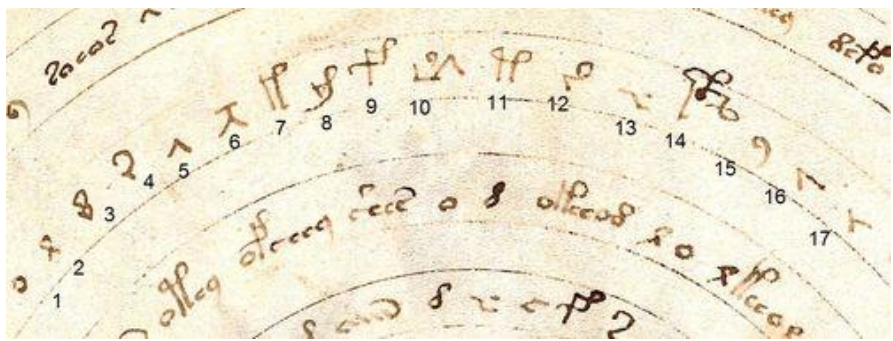

Рис. 78 Изображение страницы #57v с идентифицированными цифрами.

Мы видим такую же, как и у Тритемия, последовательность из 17 цифр (в данном случае ряд натуральных чисел от 1 до 17). Она повторяется 4 (четыре) раза и одновременно делит все окружности на 4 равные части. Так же, как и в книге Тритемия, где 17 цифр (достаточно посчитать их количество), повторяющихся 4 раза, делят окружности на равные части. У Тритемия эти цифры имеют определенные значения, отличные от цифр манускрипта.

Причины – в принципе кодирования – у Тритемия использован принцип сдвига и перестановки букв по определенному ключу, и фактически его цифры указывают на число вариаций шифра.

В манускрипте этот принцип существенно модифицирован и указание числа вариаций шифра уже не принципиально.

Диаграммой Тритемия управляют 4 «великих императора» (взятых из так называемой «Магии Соломона») – императоры Севера, Востока, Запада и Юга. Их имена *(Демониэль, Карнесиэль, Аменадиэль, Каспиэль)* прописаны на уровне 1Т.

В манускрипте, вместо имен, использованы изображения фигур 4-х «великих императоров» на уровне 5М. Пропорции изображений в манускрипте полностью соответствуют «могуществу» каждого из императоров из книги Тритемия.

Более того, с помощью выше найденных ключей вы сами можете прочитать секретное кодированное послание в диаграмме из манускрипта на уровне 4M. И, что весьма интересно, найдете инструкцию типа:

«На (стр.) 1-2-3-4-… возьми 1-2-3-4-5-… N возьми, как M …» и т.п. Последовательность 1-2-3… использована для примера, в диаграмме указаны другие цифры их приведенного набора от 1 до 17.

Последовав данным инструкциям, вы попадете на страницы, содержащие латинские буквы разного цвета, размера и написанные разными шрифтами. Диаграмма дает нам правило конструирования слов из данных букв. Это еще один шифр манускрипта. Интересным примером может послужить страница 4r – по странице разбросаны и спрятаны как минимум 5 букв. Если их читать, начиная с верхней – получится слово "AFTOR". Но это, возможно, обманка или случайное совпадение, поскольку инструкция из диаграммы диктует другую последовательность, составленную из букв с разных страниц.

Ряд страниц, на которые ссылается диаграмма, увы, отсутствует. Как, например, страница 12…

Другой сложностью является то, что часть букв написана микрошрифтом («мизерные» буквы), который при имеющемся разрешении электронных снимков, по размерам соизмерим с дефектами пергамента и может добавить ложные данные в анализ.

Ряд исследователей манускрипта предполагает, что нумерация страниц манускрипта была сделана позже, одним из последующих владельцев.

Однако наличие ссылок из диаграммы на стр. 57v скорее говорит об обратном, если, конечно, цифровые ссылки на уровне 4M не являются ссылками на внешние источники (что, на мой взгляд, маловероятно).

Итак, диаграмма в манускрипте – это модифицированная диаграмма из «Стеганографии» Тритемия. И это знание дает нам очень интересную информацию.

"STEGANOGRAPHIA", как рукописный манускрипт, была закончена Тритемием только в 1500г…

Что, в свою очередь, означает, что наш манускрипт был написан позже. И это означает, что манускрипт был написан в 16-ом столетии... Т.е. 1438г. доказано остался позади и мы потихоньку приближаемся к временам Эдварда Келли...

Авторство Келли, темнее менее, остается под вопросом. И не потому, что я, допустил какие-либо ошибки в дешифровке, а потому, что скрытую запись, включающую имя Эдварда Келли, мог сделать кто-то еще...

Возвращаясь к 1585г - в это время Келли очень тесно сотрудничал с Джоном Ди. Они даже проживали в одном доме...

Далее будем внимательно обращать внимание на появляющиеся совпадения и их количество...

## 6.1. Пример использование кодов.

На текущий момент любой из читателей данной статьи может приступить к прочтению страниц с идентифицированными кодами.

Далее будет рассмотрен ряд примеров прочтения фрагментов манускрипта.

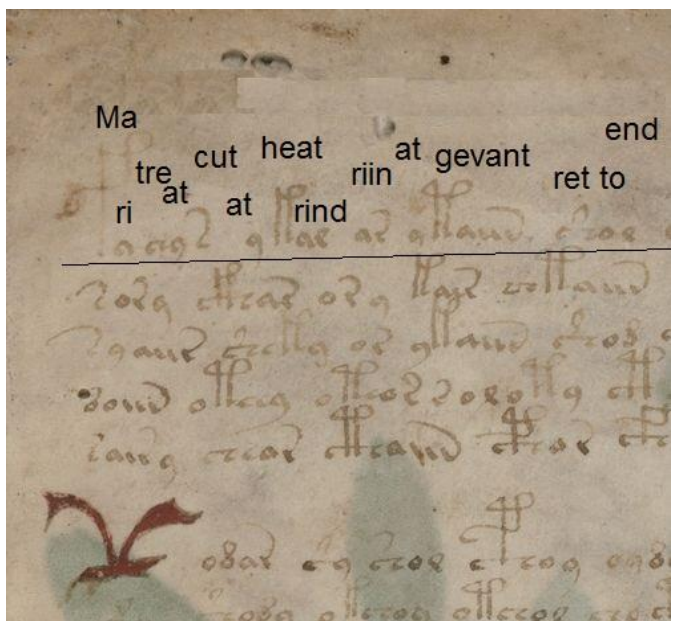

Рис. 74 Пример дешифровки первой страницы манускрипта (1r)

Книга начинается с фразы "Mari tre at cut at heat rind riin at gevant ret to end".

Собственно, первые два символа мы рассматривали ранее. Однако именно расшифровка остального текста дает нам понимание, что

<span style="color:red">полученное "Mari" означает современное "marry" – довольно типичное слово для алхимического трактата, вспомним хотя бы «Алхимическую свадьбу» Христиана Розенкрейца.</span>

<span style="color:red">На этом примере видна красота авторской идеи – "TRE" отдельно означает «дерево», "AT" означает «НА», но вместе они образуют слово "TREAT" – обрабатывать…</span>

Расшифровать начало можно следующим образом:
   "Соедини обработанные (или обрабатываемые) куски (отрезы) на горячей коре (поверхности), полей полученным красным для окончания."

## 6.2   Дополнительные данные для поиска автора и возможные выводы.

1) Итак, мы уже знаем, что манускрипт написан на староанглийском – поэтому можно предположить, что его автором вряд ли был сам Тритемий

2) Автор манускрипта вне всяких сомнений использовал труд Тритемия "Steganographia" (т.е., повторюсь – 16-ый век)

3) Автор манускрипта использовал очень большое количество данных из различных областей средневековой науки.

4) Автор манускрипта по каким-то причинам был заинтересован в создании надежного метода шифрования.

5) Автор манускрипта прямо или косвенно был связан с Алхимией (Символ красной птицы на первой странице манускрипта – возрождающийся Феникс, частично уничтоженный в правом нижнем углу первой страницы символ – Уроборос, сам манускрипт начинается с некоей «свадьбы» и т.д.)

Кстати, об упомянутом в п.5) Уроборосе – знаке, распространенном среди всех древних народов, включая аборигенов недавно (на тот момент) открытой Америки.

Приведенное (и частично уничтоженное) изображение Уробороса – копия Уробороса из орнаментов Нового Света. Т.е. опять 16-ый век.

Одна из упомянутых нами выше персон в 1570-х участвовала в целой экспедиции к берегам Северной Америки. И имя этой персоны - Джон Ди…

Итак, возвращаясь к диаграмме Тритемия:
- рукописная версия – 1500г.(16-ый век)
- первая печатная версия – 1608г.(17-ый век)

Однако мы располагаем неопровержимыми доказательствами, что Джон Ди изучил все доступные труды Тритемия, включая «Стеганографию» еще в середине 16-го века!

Согласно архивам и дневниковым записям Джона Ди известно, что рукописный вариант "Стеганографии"(естественно, это был список с оригинала Тритемия) был приобретен им во время одной из дипломатических миссий. Он был настолько потрясен книгой, что сам переписал ее, посвятив свой труд венгерскому правителю, покровительство которого пытался снискать[5]. Мы даже знаем дату завершения этой кропотливой работы - 15-е февраля 1562, а также мы знаем место, где эта работа была проделана:

[…] Antwerpia apud Gulielmum Silinum in Angelo aureo : in platea, vulgariter, Den Camer straet, vocata […]

- В Антверпене, в гостинце «Золотой Ангел», на улице под названием Den Camer straet.

Вновь обратимся к архивам Ди и обнаружим его рукописный вариант диаграммы Тритемия [6]. Этот рисунок датирован 1591г.

Заметим – именно одной конкретной диаграммы Тритемия…

Это означает, что Ди на протяжении нескольких десятилетий постоянно работал со «Стеганографией» Тритемия, начиная еще с 1562г. (или, возможно, с конца 1561 – к сожалению, мы не знаем, сколько времени Ди понадобилось на переписку манускрипта) и прекрасно разобрался в принципах шифрования Тритемия и их слабых сторонах.

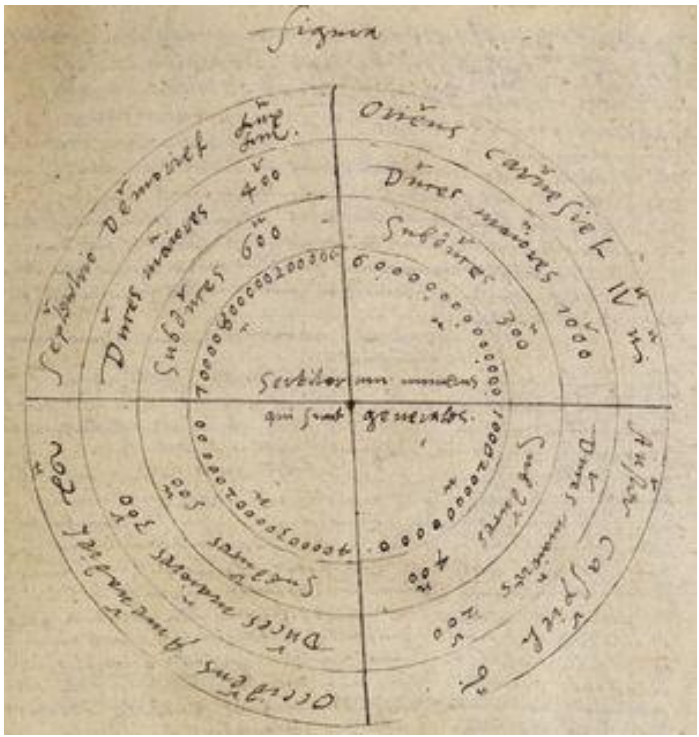

Рис. 79. Изображение диаграммы Тритемия, выполненное собственноручно Джоном Ди. 1591 [6]

Итак, еще раз подведем итог: **наш манускрипт написан на староанглийском, использует уникальные для того времени методы шифрования и содержит модифицированную диаграмму Тритемия из "Steganographia".**

Что еще мы можем сказать о манускрипте – он вполне читаем и представляет собой структурированную книгу.

Теперь попробуем, оторвавшись от поисков ключей к дешифровке, ответить на несколько важных вопросов:

- Каково было назначение манускрипта?
- Какова предыстория манускрипта?
- Какова судьба манускрипта?

Для ответа на вопросы выше, ответим сначала на более простой вопрос:

Знаем ли мы еще какие-либо шифрованные манускрипты, написанные возможно на Английском (или в Англии) в аналогичный период и прямо или косвенно связанные с упомянутыми выше возможными авторами?

Однозначно – ДА!

Это **"Book of Dunstan"** (она же «Книга Дунстана») – книга, которая постоянно находилась у Эдварда Келли во время его совместной деятельности с Джоном Ди…

Согласно официальной легенде:

- Книга была найдена Эдвардом Келли во время его путешествия (согласно Ди – по 11 английским землям). Келли принес ее в дом Ди в 1583г. до их совместного путешествия на континент (т.е. до 21 сентября 1583г.)([7] p.lxi)
- Манускрипт, который Келли называл "Book of Dunstan" якобы был найден в Гластонберийском аббатстве (Glastonbury Abbey) ([7] p.lxii). Аббатстве, в котором, согласно истории, служил Св. Дунстан в X веке.
- Другая версия говорит о том, что Келли приобрел книгу и 2 коробочки волшебного порошка по пути в Уэльс в какой-то таверне.

Но, что наиболее важно для нашего исследования (или скорее уже расследования) - единственным источником сведений о книге Дунстана и истории ее появления являются исключительно записи Джона Ди. И это факт настораживает.

Продолжим. Независимо от того, каким бы образом ни была обретена «Книга Дунстана», у этой книги не было никаких видимых причин называться «Книгой Дунстана» (за исключением весьма призрачной связи с Гластонберийским аббатством) и оба – Ди и Келли, как два образованных и умных человека – прекрасно это знали…

В противном случае (будь, обретенный таким образом, манускрипт реальным трудом Св. Дунстана) имелось бы немало независимых ссылок других авторов на этот труд, особенно после столь неожиданного обретения, поскольку такая потрясающая находка даже для того времени явилась бы великой сенсацией – найден труд одного из весьма почитаемых католических святых, посвященный "святым аспектам алхимии". Можно предположить, что большое количество ученых и священнослужителей принялись бы делать списки с этого труда. Но ничего подобного не произошло. «Книга Дунстана» так же неожиданно исчезла из истории, как и появилась, в течение всего около 20 лет (пока жив был Келли). В дневниках и воспоминаниях семейства Ди срок жизни книги с таким названием еще короче – судя по нескольким скудным упоминаниям – всего 4,5 года…

Теперь сравним истории «Книги Дунстана» и нашего манускрипта:

- Ничего не известно о «Книге Дунстана» до 1583г.
- Ничего не известно о манускрипте, носящем сейчас имя Войнича, до конца 1590-х или, возможно, до 1612г. когда умер Рудольф II
- «Книга Дунстана», согласно записям Ди, была взята в их совместное путешествие с Келли на континент
- Согласно Ди – книга принадлежала Келли
- После смерти (*или скорее казни – есть предпосылки так полагать*) Келли в конце 1597г (*хотя наиболее вероятно в период с сентября 1595г по конец февраля 1596г*) ничего неизвестно о "Книге Дунстана".
- Таинственный манускрипт, носящий сейчас имя Войнича, был приобретен Рудольфом II (1552 - 1612) после 1586г., вместе с тем достоверно известно, что в период конца 1580-х Келли (и возможно Ди) были весьма близки ко двору Рудольфа. В опубликованных дневниках Ди, описывающих период начала 1598г. говорится, что «долгожданное» (!) приглашение ко двору Рудольфа «наконец-то» (!) получил Келли.
- Достоверно известно, что «Книга Дунстана» была на руках у Келли в период 1583г - 1588г когда он был приглашен ко двору Рудольфа. Келли находился при дворе как минимум до августа 1595г.
- Родиной «Книги Дунстана» является Англия, так что существует высокая вероятность того, что книга была написана на английском.
- Манускрипт Войнича написан на английском.

Слишком много совпадений…

Еще одна важная ремарка (и к ней мы еще раз вернемся при анализе последней страницы манускрипта).

Очень интересная запись в дневнике Ди от 12-го декабря 1587г. ([8] p. 25;[9]).

И говорит она следующее:

Dec. 12th [1587], afternone somwhat, Mr. Ed. Keley his lamp overthrow, the spirit of wyne long spent to nere, and the glas being not stayed with buks abowt it, as it was wont to be ; and the same glas so flit.ting on one side, the spirit was spilled out, and burnt all that was on the table where it stode, lynnen and written bokes, as the bok of Zacharius with the Alkanor that I translated out of French for som by spirituall could not ; Rowlaschy his thrid boke of waters philosophicall; the boke called Angelicum Opus, all in pictures of the work from the beginning to the end ; the copy of the man of Badwise Conclusions for the Transmution of metals ; and 40 leaves in 4°(*), intitled, Extractiones Dunstani, which he himself extracted and noted out of Dunstan his boke, and the very boke of Dunstan was but cast on the bed hard by from the table."

*Перевод приблизительно таков:*

**«12 декабря 1587, около полудня, Мр. Эд.Келли опрокинул свою лампу, спирт широко разлился по столу, повсюду разлетелись осколки стекла[…]**

**Спирт загорелся и зажег все, что было на столе, за которым он работал - скатерть и книги, такие как книга Захарии с Алканором, которую я перевел с французского […]; третья книга Роулаши о водах философских; книга, называемая Опус Ангелилум, вся в рисунках от начала и до конца, копия книги человека из Будвайза о трансмутации металлов; 40 листов в 4-х стопках, названные «Экстрактом из Дунстана», которые он собственноручно вырвал и сделал об этом пометку в книге Дунстана, а саму книгу Дунстана резко бросил со стола на кровать.»**

Это последнее упоминание «Книги Дунстана» в записях Ди.

Из этой записи мы узнаем, что 40 листов Келли собственноручно удалил из «Книги Дунстана» и оставил в самой книге пометку об этом – это одна из очень важных зацепок, которая поможет раскрыть секрет манускрипта. И важно - мы знаем, что приблизительно такого же количества листов не хватает в манускрипте Войнича.

По легенде «Книга Дунстана» (как и наш манускрипт) была записана шифром и Келли, якобы безуспешно, пытался ее расшифровать, и, собственно, за помощью в расшифровке обратился к Ди.

Великое кроется в мелочах. Именно запись о "Extractiones Dunstani" нарушает гладкость легенды о манускрипте и говорит о том, что и Келли и Ди знали коды «Книги Дунстана», в противном случае создание любого "Extractiones Dunstani" является бессмыслицей.

Итак, сделаем предварительный вывод (который полностью докажем в соответствующей главе):

**Манускрипт Войнича и «Книга Дунстана» - один и тот же манускрипт, созданный в конце 16-го века в одном экземпляре для достижения некоей важной цели.**

Можно сказать, что у Келли и Ди были достаточно веские причины для создания шифрованного манускрипта с таким названием. Такая книга при определенных условиях могла бы быть очень полезной. И вот почему:

- Благодаря названию «Книга Дунстана» напрямую ассоциируется со Св. Дунстаном, с человеком, посвятившим всю свою жизнь служению Богу, католической вере, человеку с непререкаемой христианской репутацией, известному во всей Европе святому. Очень важный подсознательный посыл кому-то из влиятельных католиков мира сего – например, Рудольфу II.

- Если так – поскольку зашифрованная «Книга Дунстана», или лучше называть ее «Книгой Св. Дунстана», содержит вполне очевидные алхимические знаки/символы, то может послужить замечательной приманкой для того же Рудольфа II, весьма неравнодушного к алхимии. Поскольку это алхимическая «Книга Св. Дунстана», то ее содержимое автоматически благословлено к прочтению достойными людьми католической веры.

Итак, сперва в истории Ди и Келли (примерно через 1-1.5 года после их знакомства) в 1583г. появляется «Книга Дунстана», и только после этого компаньоны 21 сентября того же года отправляются на континент. Их путешествие обставлено как частная миссия двух ученых-алхимиков, организованная польским князем Альбрехтом Ласки (Olbracht Łaski).

Используя связи влиятельного Ласки, а в последствии покровительство и связи одного из самых могущественных политиков Европы – Вильяма Розенберга, компаньоны создают свою сеть тесных связей с различными влиятельными людьми Европы.

И, в конечном итоге, получают место при дворе Рудольфа II…

По моему мнению – главной целью Ди и Келли как раз и являлось получение места при дворе императора. Возможно, альтернативной целью был двор короля польского. Если говорить о причинах этой миссии, то мне они представляются вполне очевидными: создать надежную структуру получения важных данных непосредственно из первых рук, из императорского двора и приближенных ко двору лиц. При этом сама миссия проводится под видом частного путешествия двух известных алхимиков.

Таким образом, «Книга Дунстана», со всей окружающей ее легендой, была создана, как одно из средств завоевания доверия Рудольфа II и в качестве дополнительного подтверждения приватности миссии Ди и Келли.

Еще одним аргументом, подтверждающим приватный характер миссии должно было стать то, что Келли и Ди путешествовали вместе со своими семьями и порядочным объемом скарба (Ди вдобавок вез часть своей библиотеки), что весьма неудобно и накладно с одной стороны, и не характерно для дипломатической миссии с другой стороны. У Рудольфа не должно было возникнуть никаких ассоциаций с тем, что данная миссия может осуществляться по приказу королевы Елизаветы I...

Теперь зададимся вопросом: каким образом Рудольф должен был удостовериться в реальной ценности «Книги Дунстана»?

Как минимум для этого использовалось несколько приемов:

- Старинный дорогой пергамент
- Загадочные рисунки и диаграммы
- Загадочный шифр
- Алхимические знаки на первой странице (Феникс и Уроборос)
- Легенда о создании манускрипта Св. Дунстаном и о событии его обретения, подтвержденная известным авторитетным ученым (Ди)

При всем вышеперечисленном, существовал еще один очень существенный момент, на случай, если содержание манускрипта все же придется раскрыть (*еще жива была инквизиция, на магию и алхимию ревнители веры смотрели пристально*) – язык манускрипта. Английский был выбран специально (*хотя Ди, кстати, прекрасно знал еще и латынь, и французский, и греческий, и немецкий, и итальянский*) – поскольку для Св. Дунстана английский тоже был родным…

Однако, Св. Дунстан жил в X веке, и его английский существенно отличался от английского 16-го века. Этот факт был известен Ди и послужил основой идеи фонетического кода, который должен был скрыть (и успешно это сделал) некомпетентность Ди и Келли в древнеанглийском языке X века. Именно поэтому язык манускрипта, как я уже отмечал, выглядит специально модифицированным или упрощенным, а может являться имитацией некоего диалекта.

В дополнение. Это совсем не первый случай, когда Джон Ди пытался получить благосклонность королевской особы с помощью книги с алхимическим содержанием. Достаточно вспомнить «Иероглифическую монаду» Ди, которую он посвятил великому римскому императору Максимилиану II [5].

**А теперь весьма существенная информация – Ди и Келли одновременно с работой над «Книгой Дунстана», работали над созданием еще как минимум одной легендарной книги – книги по Енохианской магии и Енохианскому языку. И использовали практически те же самые приемы кодирования - с помощью неизвестных науке символов, являвшихся одновременно словами, буквами и числами. Из дошедшего до нашего времени Енохианского словаря - словообразование строилось по схожему с нашим манускриптом принципу – новые слова (благо здесь не требовалось имитировать реальный язык) образовывались склейкой или намеренным искажением слов из разных языков – практически идеальный способ доказать, что это и есть язык человечества до Библейского Вавилонского смешения языков. Более того, некоторые символы Енохианского языка, несмотря на иную графическую реализацию, имеют схожие значения, что и символы в нашем манускрипте. Например, символы UN со значением «und» в Енохианском алфавите и символ «AND» в нашем манускрипте. Использован тот же принцип сокращения или упрощения значений символов – OR=orh, UR=our, Na=nach, UN=und и т.д. Посмотрите на символы нашего манускрипта: MA = mat(?) или mature (?), RI =?, GE=GEV=give(?).**

## 6.3 Загадка с замком в манускрипте Войнича.

Одним из наиболее серьезных аргументов апологетов того, что манускрипт Войнича был написан до середины 15-го века, является изображение замка на страницах манускрипта #85r и #86v.

Основная деталь в изображении замка, которая косвенно намекает на это - зубцы на замковой стене, исполненные в дизайне «ласточкин хвост».

Считается, что такой дизайн был разработан и использовался только на территории Италии, особенно ранее 15-го века, что, в свою очередь, вполне вписывается в датировку пергамента радиоуглеродным методом. Считается, что в остальной Европе этот дизайн появился много позднее.

Не будем пока с этим спорить. Просто попробуем найти ответ на вопрос – мог ли автор манускрипта видеть аналогичный дизайн где-либо еще не посещая Италии.

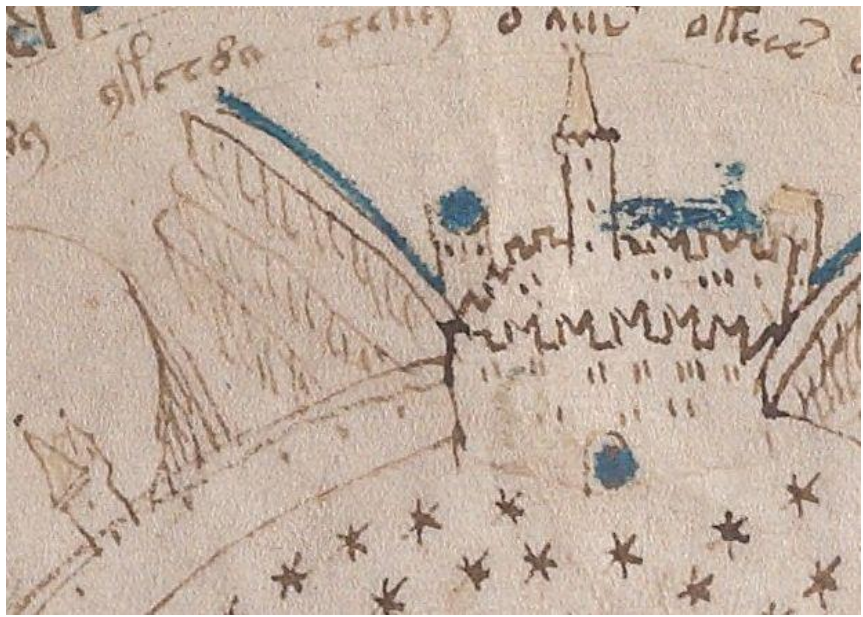

Рис. 80 Часть изображения замка со страниц #85r-86v, где хорошо видны зубцы в форме ласточкин хвост.

Для начала проанализируем изображение.

Как и во всем манускрипте – изображение аллегорично. Замок в манускрипте является частью изображения большой специфической схемы, аллегорично изображающей функциональные взаимосвязи в человеческом организме.

Обратим внимание – замок на картинке изображен весьма упрощенно. Он не содержит никаких деталей, указывающих на то, мог ли существовать подобный замок в реальности, и если – «да», то где именно. Рисунок также не содержит никаких специфических элементов конструкции – только стены и башни.

Изучив изображение еще более пристально, мы увидим:

- Слишком большое количество окон в стенах, и их расположение весьма сомнительно для фортификационного сооружения

- Не показано никакой внутренней инфраструктуры, которая позволила бы гипотетическим обитателям подобного замка добираться до этих самых окон в крепостной стене.

Возможно, сам автор манускрипта для создания своего изображения использовал увиденную где-то картинку. Вопрос – где?

Например, здесь – см. Рис. 81.

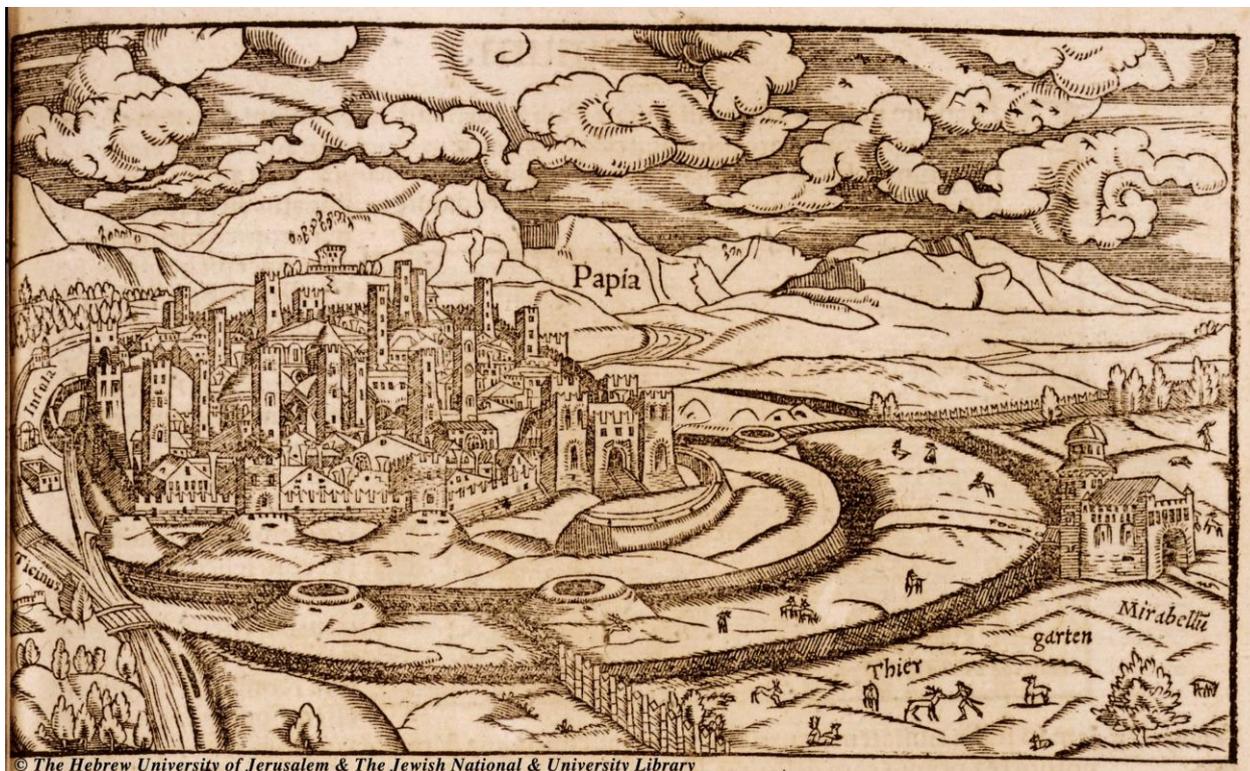

Рис. 81 Изображение замка в итальянском городе Папия (ныне – Павия (Pavia)), взятое из книги Себастьяна Мюнстера (Sebastian's Münster (1489—1552)) «Cosmographia» [10,11] с зубцами типа «ласточкин хвост» на стенах.

Труд Мюнстера «Cosmographia» был впервые опубликован в Базеле в 1544г.

Это была одна из самых популярных книга по географии для того времени, переведенная на многие языки – немецкий (оригинал), латынь, английский, итальянский, французский и даже чешский.

В период с 1544г по 1628г книга выдержала более 40 изданий.[12]

Изображение на Рис.81 впервые было добавлено в латинское издание в 1550г.

Это единственное изображение итальянского замка с зубцами типа «ласточкин хвост» во всей книге…

Но, что более важно для нас – Себастьян Мюнстер никогда не бывал в Папии. Он создал большинство своих изображений на основе рассказов путешественников, паломников и описаний и картинок из книг других авторов.

Утверждение выше достаточно легко проверить – благодаря наличию в городе великой святыни – гробнице Св. Августина в базилике Св. Петра, рисовать сюжеты, связанные с Павией было весьма распространенным занятием и существует немало картин с изображениями средневековой Папии.

И еще один немаловажный для нас факт – «Cosmographia» Мюнстера, французское издание 1573г, легко обнаруживается в списках библиотеки Джона Ди [13], сделанных им собственноручно в течение двух недель до отъезда на континент...

Как я уже отмечал – слишком много совпадений.

Еще о зубцах в виде «ласточкин хвост» - точно такая же техника использовалась при строительстве стен Пражского Града. Особенно это хорошо видно на южной стене… Тот же «ласточкин хвост». Согласно истории строительства замкового комплекса – южные стены были закончены к 1560г, т.е. Ди и Келли видели зубцовку южной стены Пражского града в начале сентября 1584, в своей поездке на аудиенцию к Рудольфу II. Таким образом, аргумент в пользу «ласточкиного хвоста», увиденного потенциальным автором манускрипта в Италии не является убедительным.

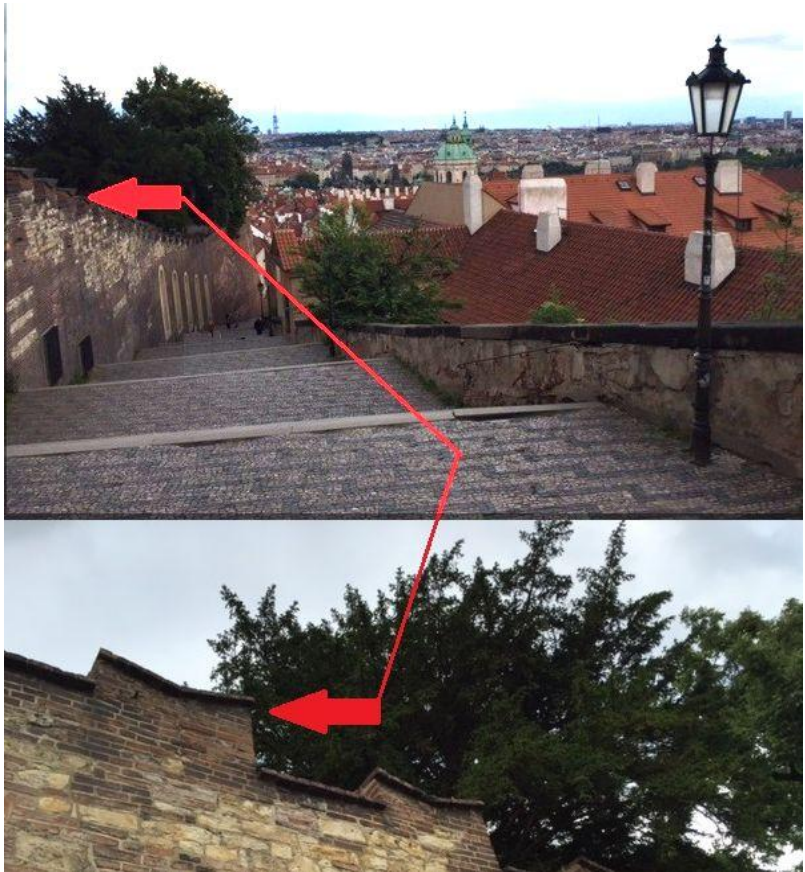

Рис. 82 Южная стена Пражского Града (фото А.У.).

Помимо прочего – купола башенок на рис.80 вполне соответствуют чешскому стилю (см. замки Праги, Крумлова, Тршебони…)

Если взглянуть на остальные детали большой картинки на тех же страницах манускрипта, то обнаружатся еще изображения фортификационных сооружений.

Пример:

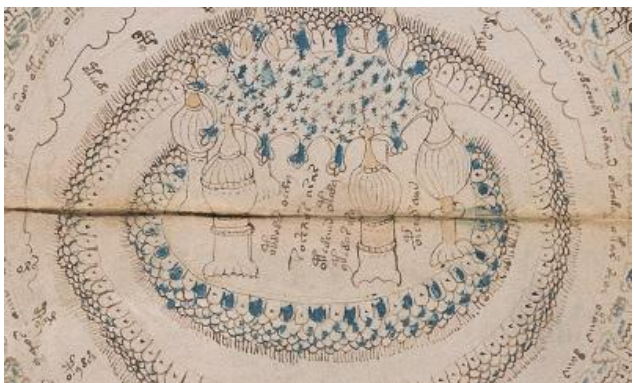

Рис. 83 Фортификационные сооружения на страницах 85r-86v. Давайте снова полистаем «Cosmographia» Мюнстера.

Как насчет Венеции?

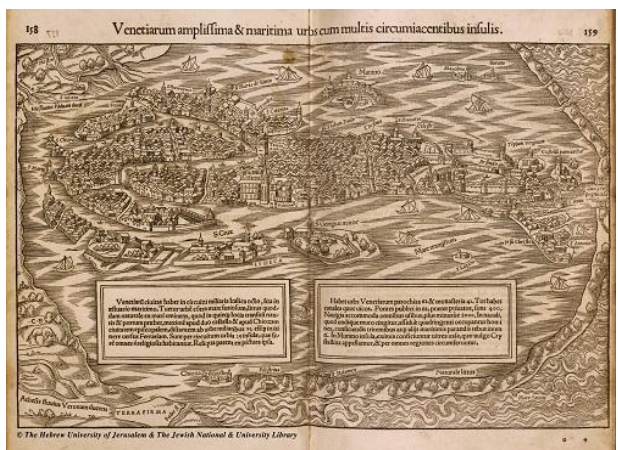

Рис.84 Изображение Венеции из книги Мюнстера (1489—1552) «Cosmographia» [10,14].

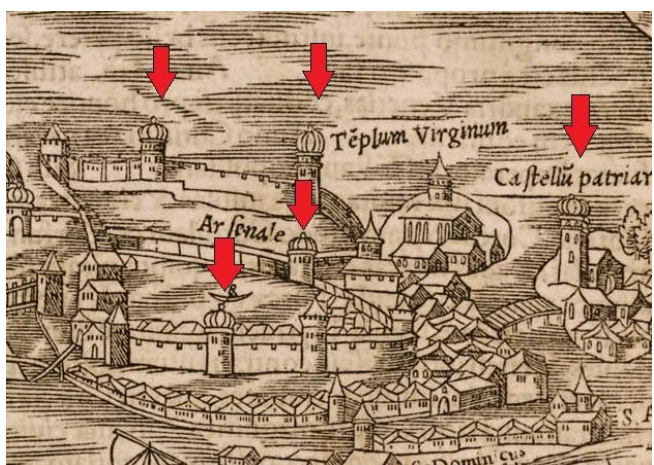

Рис.85 Увеличенный фрагмент изображения Венеции из книги Мюнстера - с цилиндрическими башнями, возможными прототипами изображений на Рис. 83

И опять - это единственное изображение в книге Мюнстера, содержащее столь необычные цилиндрические башни…
Остальные круговые диаграммы со страниц 85r-86v не менее интересны.

Нечто похожее можно найти в книге Себастьяна Серлио (Sebastian Serlio (1475 -1554)) "I sette libri dell'architettura"("Семь книг об Архитектуре") в книгах/разделах #3 и #5, посвященных строительству павильонов/беседок округлой формы - достаточно обратить внимание на проекции этих сооружений, чтобы убедиться, что могло послужить источником для фантазий автора манускрипта. Как известно, Ди очень интересовался архитектурой и был знаком с трудами Серлио.

Я склоняюсь к версии, что за основу изображения на страницах 85r-86v были взяты изображения из других книг.

Библиотека Джона Ди – необыкновенно интересный источник ответов на загадки манускрипта…

Итальянский дизайнер и профессор архитектуры из университета г. Бари – Др. Дузеппе Фаллакара(Giuseppe Fallacara), занимаясь изучением одного из самых необычных и загадочных сооружений в мире – Castel-del-Monte, независимо от меня пришел к выводу, что автор нашего манускрипта для создания одной из круговых диаграмм использовал не дошедший до наших времен уникальный манускрипт с архитектурным описанием Castel-del-Monte [???].

## 6.4 Дополнительные выводы по авторству и датировке манускрипта.

- Наиболее вероятно, что манускрипт написан Джоном Ди. Или под его непосредственным руководством. Существует вероятность того, что Эдвард Келли также был привлечен к написанию манускрипта.

- Основная часть манускрипта, возможно, была готова к концу 1585г. Возможно, окончательно работы по его созданию были завершены в конце 1587г или начале 1588г.

- Настоящее название манускрипта – «Книга Дунстана»

## 7 Примеры расшифровки.

Само по себе обнаружение кодов и ключей – это малая часть для дальнейшей работы специалистов по его переводу и интерпретации.

Для любопытствующих далее я хотел бы продемонстрировать несколько примеров, как использовать ключи для прочтения.

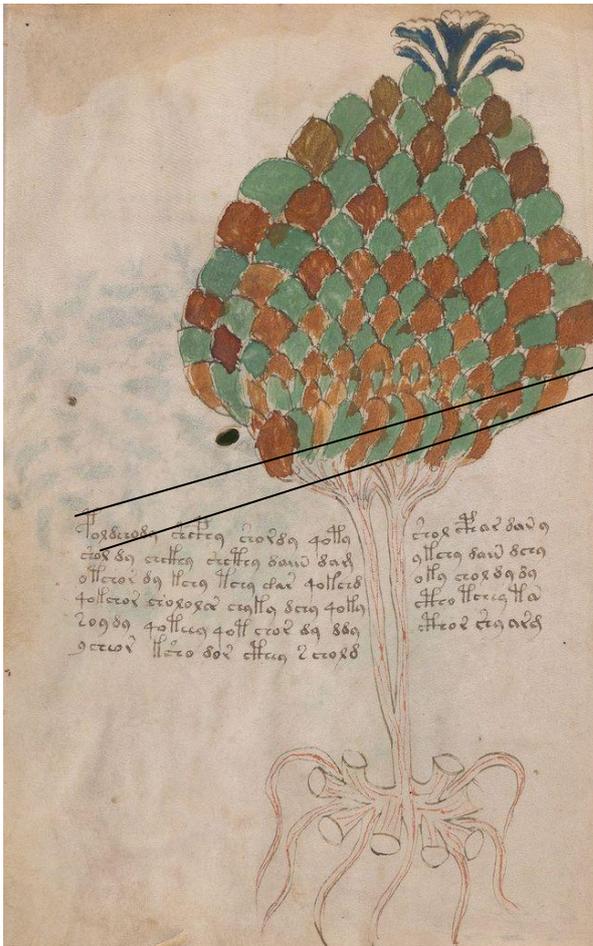

Начнем со страницы 11v

Попробуем прочесть две первые строки полностью

Рис. 86  Изображение страницы 11v

Ниже приведена реконструкция текста.

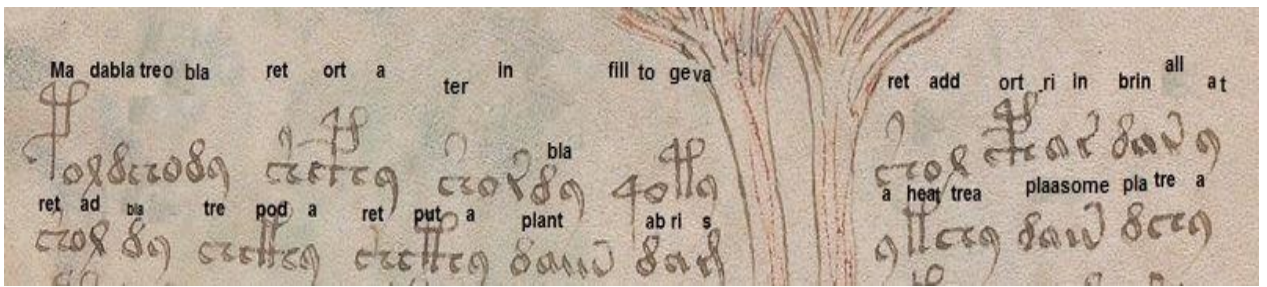

Рис. 87 реконструкция первых двух строк страницы 11v

Имеем следующую запись:"**Ma to dabla tre to plat ret ort at ter in plat fil gevat ret ad ortri in brin all, ret ad pla at tre potat ret putat plant pla riis at heat tre at plasom pla tre at**"

Реконструкция:
"**Ma to (или Mat)dabla tre to plat, retortat, tern in, plat fil, gevat ret ad ortri in, brin all, ret ad, pla at trepotat, ret putat plant, pla riis at heat treat, plasom pla treat**"

Современный вариант:

Реконструкция:
"Mat double trunk tree. Took plate and retort. Tern in trunk. Plate fill given red. Add directly in. Bring all. Red Add. Place at tripod, put red plant, place rise at heat treat, place blossom treat"

Возможно, говорит нам это следующее:
"**Приди к дереву с раздвоенным стволом** (как на картинке – А.У.) **с блюдом, ретортой, встань** (реторты (?) в источник сока – А.У.)**, наполни блюдо полученным красным** (возможно порошком, возможно соком – А.У.) **добавь прямо туда, возьми всё, добавь красный, положи на треноги, добавь красное растение, добавь рис для тепловой обработки, положи цветки для обработки...**"

Еще пример.

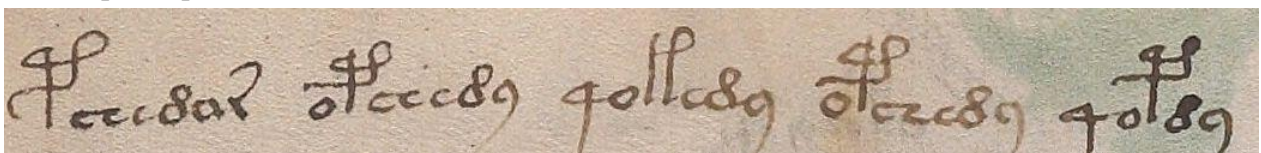

Рис. 88 Изображение символов первой строки страницы 75v
Это анатомический раздел. Соответствующее изображение на странице 75v содержит изображение человеческой трахеи. Читаем.

"**Ma tre hear pla riin to ma tre hear plat fil heat hear plat to ma tre hear fil ma plat**"

"**Ma trehear, pla riin to ma trehear, plat fil, heathear plat to ma trehear, fil ma plat**"

"**Главную трахею возьми, используй проточную воду (ручей) для главной трахеи, наполни блюдо, блюдо с нагревателя поднеси к трахее, наполни главное блюдо...**"

Важная ремарка:

В вышеприведенном случаях символ 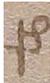 ("МА") применен, скорее всего, в одном из следующих значений – «meet» (встреть), «make» (сделай), «move (to)» (двигайся к…), «bring» (принеси), и "major" зрелый, основной) или «mature» (взрослый, зрелый).Возможно, значение определяется символом правила "о"…

Вернемся к странице 100r из фармацевтической части и попробуем идентифицировать некоторые растения. Обратим внимание, что не ко всем растениям на странице автор прописал названия, вместо названий даны инструкции, что делать с тем или иным растением. Это означает, что такие растения должны быть опознаны специалистом без дополнительных подсказок исключительно по внешнему виду на рисунке, т.е. существуют в природе именно в таком виде. Красным шрифтом я выделил свои предположения.

В манускрипте часто встречается слово «ТРАХЕЯ». И не только для обозначения человеческого органа, но и в отношении растений. Могу лишь предположить то, что в отношении растений под трахеей понимается полый стебель или стебель или ствол, как таковые.

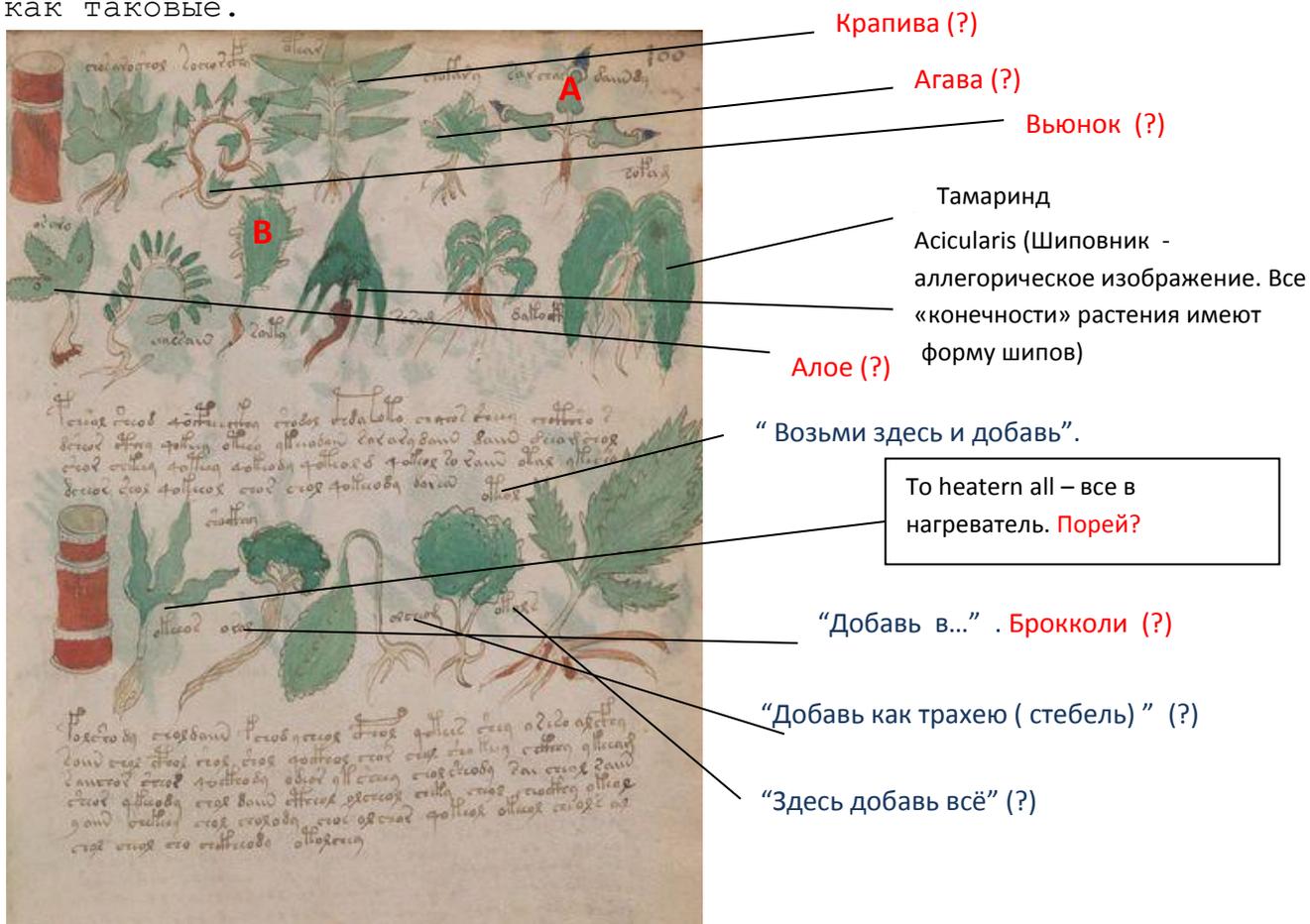

Рис. 89 Идентификация растений со страницы 100r

Два наиболее интересных экземпляра я обозначил как А и В.

Начнем с образца А.

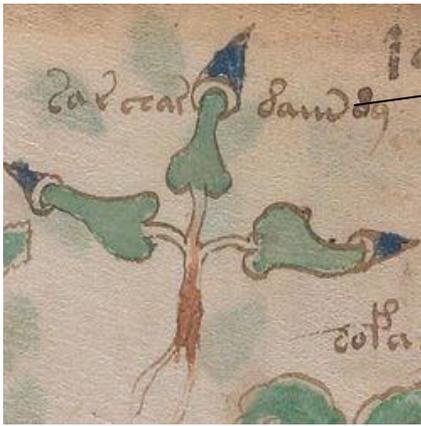

Рис. 90 Образец А

Возможное зашифрованное название

Данные экземпляр визуально никак не ассоциируется с каким-либо реальным растением. Поэтому предположим, что автор манускрипта написал рядом с растением его название. Указанная символьная группа содержит комбинацию "PLANT". Полное читаемое слово дает нам "PLANTBLA" или "PLANTAPLA".
Существует весьма известное растение с созвучным латинским названием - "Plantago" или, в современном английском – "Plantain".
Попробуем проверить, насколько наше прочтение соответствует реалиям…

Возможно, что это - Plantago lanceolata (Подорожник ланцетный)[15]

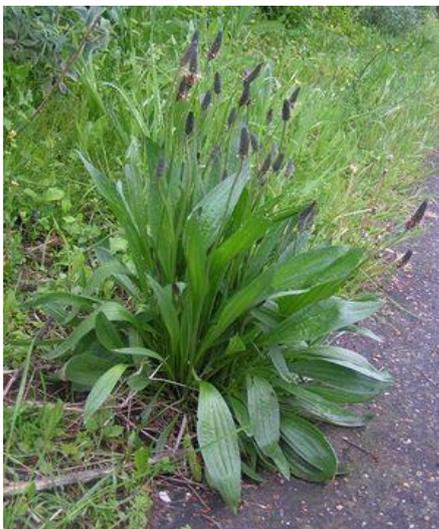

Возможно, что не все части растения следует использовать для рецепта, а только листья и "пики".

Рис. 91 Изображение Plantago lanceolata.

Рассмотрим теперь образец В. Я выбрал его для исследований, поскольку обнаружил, что в 2014г. ряд специалистов [16] идентифицировали это растение как Опунцию с американского континента. Они даже привели имя этого растения, ассоциировав подпись к нему в манускрипте с ацтекским названием Nochtll или Noshtll…

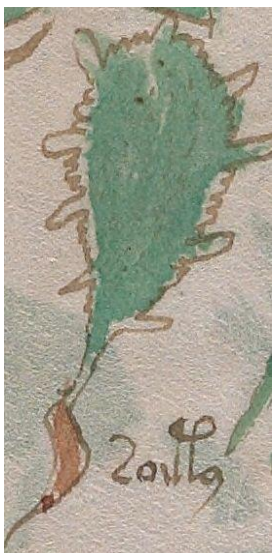

Мне удивительно, почему было выбрано именно это изображение, ибо оно может быть сравнимо с другими – с аналогичным форм-фактором. На этой же странице манускрипта присутствуют еще два кандидата, вполне подходящие под то, чтобы оказаться Опунцией (Opuntia)…

Рис. 92 Изображение образца B.

С использованием найденных ключей мы можем прочесть подпись к картинке, и звучит она так: "AL TO GEV AT" – «ВСЕ ВМЕСТЕ»…

Увы, с названием это нам не помогает, но дает понять, что имено следует взять для рецепта у данного конкретного растения.

______________________________
Два следующих примера были взяты из работы [17], содержащей множество идей по идентификации изображений и расшифровке слов. Этот многостраничный труд удивляет своей странной логикой...

Начнем со страницы 46 вышеуказанного источника [17]. Она посвящена анализу страницы # 31r, а если говорить более точно – анализу первой символьной группы первой строки, которая, по мнению авторов статьи, возможно, читается как "KOOTON" (Cotton):

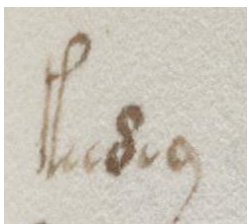

Рис. 93 Первая группа символов на странице 31r.

Если мы применим найденные ключи, то получим:

"Heat ern pla here at" – или "Heatern pla hereat", что в переводе на русский означает «Здесь (*или сначала*) поставь нагреваться (*или нагреватель*)»

Автор манускрипта использует несколько понятий для слова «нагреватель» и «нагревать» - "heathere" (непосредственно нагреватель – как целый прибор), "heatern"(вполне возможно имеется в виду «сушилка»), "heatereh" (вполне возможно – топка, духовка, емкость для температурной обработки) и еще ряд, которые будут приведены в отдельном словарике в отдельной статье.

Одни из них означают приспособления типа жаровни, другие – сушильные устройства, третьи – чаны для нагрева.

На Рис. 94 (ниже) приведен пример изображения нагревателя (скорее всего – термической сушки с топкой), который состоит из двух главных элементов: топки и нагревательной или сушильной камеры, куда устанавливаются соответствующие блюда (к каждому из этих элементов даны кодированные подписи).

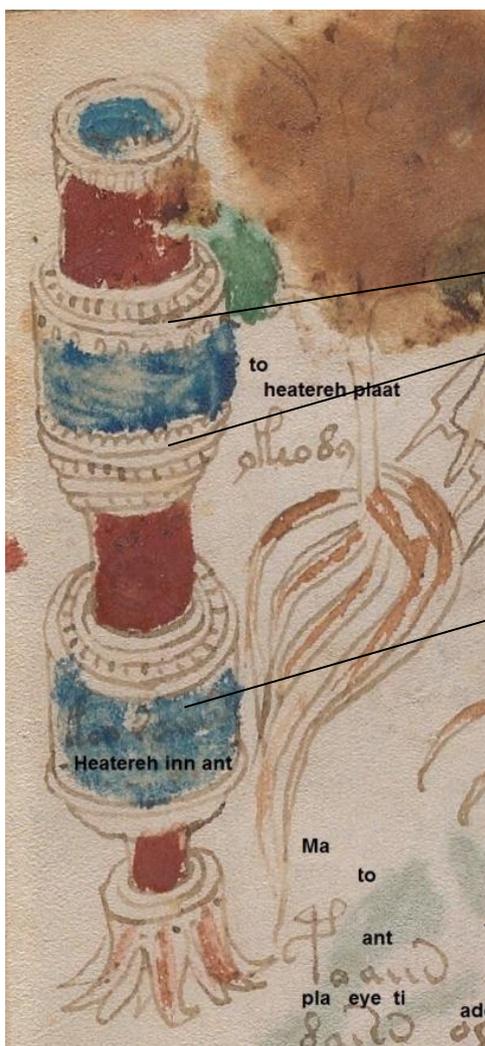

Рис. 94 Пример изображения нагревателя в манускрипте

Теперь перейдем ко второму примеру.

Обратим внимание на страницу 19 указанной статьи [17], посвященную анализу страницы манускрипта # 68r.

Автор статьи [17] сначала идентифицирует созвездие Плеяд (Pleiades) и, в качестве результата, логично предполагает, что данный сектор диаграммы посвящен знаку Тельца (TAURUS). Поэтому символьную строку над Плеядами автор статьи идентифицирует… как "Taurus".

Пример удивительной логики – Плеяды ассоциируются только с Тельцом и ни с чем больше, и без их изображения на диаграмме определить сектор Тельца было бы невозможно в принципе. Зачем автору манускрипта дважды подтверждать подлинность Тельца?

Автор статьи [17], в своем разделе "TAURUS", продолжает развивать свою мысль еще более оригинальным способом:

*It was noted at the beginning of this article that no word of the VM has been convincingly translated or glossed, but in fact there is one word which has received a degree of consensus. On page 68r, in a dramatic diagram apparently showing the moon in the heavens, a set of seven stars has been suggested to show the 'Pleiades' sometimes known as the Seven Sisters, in the constellation of Taurus (……) and the accompanying word has <u>sometimes</u> be interpreted to indicate TAURUS (Zandbergen 2004-2013)   […]*

Перевод:
«В начале этой статьи *([17] – А.У.)* было обозначено, что ни одного слова из манускрипта не было убедительно переведено или истолковано, но в тоже время существует слово, по которому достигнут консенсус. На странице 68r, на драматичной диаграмме, отображающей, видимо, луну на небе, скопление из семи звезд  было предположительно определено,  как Плеяды, известные также, как «Семь сестер» в созвездии Тельца (…)  и соответствующее слово рядом **иногда** интерпретируется как обозначение Тельца (Taurus) *(Zandbergen 2004-2013)   […]*

Честно говоря – совершенно неожиданное заключение. Существует астрономический раздел манускрипта, где Телец представлен дважды… Без попытки сокрытия, что это именно Телец. Автор манускрипта не поленился даже название месяца написать под его изображениями нешифрованной латиницей.

Но самый загадочный для меня вопрос -  что означает "**иногда**"? Если иногда – Телец, то какая интерпретация в других случаях?

Набор символов, идентифицируемый автором статьи [17] как одно слово, визуально  (как минимум - для меня) связан именно с Плеядами, равно как остальные два набора символов – один со звездой, другой с кривой линией соответственно.

Непонятно, почему для анализа набора символов был взят только один из трех, и непонятно, почему для поиска идентификации было выбрано лишь название созвездия Тельца ("Taurus"), хотя намного логичнее было бы начать хотя бы с названия Плеяд.

Сам знак Плеяд на данной диаграмме  - это мое мнение – был использован только для того, чтобы читающий мог ориентироваться в диаграмме, это маркер, от которого по известным правилам читающий смог бы найти начальный сектор диаграммы, а возможно, рассчитать корректный день месяца. Как уже говорилось ранее, у автора манускрипта не было необходимости одновременно рисовать Плеяды и писать слово «Телец» - это очевидно.

Рассмотрим страницу 68r и символьные подписи на диаграмме.

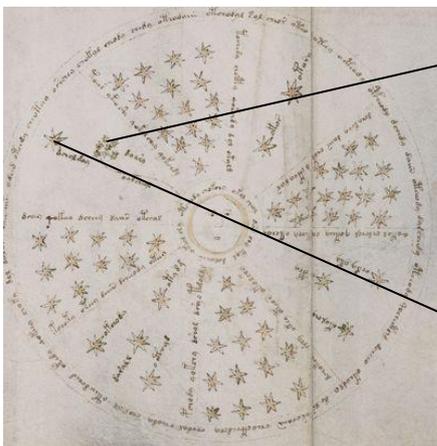

Это скопление Плеяд и чуть выше набор символов.

Опять же - если Плеяды/Телец были опознаны, это означает, что звезда, следующая за Плеядами левее, тоже должна быть идентифицирована - как Альдебаран (α Tau, α Tauri, Alpha Tauri)

Рис. 95 Изображение на странице манускрипта 68r.

Очевидно, что для Плеяд или Тельца, а также для Альдебарана должны работать одни и те же правила кодирования, вероятно, как и для надписи у «кривой линии». В данном случае - даже неважно, на каком языке написан манускрипт. Здесь важно по набору идентифицированных изображений установить корреляцию, позволяющую определить правила символьной записи, которые, в случае успеха смогли бы дать ключи к манускрипту.

Если такой корреляции нет – значит, выбран ложный путь поиска.

Но автор статьи [17] почему-то не использует такой возможности.

Конечно, мы уже можем прочесть подписи к диаграмме...

Для начала рассмотрим 12 созвездий на страницах манускрипта.

Казалось бы, что здесь представлены все знаки зодиака за исключением Козерога и Водолея. А знаки Овна и Тельца представлены дважды.

Возникает вопрос – почему одни знаки отсутствуют, а другие повторяются?

Моим единственным предположением было – Козерог и Водолей – это холодные зимние месяцы в регионе, где проживал автор манускрипта.

Допустим. Тогда зачем автор манускрипта удалил их из своего календаря?

Как минимум, в эти зимние холодные месяцы ничего не растет – нечего сажать, нечего собирать…

Соответственно, вариантов ответа совсем немного – наиболее вероятно, мы имеем дело с лунным календарем, содержащим инструкции, что и когда делать с семенами, растениями и урожаем.

Как еще можно подтвердить данное предположение? Попробуем понять, почему Овен (страницы 70v и 71r) и Телец (страницы 71v и 72r) представлены дважды.

Мои предположения касательно Овна следующие:

- Во-первых, на изображениях в календаре вовсе не Овен… А Козерог…

- Да, латинская готическая надпись под изображением Козерога безапелляционно заявляет, что это месяц – "Abril"(апрель), и это определенным образом заставляет читателя сомневаться в моем предположении. Но, я настаиваю – на изображениях именно Козерог. И все надписи латиницей под ним и остальными месяцами приведены только для необходимости обратить внимание посвященного читателя на факт, что на изображениях – Козерог. Непосвященный будет считать изображения – Овном, и, соответственно, допустит ошибки при интерпретации.

Идем дальше…
- Козерог на обоих изображениях поедает некий зеленый куст (на первом съеден кусочек меньшего размера, чем на втором)

- Если мы обратимся к лунному календарю – мы найдем разгадку возникшей нестыковки – автор манускрипта намекает, что наиболее важный для работ период наступает в апреле, когда Луна убывает, переместившись в созвездие Козерога…

- Согласно изображению - важно не просто вписаться в указанные лунные фазы (обозначенные, как не съеденные части куста), но еще и учесть определенные дни – обратите внимание на расположение копыт Козерога и холмики, на которых они располагаются.

Возвращаясь к главе 4, где производится поиск скрытых в изображениях букв. Собственно понимание, что на последней странице именно Козерог (Capricorn) пришло после анализа картинок лунного календаря манускрипта. В календаре нет изображений Водолея и Овна.

Мы не будем заниматься анализом повторяющихся Тельцов (тут никаких сомнений – в обоих случаях – Телец), но подход будет совершенно аналогичный.

Кстати, еще одна интересная загадка манускрипта – изображение созвездия Рака. Оно состоит из двух раков – красного и белого. Я не уверен в точности своих предположений, но красный рак вполне может соответствовать полному лунному затмению с эффектом «кровавой Луны».

Белый рак, в свою очередь, может символизировать некие события, связанные с солнцем… Это всего лишь идеи для дальнейших изысканий.

Вернемся к странице манускрипта # 68r и узнаем, что означают надписи к Плеядам и Альдебарану.

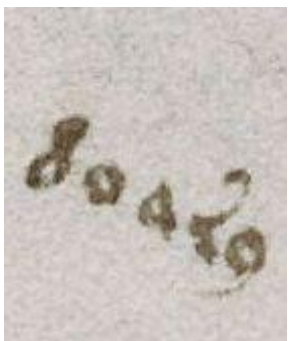
"pla to *(two)* riin at"  - «дважды полей» (?)

Рис. 96   Изображение подписи к Плеядам на странице 68r.

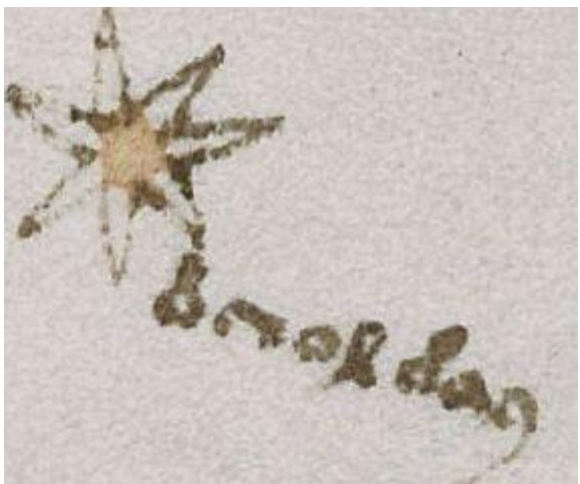

"Pla tre to end pla rin at"

"Возьми *(посади)* дерево и в конце полей"

Рис. 97    подпись к Альдебарану    на странице 68r

Думаю, всем будет интересно, что говорит подпись к «кривой линии».

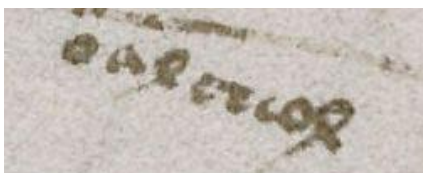

Рис. 98    подпись к «кривой линии» на странице 68r

"To rind tre here ad" – "Two tree rind add here" .
«Здесь добавь две   древесных коры»

Следует не забывать, что ряд символов используется автором в качестве обозначения чисел   (см. раздел 6.1.) и на диаграммах манускрипта есть надписи с очень высокой вероятностью содержащие сведения о количестве того или иного продукта или повторяемости действия.

В манускрипте вы найдете множество простых и часто однокоренных слов. Например: "plant, grind, rin, rina, herba, grid, grin, brid, brind, pla, apla, aplant, ortea, cut" и т.д.

## 8   Анализ текста последней страницы.

Итак, мы уже идентифицировали большую часть ключей к основному тексту манускрипта.
Но существует еще и текст последней страницы, символы которой встречаются где-либо еще крайне редко, или же – только на последней странице. Тем не менее – попробуем построить логическую схему, которая бы позволила нам расшифровать текст последней страницы.

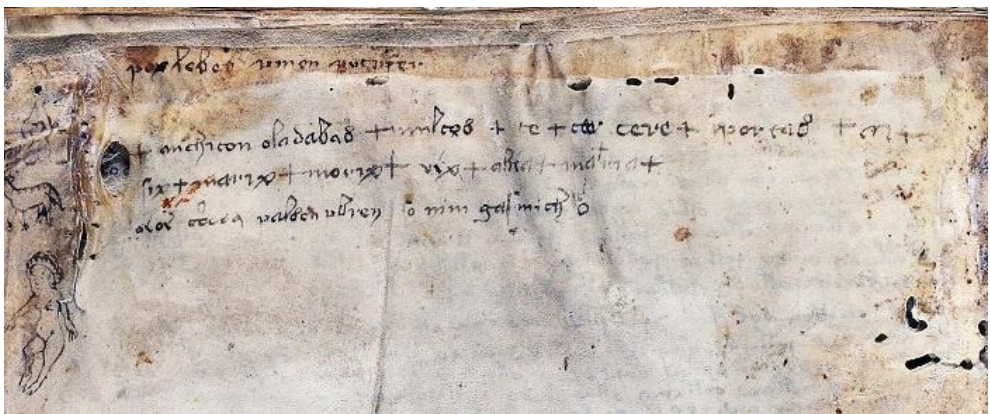

Рис.99   Верхняя часть последней страницы.

## 9  Постановка задачи.

Для начала вернемся к структуре манускрипта. Она содержит следующие разделы:

- Ботанический
- Астрономический
- Анатомический
- Астрологический
- Фармацевтический
- Рецептурный

Еще раз обратим внимание на то, что для написания манускрипта, охватывающего столь обширные разделы, автору необходимо было использовать большое количество источников информации и/или, как минимум, иметь долгосрочный доступ к фундаментальной библиотеке.

С этой точки зрения Джон Ди  – вновь представляется весьма привлекательным кандидатом на авторство.

Более того, Ди составил каталог своей библиотеки в Мортлейке, датированный 6-ым сентября 1583г (ровно за 2 недели до отбытия на континент)[???]. Так почему бы нам не попробовать сравнить данные из каталога с данными манускрипта, тем более, что определенную корреляцию мы уже обнаружили – как в случаях с Тритемием и Мюнстером.

Одним из моих самых первых предположений при визуальной оценке содержимого последней страницы было:

- Что последовательность групп символов, разделенных знаком "+" может являться списком авторов, привлеченных для написания манускрипта

Данное предположение, в принципе, не так уж и далеко от финальной идеи – почему бы этой символьной последовательности не быть списком использованных для написания книг?

Итак, текущее предположение – текст последней страницы содержит список книг библиотеки Джона Ди.

## 10 Начальная попытка дешифровки текста последней страницы.

В первую очередь следует отметить, что текст последней страницы содержит ряд ранее идентифицированных символов.

По сравнению с остальным манускриптом почерк на последней странице выглядит иным, выполнен небрежно, видимо другим человеком или… Другой рукой.

Это создает определенные трудности при идентификации даже ранее найденных символов. Поэтому часть ранее найденных ранее символов решено было пока не использовать, во избежание существенных ошибок.

На странице # 66r манускрипта мы видим ряд символов, использованных также и для написания текста последней страницы. Это подсказка, оставленная автором.

В данный момент эти символы еще сложно идентифицировать, но, возможно, позднее они нам помогут с верификацией расшифрованного текста.

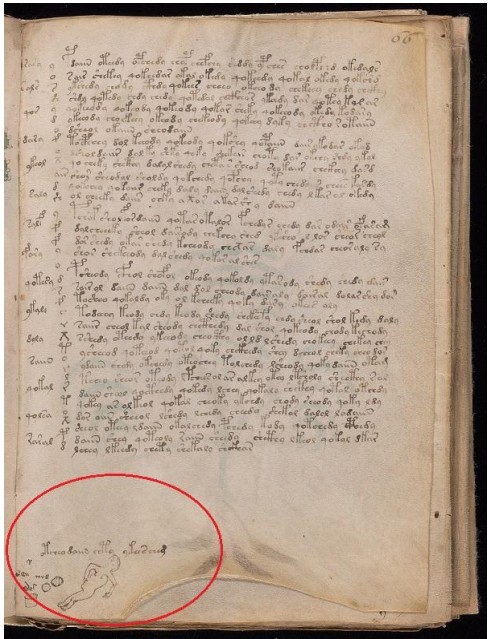

Рис.100 Страница 66r – локализация подсказки выделена красным.

Итак, начнем наш поиск.

Первым кандидатом на дешифровку выберем следующий набор символов:

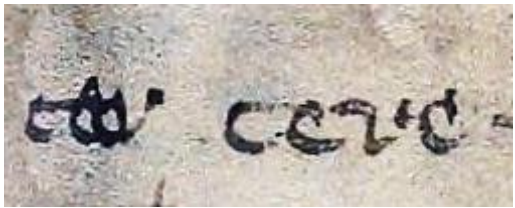

Рис. 101 – символьный набор для начального поиска ключей

Итак, мы знаем, что автором основного метода кодирования манускрипта – «Стеганографии», был Тритемий (Trithemius или Trithemi). Выбранный набор символов содержит символ "TRE", спеллинг которого может быть также - [tri:]. Попробуем это использовать.

Предположим, что более длинная часть набора символов, начинающаяся с символа "TRE" является частью написания слова "TRITHEMI".
Кстати, короткий набор символов, также выглядит начинающимся с символа "TRE" или его некоей модификации (допустим, здесь имеет место накладка 2-х или более символов, включая "TRE")

Получим возможный результат.

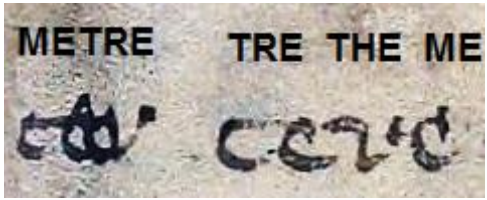
Рис. 102   Возможная расшифровка: "Metre Trithemi"

Используем ее для идентификации новых ключей (но, пока нет уверенности в их корректности, не будем их относить к списку окончательно идентифицированных).

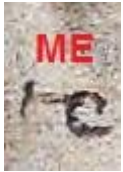

Рис. 103 -   Символ "ME" ([mi:] и/или [mə])

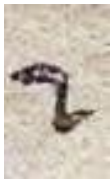

Рис. 104 -   Символ "THE"

Касательно слова "METRE". Я использовал предположение о том, что в данном наборе символов имеет место наложение "TRE" и "ME", и возможным правилом прочтения таких конструкций может быть следующее: в случае, когда символ стоящий правее (в данном случае "ME") накладывается на символ, стоящий перед ним (в данном случае "TRE") – порядок прочтения меняется
  - Сперва читается правый символ – "ME"
  - Затем читается предстоящий символ – "TRE"

Скажу сразу – предположение оказалось верным и автор манускрипта использовал его еще раз далее.

На данный момент наше исследование укладывается в рамки перечисления книг или их авторов, использованных для написания манускрипта. Книги и манускрипты Тритемия присутствуют в каталоге Ди.

Следующее предположение, вытекающее из базового подхода – текст последней страницы может содержать или содержит слово/символ "BOOK" или, что может быть более корректным для времен Джона Ди, слово/символ "BOK" (в дневниках Ди употребляются оба слова, но "BOK" встречается чаще).

Как мы можем идентифицировать локализацию данного символа?

Как минимум, в ряде случаев, он должен находиться в комбинации с предшествующим ему символом "THE" (что многократно наблюдается в дневниковых записях Ди).

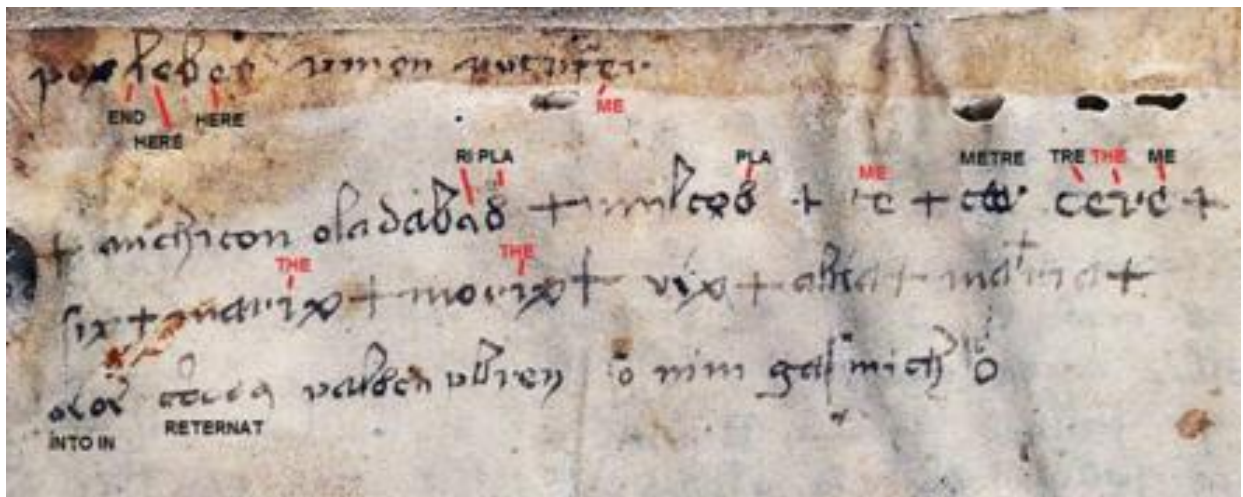

Рис. 105 Идентифицированные символы с подстановкой в текст последней страницы(черным цветом – найденные ранее, красным – идентифицированные в данной главе).

Итак, мы отметили расположение символа "THE". Кроме случая с символическим написанием Тритемия, две новые локации "THE" связаны с одним и тем же символом, который идентифицируем, как предполагаемый символ "BOK" или "BOOK".

Предполагаемый кандидат:

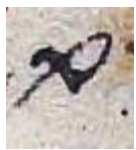

Рис. 106 возможный символ "BOK" или "BOOK"

Теперь имеем:

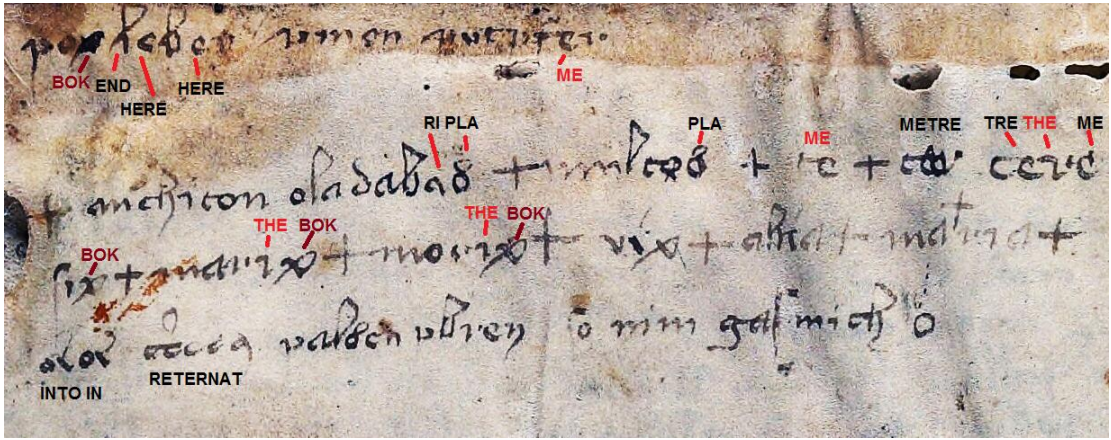

Рис. 107 Идентифицированные символы текста последней страницы.

Следующее мое предположение, подсознательно ассоциирующееся с содержимым манускрипта и списком библиотеки Джона Ди – где-то, возможно, есть символьная запись – Авиценна (Avicenna) и/или Вилланова (Villanova).

Потенциальный кандидат для символьной записи слова «Villanova»

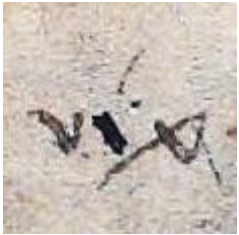

Рис. 108 Возможная символьная запись - VILANOVA

Выделим из нее новые ключи:

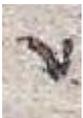

Рис. 109 Символ VI [как современное английское "we"]

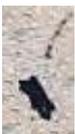

Рис. 110  Символ "IL" или "ILA" [ila:]

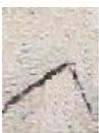

Рис.111  Символ "NO"

Здесь мы опять наблюдаем случай, аналогичный "METRE" – накладка символов "NO" и "VI".

Если мы правы с идентификацией – финальная запись выглядит так: "VIILANOVI", точнее - по базовым правилам конструирования слов (одинаковые буквы не удваиваются) - "VILANOVI"

Внесем новые ключи в текст последней страницы.

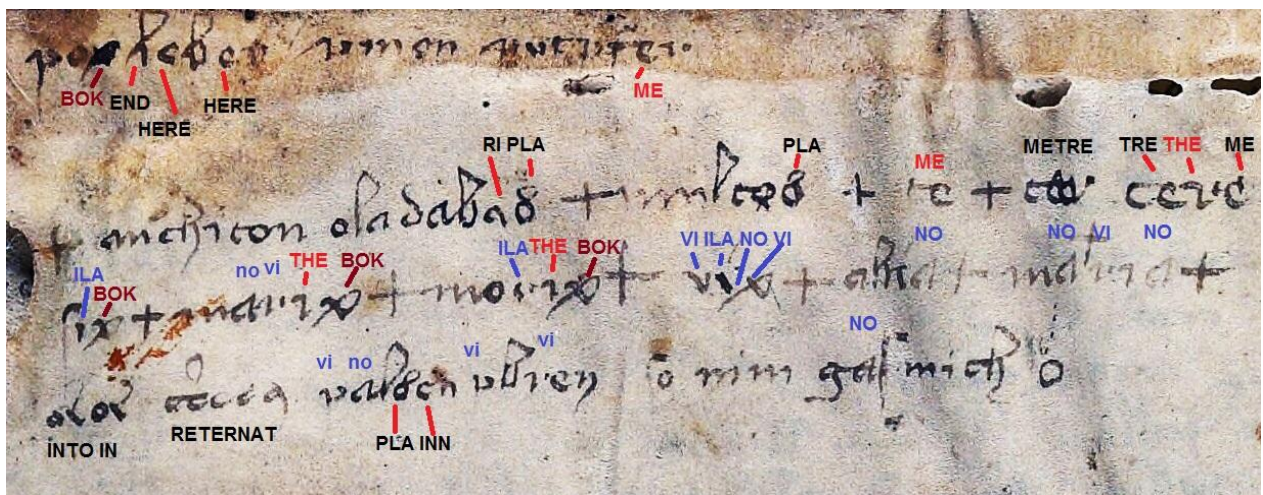

Рис. 112 Последняя страница с идентифицированными символами.

Мы пришли к моменту, когда требуется поискать дополнительные подсказки для дальнейших изысканий.

Вернемся на страницу манускрипта 66r и попробуем идентифицировать ключи в подсказке.

Возможно, что изображения под символами смогут нам помочь.

Цилиндр и кружочки слева от лежащей дамы лично у меня четко ассоциируются с изображением кусочков костей:

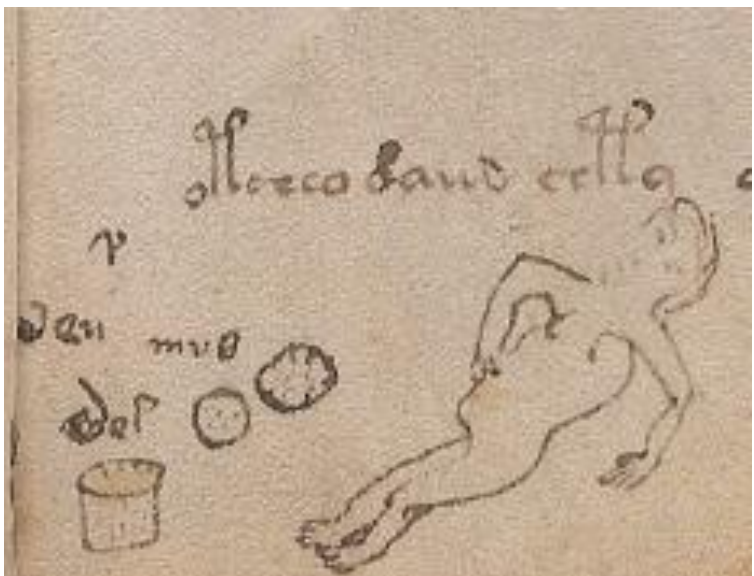

Рис.113 Увеличенное изображение возможных подсказок на странице 66r

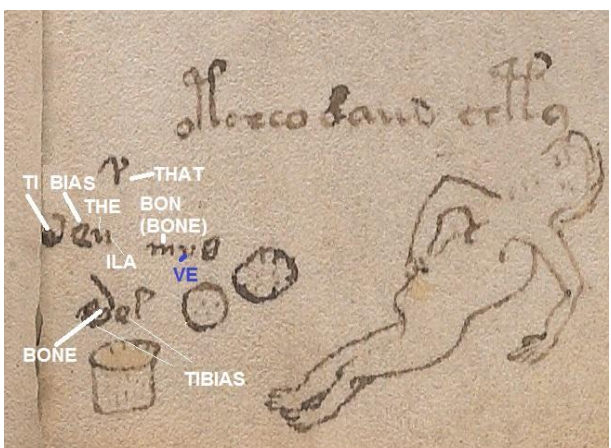

Рис. 114  Подсказки со страницы 66r – возможная идентификация.

Итак… Имеем следующие идентифицированные символы:

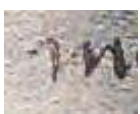

Рис. 115   Символ "BONE"

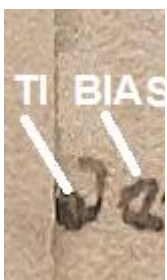

Рис. 116 Сочетание символов "TI"[ti:] и "BIAS" (требует дополнительной проверки).

Комбинация возможно читается, как "TIBIA" (bone) – «берцовая кость».

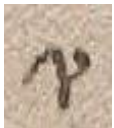

Рис. 117 Символ "THAT"

Следующий символ требует существенной верификации, но пока предположим, что это символ означает "SIC":

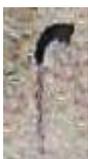

Рис. 118 Символ "SIC"

Почему "SIC"? Пока что такая идентификация ложится в структуру базовых предположений (простое слово, имеет смысловое прочтение слева-направо и справа-налево, имеет латинское значение, может являться как началом словесной конструкции, так и его концовкой.

Тем не менее, это предположение требует тщательной проверки, поскольку мы имеем ранее идентифицированный символ "ILA" и наличие второго символа с похожим значением (хотя, в те времена эти символы вполне могли разное значение) вызывает сомнения.

Обратим внимание на символьную строку #3 на последней странице.

Она начинается символом как на Рис.118 и продолжается символами "ILA" и "BOK".

Если данный символ действительно означает "SIC" - мы получаем SICILA BOK.

Заглянем в каталог Джона Ди.

Находим следующее:
"Joh: de Sicilia in canones Arzachelis de tabulis Toletanis.— Quaestiones mathematicales."

Вполне возможно, что имелся в виду именно John of Sicily, но существует также вероятность, что речь идет о книге сицилианского происхождения, сборнике без конкретизации его авторов.

Пока остановимся на этом и подставим все найденные значения в текст.

Имеем следующее:

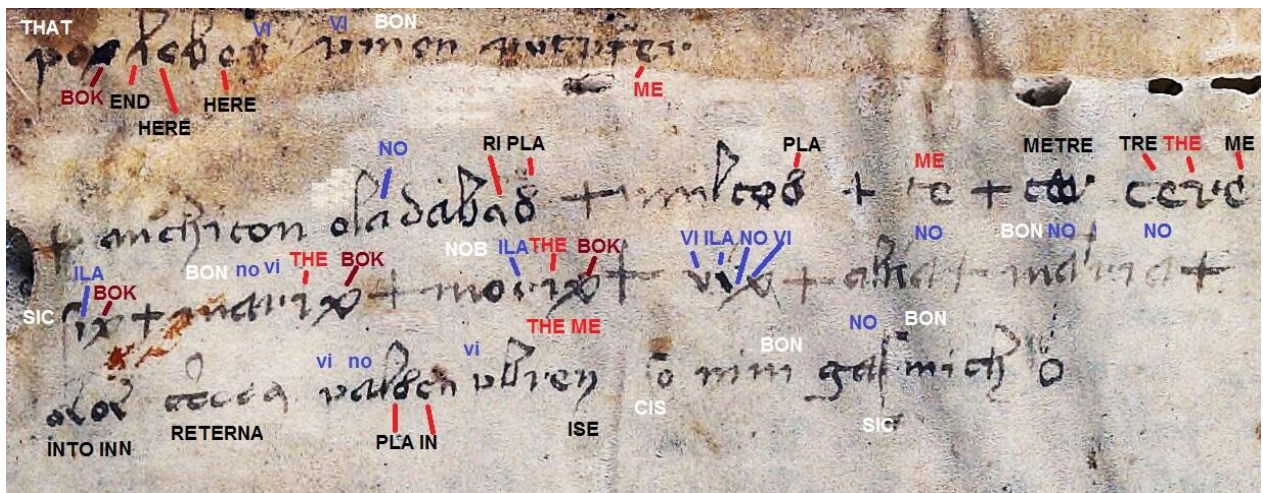

Рис. 119 Последняя страница с возможной расшифровкой.

1) Первая строка:
THAT BOOK END HERE HEAVY. WE BON ……………… ME….

2) Вторая строка:
…………   …NO (использован символ правила) … RI PLA + (может подразумевалось – R.I.P?)…. ANY …. PLA   + ME + METRE TRITHEMY + ……...PLA + … + *(два последних символ этой страницы не показаны на  Рис.119)*

3) Третья строка:
SICILA BOOK + BONNOVI THE BOOK + NOBILA (использован символ правила)   THE BOOK + VILANOVA + …NO  + BONNO + VI..NO

4) Четвертая строка:

INTO INN RETERNA WE NO … PLA IN (возможно PLAIN или PLAN)  … WE…… THEM ISE…  CIS (использован символ правила)… BON …NO SIC BON ….. 

## 11 Настало время  сказать :"Прощай, манускрипт Войнича!" Давайте скажем: "С возвращением, "Книга  Дунстана!".

По ходу моих размышлений в данной статье, читатель вполне смог понять, что «Книга Дунстана», как первоначальное название манускрипта и была моей главной версией и, наконец, она подтвердилась...

Работа шла с последней страницей манускрипта, закодированной, в большинстве своём, шифром, который практически нигде более не использовался.

Помимо прочего, почерк на последней странице сильно отличается от всего остального текста - скачущий, небрежный, плохо разборчивый...

Для начала, вновь обратимся к дневнику Джона Ди, к записи от 12 декабря 1587г. ([8] p. 25;[9]). В оригинале она звучит так:

"Dec. 12th [1587], afternone somwhat, Mr. Ed. Keley his lamp overthrow, the spirit of wyne long spent to nere, and the glas being not stayed with buks about it, as it was wont to be ; and the same glas so flit.ting on one side, the spirit was spilled out, and burnt all that was on the table where it stode, lynnen and written bokes, as the bok of Zacharius with the Alkanor that I translated out of French for som by spirituall could not ; Rowlaschy his thrid boke of waters philosophicall; the boke called Angelicum Opus, all in pictures of the work from the beginning to the end ; the copy of the man of Badwise Conclusions for the Transmution of metalls ; and 40 leaves in 4°, intitled, Extractiones Dunstani, which he himself extracted and noted out of Dunstan his boke, and the very boke of Dunstan was but cast on the bed hard by from the table."

«12 декабря 1587, около полудня, Мр. Эд.Келли опрокинул свою лампу, спирт широко разлился по столу, повсюду разлетелись осколки стекла[...]

Спирт загорелся и зажег все, что было на столе, за которым он работал - скатерть и книги, такие как книга Захарии с Алканором, которую я перевел с французского [...]; третья книга Роулаши о водах философских; книга, называемая Опус Ангелилум, вся в рисунках от начала и до конца, копия книги человека из Будвайза о трансмутации металлов; 40 листов в 4-х стопках, названные «Экстрактом из Дунстана», которые он собственноручно вырвал и сделал об этом пометку в книге Дунстана, а саму книгу Дунстана резко бросил со стола на кровать.»

Обратимся также к записи от 11 декабря 1587 года:

"Dec. 1st to 11th, my Lord lay at Trebon and my Lady

*all this tyme[…]"*

**«1 - 11 декабря, мой Господин лежит в Требоне и моя Госпожа всё это время».** (Возможно Ди говорит о себе и своей жене, поскольку часто пишет о себе от третьего лица. Второе предположение – подразумевается Вилем из Рожмберка (он же Вильгельм фон Розенберг, он же William of Rosenberg ) и его супруга, под покровительством которых в Требоне живут Ди и Келли. «Лежит» в данном случае означает «болеет».

## 11.1 Расставим акценты.

Теперь для нас будет важна каждая мельчайшая деталь.

Итак, что мы узнали из этих 2-х записей?

(1) <u>Зима</u>, почти середина декабря 1587 (возможно, лежит снег…)

(2) Ди и Келли находятся в Требоне (совр. – Тршебонь, Чехия)

(3) Ди и его жена, возможно Келли и его жена, возможно все они - болеют. Моё предположение – Ди говорит о себе и своей жене, поскольку часто в своих дневниках пишет о себе от третьего лица.

(4) Большое возгорание произошло в комнате, где Ди и Келли работали с книгами. Т.е. Ди и Келли пришлось тушить начинавшийся пожар.

(5) Келли разбил горящую спиртовую лампу, спирт разлился и воспламенил, все, что было на столе.

(6) Одна из книг на столе - книга Дунстана

(7) Кроме неё, помимо других книг, на столе 4 стопки по 10 листов, «экстрагированных» Келли из книги Дунстана.

(8) Келли оставил пометку в книге Дунстана о 40 «экстрагированных» листах.

Здесь опять важно обратить внимание, что последняя страница написана как бы другой рукой.

Что же, попробуем прочесть последнюю страницу строку за строкой (Рис. 120)

Здесь важно сделать еще одну ремарку – как читать символы.

Для меня, как для человека, для которого английский язык не родной, в некотором смысле легче адаптироваться под английский манускрипта с его фонетическим кодом, звучащий по иному для современных англичан.

Вот пример:
a) "ve" = [vi:] = современное "we", так же как "vy"(в окончаниях слов) – это базовый звук для символа, использованный автором манускрипта.
Но в некоторых случаях может читаться как [və]
b) "me" = [mi:] = современное английское "me"
В некоторых случаях [m] или [mæ] – в зависимости от места символа в слове или строке.
c) "ni" = "ny" = [ni:]
d) "ri" = "ry" = [ri:]
e) более сложный случай:
"AND" также созвучно "END" = [ənd].
В начале или в середине предложения оно означает "И", в конце – означает "КОНЕЦ".

Это правило относится ко всем ключам манускрипта. Автор или авторы, как бы забыв про грамматику, играют звучанием слов, и им удалось добиться результата, который практически во всех случаях позволяет однозначно определить значение построенного слова – применяется т.н. «фонетический код».

## 11.2 Расшифровка.
Давайте начнем.

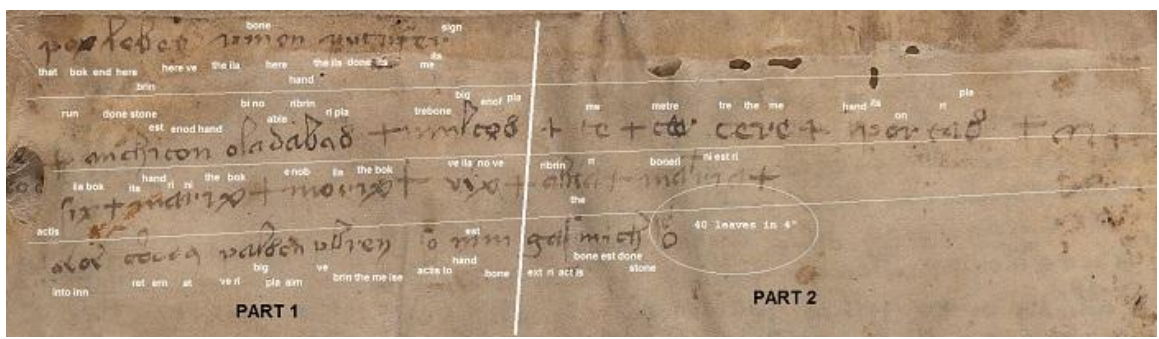

Рис.120. Текст последней страницы манускрипта.

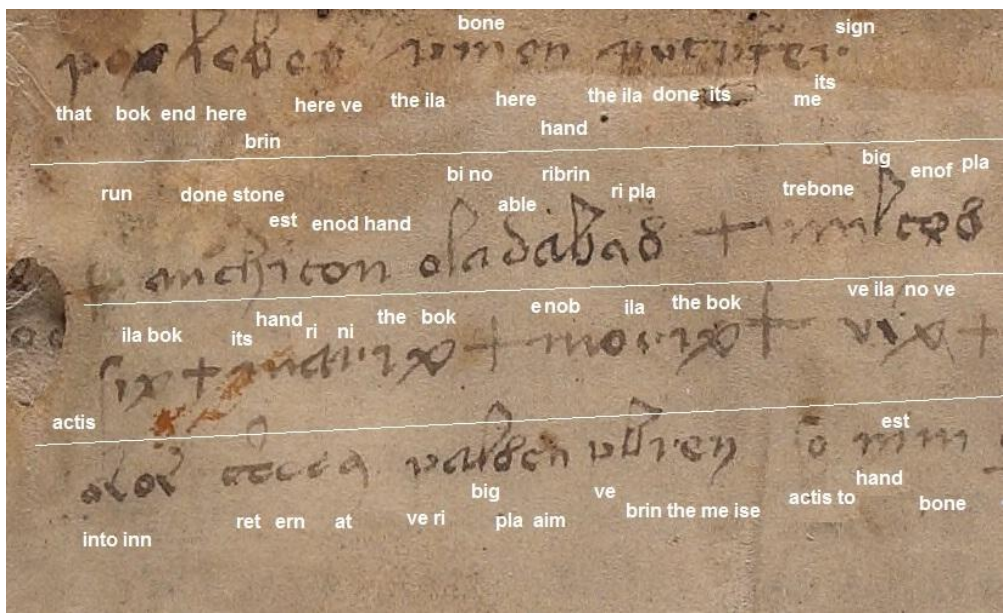

Рис.121 Последняя страница Часть 1

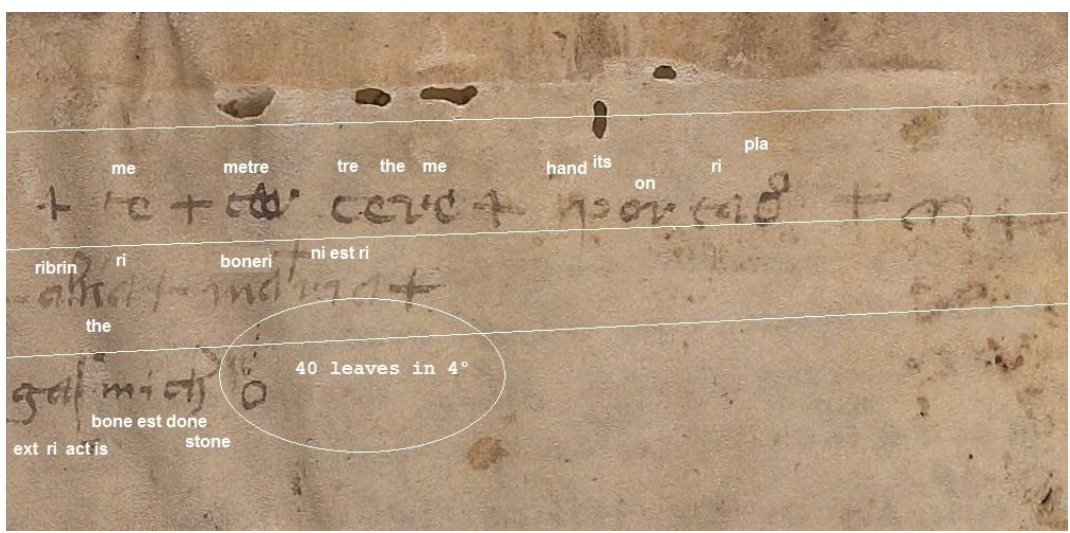

Рис.122 Последняя страница Часть 2

**СТРОКА #1.**

**"That bok end here. Brin here ve the ila bone here hand. The ila done its me its sign"**

В переложении на современный английский (исходя из спеллинга вышеприведенных слов):

**"That book ends here. Bring heavy illness of the bone of the hand. The illness donates me its sign"**

*"Эта книга здесь заканчивается. Получил тяжелую болезнь сустава руки. Эта болезнь награждает меня своими знаками."*

Помните, Ди пишет, что болен.

### СТРОКА #2.

**"Run done stone est enod** (использован символ правила "о" после символа "done") **hand be** (символ правила "о" использован перед символом "bee") **no able ribrin ripla + trebone bee enof pla + me + metre tretheme + hand its on + ripla + ….."**

В переложении на современный английский:

**"Run Dunston asthenic** *(возможно «est enod» - подразумевает «another»)* **hand. Be no able to re-bring or replace. Lay in Trebone big enough. Me. Metre Trithemij. His** *(возможно Келли)* **hand works. Replace***(AU: возможно - replaced)….. "*

**"Пишу Дунстана астенической (другой) рукой. Не имею возможности даже взять (поднять) чего-нибудь или переложить. Лежал в Требоне достаточно долго. Я. Мэтр Тритемий. Его рука работает. Переложил."**

Здесь автор сообщает, что пишет книгу Дунстана «астенической» - т.е. плохо послушной, больной или «другой» рукой. Здесь же автор указывает на свою локализацию в Требоне. Там же, где согласно записям Ди и происходят события 1-12 декабря 1587г. Здесь же автор записи прямо говорит о своем авторстве книги Дунстана и называет Тритемия учителем. Кроме того - теперь понятно, почему почерк последней страницы выглядит сделанным другой рукой – больная рука.

Вообще, такой стиль подачи материала – спонтанность, с акцентами, понятными только ему самому, - характерен для дневниковых записей Ди.

### СТРОКА #3.

**"actis ila bok + its hand ri ni the bok + enob ila the bok + velanove + ribrin the ri + its hand ri + ni est ri + "**

В переложении на современный английский:

**"Working ill with book. His hand running the book. A nobili the book. Vilanova. Re-bring very. Boneri**(*возможно, имеется в*

*виду один из друзей Ди – Edmund Bonner – Епископ Лондонский (к тому времени уже умерший (1569)) – AU)* **. Niestri** *(возможно – "nostri" – лат «наш» – AU)"*

*"Болею, но работаю с книгой. Его рука пишет книгу. Знатную книгу. Вилланова. Переносить очень… Бонэри. Наш."*

Наш автор все еще тяжело болен. «Его рука» - возможно речь о Келли, но Ди иногда в дневниках тоже говорит о себе от третьего лица.

Еще одна характерная черта дневниковых записей Ди – писать на разных языках, даже в рамках одной заметки, преимущественно примешивая к английскому латынь.

Вторая характерная черта дневниковых записей Ди – сокращения слов. В большинстве случаев - понятные.

Здесь мы наблюдаем все эти признаки.

### СТРОКА #4.

**"into inn reternat ve ri big pla aim we brin the me ise actis to handest bone  ext ri actis bone est done stone  . . . . o"**

В переложении на современный английский:

**"After coming back - very big flame.  We bring them ice. Working on hand's bone.  Extractis  bonest Dunstone. 40 leave in 4°"**

*"По возвращении внутрь – очень большой огонь. Мы принесли (использовали для тушения или, возможно, Келли, спасая книгу – получил ожег, и лёд предназначался ему) ему лёд (снег). Разрабатываю сустав руки. Экстракт из благородного Дунстана. $4^0$ (или 40 листов в 4-х стопках)."*

Мы помним, что в записи Ди говорится о том, что Келли оставил заметку в книге Дунстана о вырванных или вырезанных 40 листах.

Так вот, только что мы эту заметку нашли…

Несмотря на все уверения Ди, что манускрипт написан Келли, я продолжаю быть уверен в авторстве Ди. Келли выступал помощником.

## 11.3 Ключи к кодам последней страницы.

**Теперь о ключах последней страницы.** Всего их здесь 35(один пока мне не удалось расшифровать) из которых - 23 новых ключа (!), уникальных только для данной страницы и 1 ключ, пересекающийся со страницей 66r [1]… При том, что для кодирования остальных частей манускрипта, автору понадобилось немногим более 30 ключей.

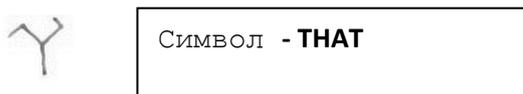

Рис.123 Символ # 31 "THAT"

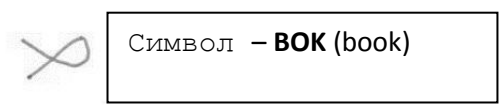

Рис.124 Символ # 32 "BOK"

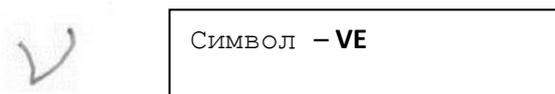

Рис.125 Символ # 33 "VE"

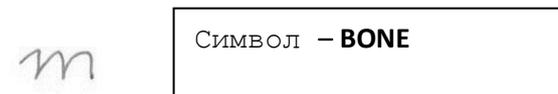

Рис.126 Символ # 34 "BONE"

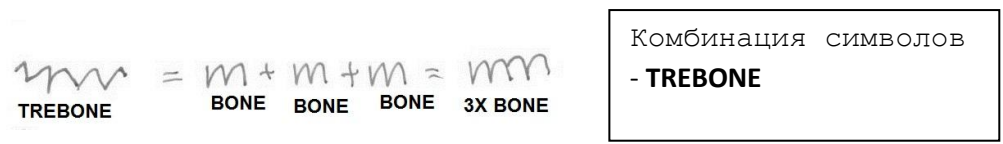

Рис.127 комбинация символов "TREBONE"

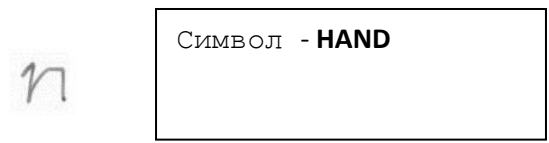

Рис.128 Символ # 35 "HAND"

Символ – **BRIN** (bring)

Рис.129 Символ # 36 "BRIN"

Символ – **NI** [ni:]

Рис.130 Символ # 37 "NI"

Символ – **ITS**

Рис.131 Символ # 38 "ITS"

Символ – **DONE**

Рис.132 Символ # 39 "DONE"

Символ – **EST**

Рис.133 Символ # 40 "EST"

Символ – **THE**

Рис.134 Символ # 41 "THE"

Символ– **ME**

Рис.135 Символ # 42 "ME"

Символ – **STONE** или **STON**

Рис.136 Символ # 43 "STONE"

Символ - **NO**

Рис.137 Символ # 44 "NO"

Символ - **ILA**

Рис.138 Символ # 45 "ILA"

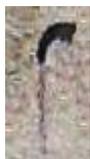
Символ - **ACTIS, ACTUS, ACT, ACTING** – смысл один и тот же при любом варианте

Рис.139 Символ # 46 "ACTIS".

Символ - **EXT**

Рис.140 Символ # 47 "EXT"

Символ- **SIGN**

Рис.141 Символ # 48 "SIGN"

Символ - **ENOF** ( современное - enough)

Рис.142 Символ # 49 "ENOF"

Символ - **ABLE**

Рис.143 Символ # 50 "ABLE"

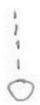 Символ — 40 и/или 4⁰

Рис.144   Символ # 51 "40" и/или "4⁰"

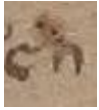 Символ — **AIM**

Рис.145   Символ # 52 "AIM"

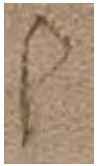 Символ — **BEE** или **BI** или **BIG**

Рис.146   Символ # 53 "BEE"

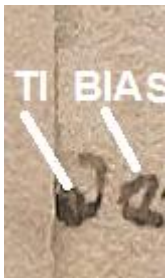 Symbol — "TI"

Symbol — "BIAS"

Рис. 147 Символ #54 "TI"[ti:] и  символ # 55 "BIAS", идентифицированные в главе 10.

Таким образом, теперь подтверждено, что:

**- Манускрипт Войнича прежде назывался книгой Дунстана.**

**- Написан Джоном Ди и Эдвардом Келли в период с 1583г. по 12 декабря 1587г.**

Кстати, касательно "Extractiones Dunstani" – в конце своей работы я сомневаюсь, что листы пергамента содержали что-либо вразумительное.  Скорее всего, это были не удаленные из манускрипта листы, а еще не пошедшие в работу, и при пожаре испорченные огнем. От их использования Ди, вероятнее всего, просто отказался.

## 11.4 Загадка с интересным ответом.

Вернемся еще раз к странице 66r.

Обратим более пристальное внимание на изображение женщины внизу страницы.

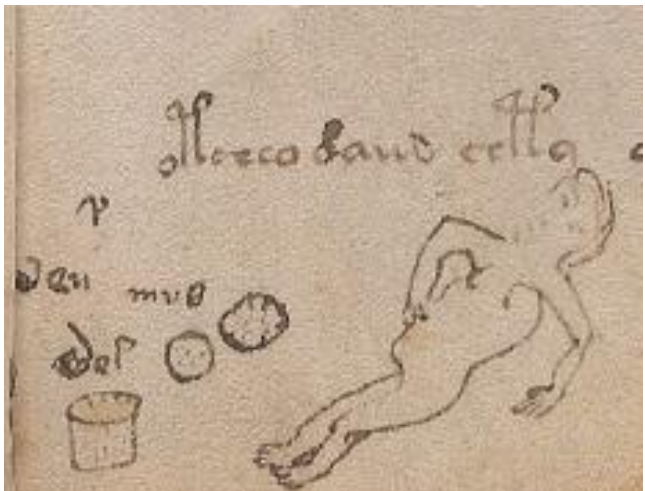

Рис.150 Увеличенное изображение женщины на странице 66r

Казалось бы – женщин (нимф) на страницах манускрипта множество…
Но это, уважаемые читатели, особенная женщина.

Что мы можем о ней сказать?

– Во-первых, для полноты картины нехватает ложа, постели или дивана, на котором возлежит эта женщина. Мысленно добавьте диван с подушками и получите… Вариант обнаженной Венеры работы старых мастеров. Как, например, на Рис. 150 [18].

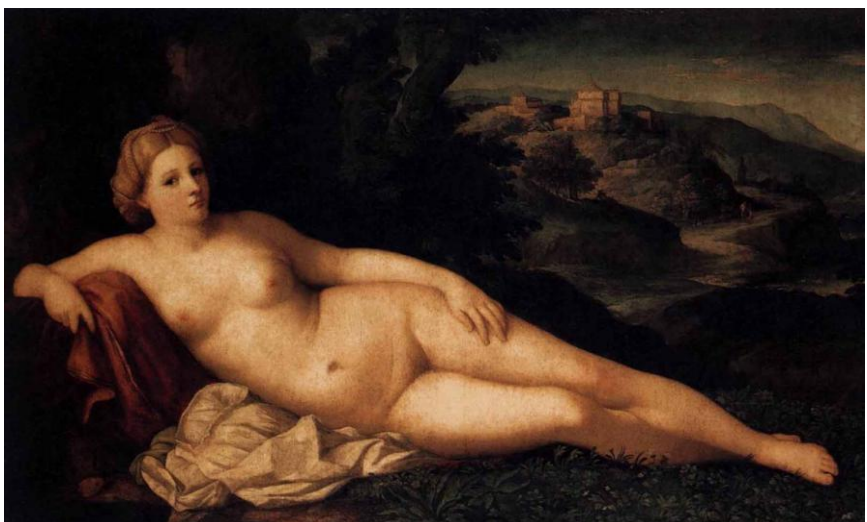

Рис. 151 Пальма иль Веккьо "Венера" [18]

- во-вторых, женщина на Рис. 150 выглядит беременной…
- в-третьих, у этой женщины, несмотря на всю простоту этого наброска - живое, с грустинкой лицо, как будто автор манускрипта рисовал кого-то с натуры.

Похоже, перед нами настоящая женщина. Попробуем вычислить, кто же на рисунке.

Для этого вновь обратимся к дневникам Джона Ди [8].

Запись от 28 февраля 1588г гласит следующее:

«Feb. 28th, mane paulo ante ortum solis natus est Theodorus Trebonianus Dee […]»
«Фев. 28-е, утром, незадолго до восхода солнца родился Теодор Требонариус *(Trebonianus – родившийся в Тршебоне)* Ди […]»

У Джона Ди родился сын.
Мы помним, что последняя запись в дневниках Ди касательно «Книги Дунстана» датирована 12 декабря 1587г… На тот момент мистрис Джейн Ди была уже на седьмом месяце беременности…
Да, уважаемый читатель, это, вероятней всего, рисунок Джейн Ди (Jane Dee) – супруги Джона Ди, исполненный рукой её мужа. Старинного портрета Джейн мне, к сожалению, в интернете найти не удалось. Но удалось найти рисунок современного художника, который, очевидно имел возможность ознакомиться с оригинальными портретами Джейн.
Сходство невероятное.

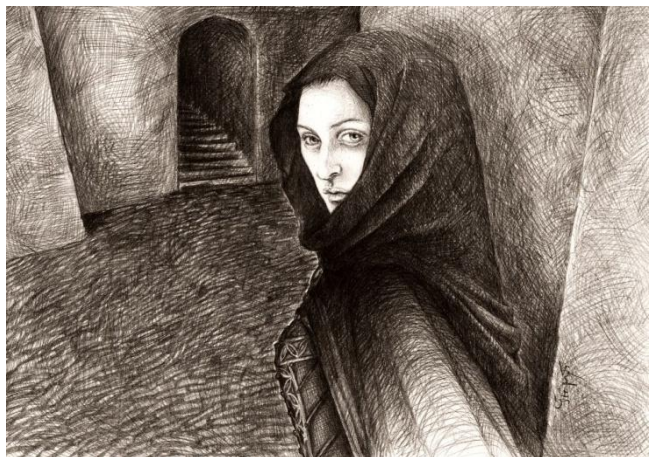

Рис. 152 Джейн Ди работы иллюстратора Meisiluosi [19] – иллюстрация к трейлеру фильма по новелле Люции Лукачовичовой (Lucie Lukačovičová) «Зеленый дракон, красный лев» (Zelený drak, karmínový lev) - исторической фантазии о жизни алхимиков в Праге при Рудольфе II.

Здесь хотелось бы еще раз обратить ваше внимание на «семантический код» подсказки со страницы странице 66r. Подсказка не совсем прямая, но тем не менее:

- рисунки чего-то цилиндрического и окружностей с характерным сечением были определены, как изображения кости – «bone».

- беременная женщина должна, по логике, родить ребенка. Т.е. ребенок родится – baby will "born".

"bone" и "born" – пишутся по разному, но произносятся схоже. Фонетический код…

## 12 Итоги работы.

В этом разделе я лишь хочу отметить наиболее важные детали результатов данной работы, которые могут пригодиться для дальнейших исследований, касающиеся приемов Ди, использованных при создании манускрипта («Книги Дунстана»):

1) "простая стеганография" - использование изображений с маскированными буквами.

2) "модифицированная стеганография" для кодирования основного текста. Вместо букв, как в книге Тритемия, Ди использовал созданные им самим символы, каждый из которых имеет значение простого слова: ("ret", "ort", "as", "key", "ant" etc). Символьная запись позволила усилить шифрование тем, что непосвященный не мог опознать язык манускрипта, и, таким образом вариативность шифра многократно возрастала. Ди ввел символ правила "o", указывавший, как правильно читать соседние символы.

3) Ди использовал случайны метод группировки символов, делая их группы похожими на отдельные слова, что направляло всех будущих дешифровщиков по ложному следу. Особенно это сказалось на проведении статистических исследований текста – они заведомо были ошибочны, ибо использовали неверные начальные данные. Основным правилом при разбиении символов на группы была их последующая однозначная интерпретация и отсутствие искажающего воздействия символов правила на прочтение.

4) Ди, вероятно, первым применил семантическую кодировку.

5) Символ "a" (RI, RY - [ri] & [rri]) – вспомогательный символ правила. Если основной символ "o" имел главное назначение задавать правила для основного текста, Символ "a" – предназначался для маскировки подсказок, оставленных Ди. С использованием данного символа в последовательности "a"+ "другой символ"+ "a"  - оба символа "a" не читаются, а "другой символ" сокращается до последней согласной буквы к которой добавляется буква "i" ([i:]). Пример: "INN" между "RI" читается как "NI" ([ni:]), "ALL" между "RI" читается как -"LI" и т.п.

6) Правило конструирования и прочтении слов – при комбинации символов, первый из которых оканчивается на ту же букву, на которую начинается следующий – эта буква не удваивается -  "VI ILA NOVI" – записывается как "VILANOVI"

7) Ди использовал аллегоричные изображения растений (как на Рис.1 – где основным признаком пригодности растения по отношению к рецепту – раздвоенный ствол, а отнюдь не пестрые листья, которые так и притягивают внимание. Вторым признаком пригодности в ряде случаев является красный цвет той или иной части растения.

8) Ди использовал инвертированные цвета – как в случае с подсолнухом на странице 33v. Прекрасный прием для того, чтобы замаскировать локализацию подсказки.

9) Ди использовал несколько уровней кодирования, как например, латинские буквы разных размеров, цветов и шрифтов, разбросанные по страницам манускрипта, например, странице 4r. Для правильного конструирования и прочтения этих кодов нужно использовать инструкции из модифицированной диаграммы Тритемия со страницы 57v. К сожалению, некоторые страницы, на которые ссылается диаграмма утеряны. А возможно их и не существовало – определенная степень предосторожности Ди на случай разоблачения.

10) Значение цвета разбросанных букв тоже влияет на порядок прочтения – это ясно из текста к диаграмме на стр. 57v.

11) Ди использовал специальные символы, как на странице 8v – в ее правом нижнем углу.

Их появление подчиняется определенному порядку – страницы маркированные данными символами: 8,16,24,32, 42, 58 и т.д. – остальное вы достаточно легко найдете сами.

Мое предположение – эти символьные записи разбивают календарь и гербарий манускрипта на определенные периоды, в которые предполагается проводить работы только с определенным набором растений.

12) Интересна символическая подпись на стр. 11v - по центру слева, выглядящая как "88". В буквенной расшифровке это звучит как "PLAPLA", в цифровой, согласно диаграмме на стр. 57v означает "33". Возможно, это пометка оставленная Келли, при появлении на свет «Книги Дунстана» в 1583г..

13) Изображения (как на последней странице) несущие в себе дополнительный код – римские цифры.

14) Изображения нимф - аллегорическое изображение болезней. На одном из рисунков, отображающих кишечник, одна из нимф держит в руках острый предмет – аллегоричное изображение кишечных колик при проблемах с пищеварением. На рисунке с желчным пузырем – нимфы символизируют камни в желчном пузыре. И т.п.

15) Аллегоричные изображения человеческих органов.

16) Большое количество аллегорий в астрономической и астрологической частях (как в случаях с Овном/Козерогом и Плеядами)

17) Аллегорические изображения в фармацевтической части.

18) Использование закодированных арабских цифр, как на стр. 57v.

19) Специальные отвлекающие ловушки, для направления непосвященных по ложному пути, как, например, на

странице 49v, где Ди специально задает ошибочную визуальную корреляцию между нормальными арабскими цифрами и своим цифровым шифром.

**И еще раз:**
1) Манускрипт Войнича (VMS) = имеет изначальное название «Книга Дунстана» ("BOOK OF DUNSTAN")
2) Манускрипт не является фейком, как таковым, поскольку содержит достаточный объем читаемой структурируемой информации.
3) Тем не менее этот манускрипт – фальшивка, созданная Джоном Ди и, возможно, с участием Эдварда Келли, в период с 1583г по 1587г (возможно 1588г).
4) Написание манускрипта было начато в Англии, закончено в Тршебони (Чехия)

Хочу добавить, что Ди, с высокой вероятностью, не продавал манускрипт Рудольфу и даже не выступал посредником. Манускрипт либо стал подарком императору от Эдварда Келли, либо попросту был конфискован Рудольфом для императорской библиотеки после расправы над Келли.

# 13. Приложение. ВАЖНЫЕ ДАННЫЕ

Высокоразрешающие снимки страниц манускрипта, доступные на сайте библиотеки дали возможность изучить мельчайшие детали некоторых изображений. Ниже представлена одна из наиболее интересных находок.

## 3 КОРОЛЕВЫ

В астрологической части манускрипта (лунном календаре) содержащей 12 круговых диаграмм с изображениями знаков зодиака мы можем выделить 3, отличающиеся от остальных наличием изображений «нимф» с коронами на головах.

1) CANCER / РАК

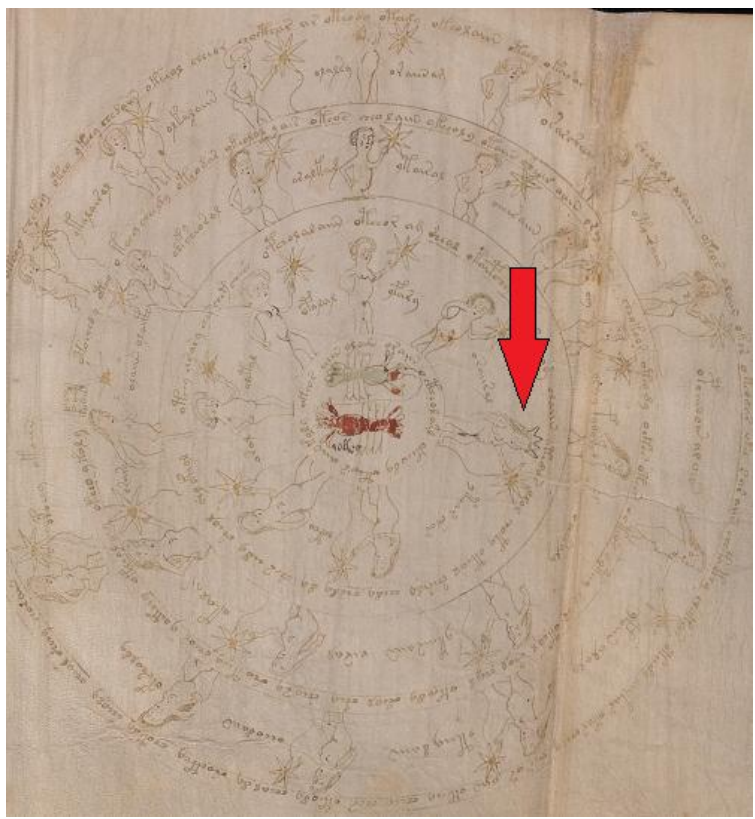
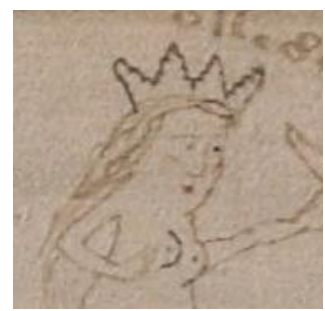

Рис. 154 "Королева" Рака (№1)

Рис. 153 Диаграмма со знаком зодиака Рак и локализацией «королевы».

2) ЛЕВ

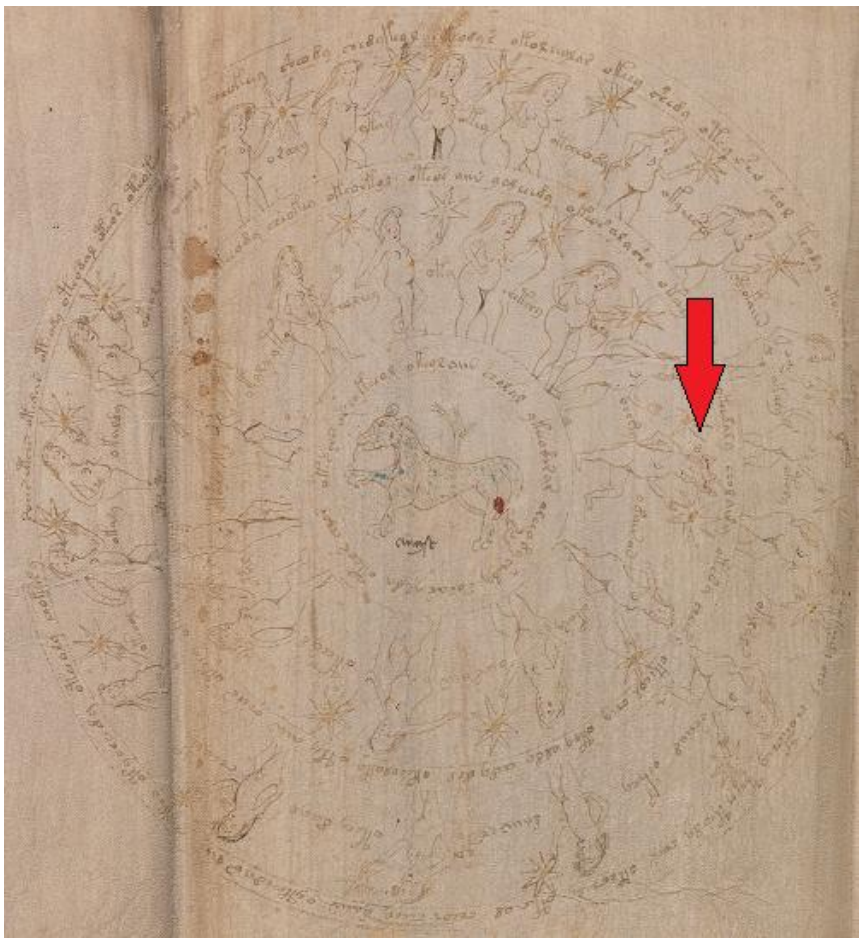

Рис. 155 Диаграмма со знаком зодиака Лев и локализацией «королевы».

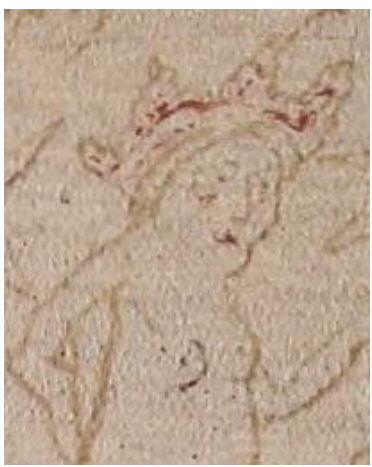

Рис. 156 "Королева" Льва (№2)

И, наконец, самая впечатляющая «королева»…

3) ВЕСЫ

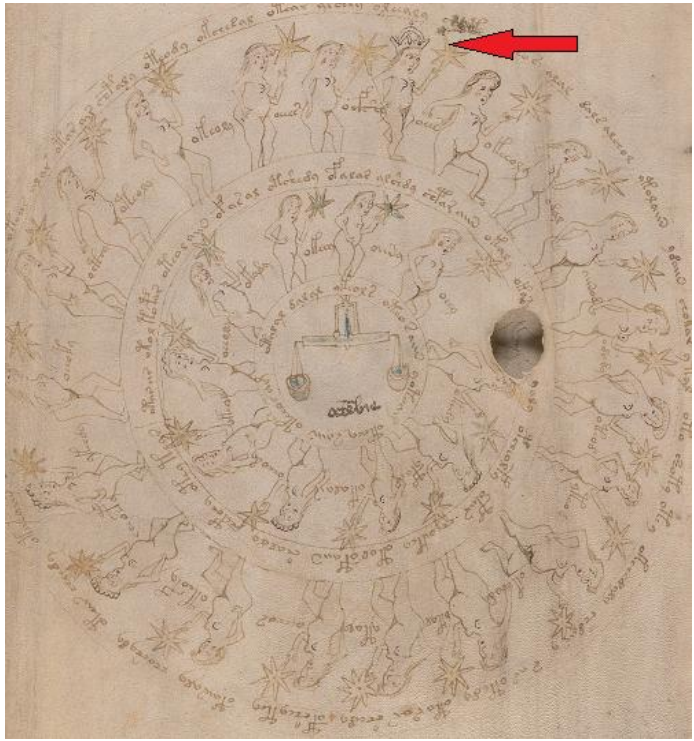

Рис. 157 Диаграмма со знаком зодиака Весы и локализацией «королевы».

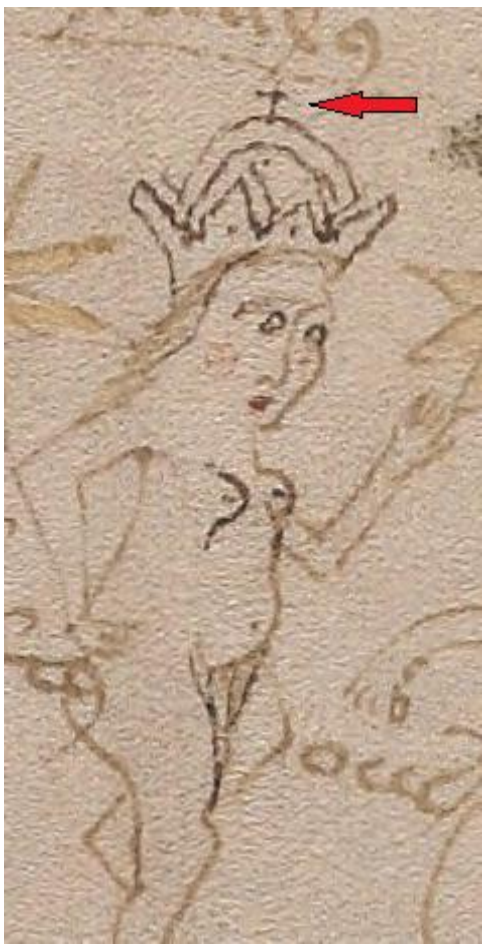

Рис. 158 "Королева" Весов (№3)

Мы видим христианскую корону! Тем самым подтверждая, что автор манускрипта был христианином!

Это однозначно подтверждает, что версии о происхождении манускрипта базирующиеся на идеях про иврит, авторов иудеев, а также арабском, исламском, азиатском и американском происхождении – заведомо маловероятны.

Интересен вопрос – какая корона послужила прототипом изображенной выше.

Ответ достаточно прост – это либо копия с другого изображения, или одна из корон увиденных авторами манускрипта Ди и Келли во время путешествия по континенту – в Польше (корона Батория) или в Праге (корона Рудольфа II).

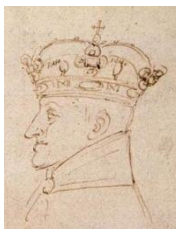

Рис. 159 Корона Рудольфа II на одном из его портретов руки Джузеппе Арчимбольдо (1575, из коллекции Национальной галереи в Праге). [20]

**НАИБОЛЕЕ ИНТЕРЕСНО - КТО ЭТИ 3 КОРОЛЕВЫ и почему их только 3?**

1) Королева РАКА – **Леди Джейн Грей** из рода Тюдор (1537 – 12.02.1554)! Также известная, как Леди Джейн Дадли (после замужества) и как **«королева 9 дней»** - являвшаяся королевой Англии и Ирландии с 10 по 19 июля 1553г.

2) Королева ЛЬВА в <span style="color:red">красной короне</span>, <span style="color:red">красным ртом и красным пятном на уродливом лице</span> это – Мария I Тюдор (18.02.1516 – 17 .11.1558), являвшаяся королевой Англии и Ирландии с июля 1553 до своей смерти в 1558. Она осталась в истории, как «КРОВАВАЯ МЭРИ» (вот почему корона красная и красные пятна на уродливом лице). Джон Ди серьёзно пострадал из-за неё.

3) КОРОЛЕВА ВЕСОВ – это Королева **Елизавета I** (07.09.1533 – 24.03.1603), **Королева-дева, Добрая королева Бесс** - последняя правительница из рода Тюдоров. Истинно Христианская Королева, которая была лояльна ко всем Христианским конфессиям.

Если копнуть глубже, то все окажется еще интереснее. И поможет нам в этом астрономия.

Итак, Джон Ди, в первую очередь, был астрономом. Для данного кодирования он использовал не атрологические понятия знаков зодиака, а информацию о фактическом нахождении солнца в том или ином созвездии на момент вступления во власть каждой из королев. Итак…

**Леди Грей** – вступила на трон, когда солнце физически находилось в созвездии Рака…
<span style="color:red">Более того, на диаграмме мы видим двух раков – КРАСНОГО и БЕЛОГО. Это может символизировать затмение… Леди Грей вступила на трон 10 июля 1553 года. В день КОЛЬЦЕОБРАЗНОГО Солнечного затмения, когда солнце (красный рак) было частично закрыто луной (белый рак)!</span>

**Мария I** – пришла к власти 9 дней спустя, когда солнце физически уже находилось в созвездии Льва…

**Елизавета I** - сменила Марию I 17 ноября 1558г, когда солнце физически находилось в созвездии Весов!

Ровно то, что мы наблюдаем на диаграммах…

<span style="color:red">16 век выделяется этой своей уникальностью, поскольку только в этом веке 3 королевы из рода Тюдоров последовательно сменили друг дружку.</span>

Это еще один факт, подтверждающий авторство Джона Ди.

## 14  Важные замечания.

i) Интерпретация найденных кодов требует дополнительной проверки у лингвистов, специализирующихся на древне- и староанглийском языках и диалектах.

ii) Все приведенные в тексте данной статьи предположения, рассуждения и выводы – являются частным мнением автора.

iii) Касательно более детальных аспектов связи "Стеганографии" Тритемия и манускрипта – я предлагаю читателю самостоятельно обратиться к соответствующим источникам, ибо вникать в аспекты Магии Соломона и соответствующей высшей нумерологии никак не водит (и не входило) в планы автора, ибо это задача далека от целей и смысла данной статьи.

iv) Некоторые примеры декодирования были специально взяты из цитируемых источников, заявлявших об обнаружении способов декодирования отдельных слов или даже страниц.

**Мне часто задают вопросы и присылают просьбы перевести несколько десятков страниц манускрипта.**

**Скажу сразу – я не собираюсь этого делать по следующим причинам:**

**- исходя из того, что манускрипт, как бы он не назывался – «Книга Дунстана» или манускрипт Войнича – подделка, выполненная Ди;**

**- исходя из вышесказанного – текст манускрипта не несет в себе более полезной информации, чем книги, с которых он был списан. А это книги, написанные на нормальном языке, книги, содержащие основы фундаментальной науки. Зачем тратить время на заведомо никому ненужный труд?**

**- найденные коды работают на любой странице манускрипта, каждый может самостоятельно поупражняться в переводе;**

**- содержание манускрипта в целом – весьма унылое и однотипное, отличающееся только принадлежностью к тому или иному разделу. Каждый из разделов представляет собой один сплошной рецепт с монотонной последовательностью – возьми,**

посади, полей, положи в нагреватель, добавь цветки и т.п. – для ботанического; отрежь, помой, положи, нагрей и т.п. – для анатомического и т.д.

## 15 Референции (практически все - англоязычные источники).